\documentclass[a4paper,12pt, epsfig]{article}
\usepackage{epsfig}
\usepackage{amssymb}
\usepackage{amsfonts}
\newskip\humongous \humongous=0pt plus 1000pt minus 1000pt

\newif\ifdtup

\catcode`\@=11

\@addtoreset{equation}{section}
\def\theequation{\thesection.\arabic{equation}}

\def\@normalsize{\@setsize\normalsize{15pt}\xiipt\@xiipt
\abovedisplayskip 14pt plus3pt minus3pt%
\belowdisplayskip \abovedisplayskip
\abovedisplayshortskip \z@ plus3pt%
\belowdisplayshortskip 7pt plus3.5pt minus0pt}

\def\small{\@setsize\small{13.6pt}\xipt\@xipt
\abovedisplayskip 13pt plus3pt minus3pt%
\belowdisplayskip \abovedisplayskip
\abovedisplayshortskip \z@ plus3pt%
\belowdisplayshortskip 7pt plus3.5pt minus0pt
\def\@listi{\parsep 4.5pt plus 2pt minus 1pt
     \itemsep \parsep
     \topsep 9pt plus 3pt minus 3pt}}

\relax

\catcode`@=12

\catcode`\@=11

\def\section{\@startsection{section}{1}{\z@}{3.5ex plus 1ex minus
   .2ex}{2.3ex plus .2ex}{\large\bf}}

\def\thesection{\arabic{section}}
\def\thesubsection{\arabic{section}.\arabic{subsection}}

\def\appendix{\setcounter{section}{0}
 \def\thesection{Appendix \Alph{section}}
 \def\thesubsection{\Alph{section}.\arabic{subsection}}
 \def\theequation{\Alph{section}.\arabic{equation}}}
\def\SymBoxes#1#2#3#4{\newdimen\un@t \un@t#3%
\raisebox{#1}{\rule{#2\un@t}{#4}\hskip-#2\un@t% lower horizontal
\@tempdimb\un@t \advance\@tempdimb by-#4\@tempcntb#2\relax%
\@whilenum{\@tempcntb>0}\do{%                         % #2 vertical lines
\rule{#4}{\un@t}\hskip\@tempdimb \advance\@tempcntb by\m@ne}%
\hskip-#2\un@t \rule[\un@t]{#2\un@t}{#4}%
\rule[\un@t]{#4}{#4}\hskip-#4%             % upper horizontal line
\rule{#4}{\un@t}}\hskip-#4}                % rightest vertical line
\begin{document}
%\begin{letter}{~}

%%%%%%Define some new commands and  macros
\newcommand{\beq}{\begin{equation}}
\newcommand{\eeq}{\end{equation}}
\newcommand{\bea}{\begin{eqnarray}}
\newcommand{\eea}{\end{eqnarray}}
\newcommand{\beas}{\begin{eqnarray*}}
\newcommand{\eeas}{\end{eqnarray*}}
\newcommand{\defi}{\stackrel{\rm def}{=}}
\newcommand{\non}{\nonumber}
\newcommand{\bquo}{\begin{quote}}
\newcommand{\enqu}{\end{quote}}
%%%%%%%%%%%%%%%%
\renewcommand{\(}{\begin{equation}}
\renewcommand{\)}{\end{equation}}
%%%%%%%%%%%%%%%%%%%%%%%%%%%%%%%%%% definitions
\def\de{\partial}
\def\Tr{ \hbox{\rm Tr}}
\def\H{ \hbox{\rm H}}
\def\HE{ \hbox{$\rm H^{even}$}}
\def\HO{ \hbox{$\rm H^{odd}$}}
\def\K{ \hbox{\rm K}}
\def\Im{ \hbox{\rm Im}}
\def\Ker{ \hbox{\rm Ker}}
\def\const{\hbox {\rm const.}}
\def\o{\over}
\def\im{\hbox{\rm Im}}
\def\re{\hbox{\rm Re}}
\def\bra{\langle}\def\ket{\rangle}
\def\Arg{\hbox {\rm Arg}}
\def\Re{\hbox {\rm Re}}
\def\Im{\hbox {\rm Im}}
\def\exo{\hbox {\rm exp}}
\def\diag{\hbox{\rm diag}}
\def\longvert{{\rule[-2mm]{0.1mm}{7mm}}\,}
\def\a{\alpha}
\def\dag{{}^{\dagger}}
\def\tq{{\widetilde q}}
\def\p{{}^{\prime}}
\def\W{W}
\def\N{{\cal N}}
\def\hsp{,\hspace{.7cm}}
\newcommand{\C}{\ensuremath{\mathbb C}}
\newcommand{\Z}{\ensuremath{\mathbb Z}}
\newcommand{\R}{\ensuremath{\mathbb R}}
\newcommand{\rp}{\ensuremath{\mathbb {RP}}}
\newcommand{\cp}{\ensuremath{\mathbb {CP}}}
\newcommand{\vac}{\ensuremath{|0\rangle}}
\newcommand{\vact}{\ensuremath{|00\rangle}                    }
\newcommand{\oc}{\ensuremath{\overline{c}}}
\begin{titlepage}
\begin{flushright}
ULB-TH/06-24\\
hep-th/0610328\\
\end{flushright}
\bigskip
\def\thefootnote{\fnsymbol{footnote}}

\begin{center}
{\large {\bf
What Does(n't) K-theory Classify?
 } }
\end{center}

\bigskip
\begin{center}
{\large  Jarah EVSLIN\footnote{\texttt{ jevslin@ulb.ac.be}}}\\
\end{center}

\renewcommand{\thefootnote}{\arabic{footnote}}

\begin{center}
\vspace{1em}
{\em  { International Solvay Institutes,\\
Physique Th\'eorique et Math\'ematique,\\
Statistical and Plasma Physics C.P. 231,\\
Universit\'e Libre
de Bruxelles, \\ B-1050, Bruxelles, Belgium\\}}

\end{center}

\noindent
\begin{center} {\bf Abstract} \end{center}
We review various K-theory classification conjectures in string theory.  Sen conjecture based proposals classify D-brane trajectories in backgrounds with no $H$ flux, while Freed-Witten anomaly based proposals classify conserved RR charges and magnetic RR fluxes in topologically time-independent backgrounds.  In exactly solvable CFTs a classification of well-defined boundary states implies that there are branes representing every twisted K-theory class.  Some of these proposals fail to respect the self-duality of the RR fields in the democratic formulation of type II supergravity and none respect S-duality in type IIB string theory.  We discuss two applications.  The twisted K-theory classification has led to a conjecture for the topology of the T-dual of any configuration.  In the Klebanov-Strassler geometry twisted K-theory classifies universality classes of baryonic vacua.

%We review various K-theory classification conjectures in string theory.  Twisted K-theory classifies conserved D-brane charges in type II string theories on spacetimes in which the topology of each spatial slice is fixed and noncompact, while it classifies magnetic RR fields when each slice is compact.  It classifies symmetric D-branes in WZW models and also in certain 2d topological theories.  In the time-dependent case it classifies D-brane trajectories modulo a class of deformations which generalize tachyon condensation.  In the case of the Klebanov-Strassler cascade, these equivalence classes of configurations are the universality classes of the gauge theory vacua.  We review two problems that plague many K-theory classification schemes, the failure of Hodge self-duality of the RR fields and the lack of S-duality covariance, and we discuss progress towards their resolutions.

\begin{center}
\vspace{.4cm}
{Prepared for the Second Modave Summer School in Mathematical Physics}

{\hspace{6.9cm}Halloween, 2006}
\end{center}

%We argue that, in 
%type II string theory with a trivial NSNS $H$-flux, a brane wrapping a cycle represents a class in K-theory if and 
%only if it is free from the Freed-Witten anomaly and the cycle if representable as a smooth manifold.  Further we 
%claim that there is such a brane for every K-theory class.  We provide an example on a D6-brane that neither corresponds to a K-theory class nor suffers from a Freed-Witten anomaly, and we show that the wrapped 
%7-cycle cannot be represented by a smooth manifold.  While the spacetime is smooth, representatives of the wrapped 
%7-cycle contain a 2-dimensional singularity which is a real cone over $\cp^2$.  We argue that even in noncritical 
%CFTs all obstructions to lifting a class to K-theory are either Freed-Witten anomalies or obstructions to the 
%existence of a smooth representative.

\vfill

%\begin{flushright}
%\today
%\end{flushright}
\end{titlepage}
\bigskip

\hfill{}
\bigskip

\tableofcontents

\setcounter{footnote}{0}
\section{Introduction}

\noindent
In 1997 Minasian and Moore suggested that Ramond-Ramond charges in type II string theories are classified by K-theory \cite{MM}, which is a group that characterizes the complex vector bundles on a given spacetime.  Ramond-Ramond (RR) fluxes are sourced by RR charges, in the same way that the magnetic field is sourced by magnetic monopoles in QED.  In QED the well-definedness of the partition function of the dual electric charges implies that the magnetic field is quantized. Combining this quantization condition with Gauss' Law one finds that magnetic charge cannot be smeared at will but must instead be localized on a codimension three surface in spacetime, which is the trajectory of a magnetic monopole.  Similarly in type II string theory the well-definedness of the partition function of dual RR charges, which are also RR charges, imposes that the RR field strengths are quantized.  This in turn implies that RR sources are localized on submanifolds, which are called D-branes.  And so Minasian and Moore's conjecture is that D-brane configurations are classified by K-theory.

Since 1997 many versions of this conjecture have appeared, classifying various features of D-brane configurations in terms of different K-theories on distinct subspaces.  Some use the K-theory of the spacetime, some use the K-theory of a spatial slice, some use relative K-theory.  In these lectures we will review several of these proposals, their physical motivations, and their ranges of validity.  

We begin in Sec.~\ref{JERISCALDA} with the older, homology classification of D-branes.  In Sec.~\ref{JESUGRA} we provide some background material on charges and fluxes in type II supergravity theories and explain why the homology classification fails.

Finally we come to K-theory in Sec.~\ref{JESEN}.  We first describe Witten's proposal \cite{WittenK} for a classification of D-brane trajectories in type IIB string theory.  This classification scheme is a consequence of the Sen conjecture \cite{Sen}, which states that all D-brane configurations in type IIB can be realized as gauge field configurations on a stack of spacefilling branes.  Distinct D-brane trajectories are identified if they are related by tachyon condensation.  Here tachyon condensation is not interpreted as a physical process, as it relates different trajectories and not states on different time slices.  Instead in concrete realizations, configurations before and after tachyon condensation are related by an RG flow.  Thus K-theory classifies universality classes of the theories which are described by various vacua of the UV theory.  While this physical interpretation of the K-theory classification conjecture is perhaps the most popular to date, its range of validity is limited.  For example, spacefilling branes are inconsistent in backgrounds with nontrivial $H$ flux \cite{Kapustin}.  In addition the spacetime is taken to be compact, but then in the case of nontorsion K-theory classes, which are the branes that continue to be stable in the classical limit, there is the usual problem that the sourced fluxes have nowhere to go and so no corresponding supergravity solution exists.  

Instead of classifying D-brane trajectories, many authors use K-theory to classify D-brane charges at a fixed moment in time.  The first were Moore and Witten \cite{MW}, who used relative K-theory to classify the charges on a noncompact spatial slice.  Diaconescu, Moore and Witten \cite{DMW} found that some D-branes which wrap nontrivial homology cycles but carry no K-theory charge can decay.  This violation of D-brane charge conservation is a result of the Freed-Witten anomaly, which we describe in Sec.~\ref{JEFWSEZ}.  As we will review in Sec.~\ref{JEKFW}, using this idea Maldacena, Moore and Seiberg \cite{MMS} (MMS) were able to interpret the K-theory classification as a classification of stable D-branes, in other words, they argue that K-theory classifies the set of conserved RR charges.  The naive group of conserved charges, the homology of a spatial slice, contains charges whose conservation is violated by the Freed-Witten anomaly.  MMS were therefore able to use the Freed-Witten anomaly to test the K-theory classification in several cases, some of which will be reviewed in Sec.~\ref{JEWZW}.  This strategy has an advantage over the tachyon condensation strategy in that it can be extended to configurations with nontrivial $H$ flux, where one finds twisted K-theory \cite{BM}.  Ramond-Ramond fluxes are sourced by D-branes, and so Moore and Witten argued \cite{MW} that these fluxes should also be classified by twisted K-theory.

The above classification schemes suffer from two major shortcomings, which are the subject of section \ref{JEPROBLEMI}.  The first is that in the democratic formulation of type II supergravity \cite{Townsend,GHT,VanProeyen} the set of Ramond-Ramond field strengths is self-dual.  More precisely, the $p$-form RR improved field strength is the Hodge dual of the $(10-p)$-form improved field strength.  The Hodge duality operator, which is called the Hodge star, is a matrix which depends continuously on the metric and so its components are in general irrational.  However according to the K-theory classification, the RR field strengths are Chern characters which are rational.  Therefore the Hodge dual of a rational field strength component is an irrational number, which is in contradiction with the fact that RR field strengths should be rational.  The solution to this problem is to choose half of the field strengths to be interpreted by K-theory, and to let the other half not be quantized.  This is the way in which, for example, the chiral scalar field theory is defined.  When the topology of each spatial slice of spacetime is time-independent there is a natural choice of a half of the fields to be classified by K-theory, one can choose the magnetic half.  The cure to the second shortcoming is still unknown.  We will see that all of the schemes relevant to type IIB superstrings suffer from a lack S-duality covariance, and so must at some level fail.

Twisted K-theory also classifies branes in some string theories besides type II.  The twisted K-theory classification of D-branes in WZW models is already well established.  Recently Moore and Segal \cite{MS} have also found a K-theory classification of branes in the target space of some simple 2-dimensional topological gauge theories, although the conjecture for A and B model topological string theory branes and pure spinors has not yet formally appeared.

Finally in section \ref{JEAPPLICAZIONI} we will describe two applications of the K-theory classification.  First, we will see how it led to a formula for the topology of the T-dual of any spacetime.  Next we will calculate the twisted K-theory of the Klebanov-Strassler geometry and find that the S-dual twisted K-classes correspond to universality classes of worldvolume gauge theories on  D-branes.

\section{Warmup: The Homology Classification} \label{JERISCALDA}

In these lectures we will review some of the most common K-theory classification schemes.  We begin with an older classification scheme, the homology classification of D-branes, which is still widely used in the literature despite the fact that it includes some unphysical and some unstable branes.  The failure of this scheme to reproduce the known gauge theory operators in the case of D-branes on AdS${}^5\times\rp^5$ was observed by Witten in \cite{BBA} and led to the discovery of the Freed-Witten anomaly \cite{FW}, which is the basis of the modern understanding of the K-theory classification scheme that will be discussed in Sec. \ref{JEKFW}.

Like the K-theory classification of D-branes, there are many variations of the homology classification scheme, some of which classify D-brane charges, some of which classify D-brane trajectories and some of which classify RR fluxes, which are the fields sourced by D-branes.  For concreteness we will restrict our attention to the homology classification of D-brane charges.  In particular, we will consider type II string theory on a spacetime whose topology is of the form $\R\times M$ where $\R$ is the time direction and $M$ is a compact 9-manifold.  Notice that this choice implies that the topology of each spatial slice $M$ is time-independent, and so we are not allowing processes in which the topology of spacetime changes.  However the metric of $M$ is allowed to change, so the universe can be expanding for example.  Thus eternal inflation may be allowed, but a big bang which starts with the universe at a point is not contained in our ansatz.  A big bang would be allowed in a classification of D-brane trajectories, for example in the K-theoretic classification scheme presented in Sec.~\ref{JESEN}.

We will not consider S-branes, as the full spectrum of D-brane charges is carried by ordinary D-branes.  It will instead suffice to consider D$p$-branes that extend along the time direction and also wrap a $p$-dimensional submanifold of the 9-manifold $M$.  Each D$p$-brane will then carry a charge which depends on the cycle that is wrapped.  

\subsection{The Homotopy Classification}

Following the basic strategy of \cite{MMS}, we are interested in classifying charges that satisfy two conditions.  First, we impose that some D$p$-brane that carries the charge be consistent.  In the case of the homology classification it suffices to impose that the brane has no boundary, and so the wrapped $p$-submanifold has no boundary.  $p$-submanifolds without boundaries are called $p$-cycles.  Second, we impose that the charge is conserved.  This implies, for example, that two $p$-cycles which are related by a small, continuous deformation should correspond to the same charge, since a D-brane wrapping one could move to the other and so while the number of branes wrapping either cycle individually is not conserved by this process the sum of these numbers is conserved.  Thus D$p$-brane charges appear to be classified by $p$-cycles where two $p$-cycles correspond to the same charge if one can be deformed to another.  Such cycles are said to be homotopic, and this set of charges is called the set of homotopy classes of $p$-cycles.

\begin{figure}[ht]
\begin{center}
\leavevmode
\epsfxsize 11   cm
\epsffile{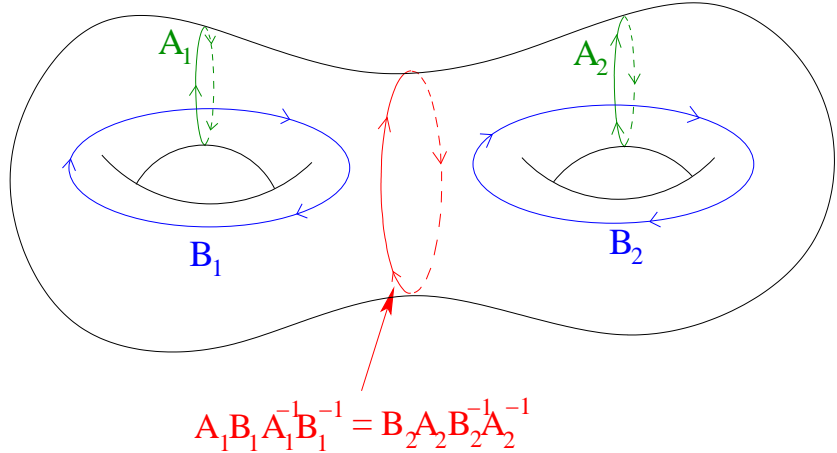}    
\end{center} 
\caption{Loops on the genus 2 Riemann surface.  A D1-brane wrapping the red loop in the middle can decay by moving steadily to the right, but it will need to split into two loops wrapping $A_2$ with opposite orientations to pass the handle, which then rejoin once the handle has been passed.  After the rejoining the brane wraps a contractible cycle and it can shrink into nothingness.  The nontrivial homotopy charge of the brane is not conserved by this process.}
\label{JERIEMANN} 
\end{figure}

In fact we know that the charges corresponding to homotopy classes of $p$-cycles are not always conserved.  Consider a simplified case in which $M$ is the product of a 7-manifold and a 2-manifold.  Let the 2-manifold be the Riemann surface $\Sigma_2$ of genus 2 and ignore the 7-manifold.  We can wrap a D1-brane around any loop on the Riemann surface.  In particular, we can wrap a D1-brane around the red loop in the center of Figure~\ref{JERIEMANN}.  The homotopy classes of loops on the genus 2 Riemann surface are elements of a nonabelian group called the fundamental group of the surface.  The fundamental group is generated by four elements $A_1, A_2, B_1$ and $B_2$ which satisfy a single relation: the product $A_1B_1A_1^{-1}B_1^{-1}A_2B_2A_2^{-1}B_2^{-1}$ is equal to the identity element
\beq
\pi_1(\Sigma_2)=<A_1,A_2,B_1,B_2|A_1B_1A_1^{-1}B_1^{-1}A_2B_2A_2^{-1}B_2^{-1}=1>.
\eeq  

The loop in Figure~\ref{JERIEMANN} represents the nontrivial element $A_1B_1A_1^{-1}B_1^{-1}$ of the fundamental group.  Nonetheless, a brane wrapping this loop does not carry a conserved charge because it can decay via the following process.  At each moment in time, the brane can move a little to the right.  Once it reaches the handle it will need to divide into two loops, one wrapping $A_2$ with each orientation.  Then, further into the future, once they have passed the handle completely, the two loops will coalesce and the brane will wrap a contractible loop on the right hand side.  This loop will then shrink into nothingness and the brane will decay.

Thus branes wrapping some nontrivial homotopy classes do not carry conserved charges because the branes can decay after some processes in which they change their topology, like the splitting of one loop into two which then coalesce that occurs in this example.  To obtain a classification of conserved charges, one must identify the unstable brane charges with the zero element of the charge group.  Unstable branes, at least D$p$-branes that can decay via the above process, are those which wrap $p$-cycles that are boundaries of $(p+1)$-dimensional submanifolds of $M$.  The $(p+1)$-dimensional submanifold is the space swept out by the decaying brane, for example the right hand side of the Riemann surface $M$ is a 2-dimensional submanifold of $M$.  The group of $p$-cycles modulo cycles which are boundaries of $(p+1)$-submanifolds is called the $p$th homology group of $M$, and is often denoted $\H_p(M)$.

\subsection{The Homology Classification}

In the above example, D$1$-branes are classified by the first homology group of the genus 2 Riemann surface, which is
\beq
\H_1(\Sigma_2)=\Z^4
\eeq
the additive group of quadruples of integers.  This is an abelian group which is generated by the same generators $A_1, A_2, B_1$ and $B_2$ as the homotopy group.  In the abelian group, the element $A_1B_1A_1^{-1}B_1^{-1}$ is trivial, as the $A_1$ and $B_1$ commute and so the $A_1$ can be moved one space to the right where it annihilates its inverse.  Thus the D1-brane wrapping the cycle in Figure~\ref{JERIEMANN} represents the trivial element of $\H_1(M)$ and so carries no conserved homology charge, which is consistent with the fact that it can decay.

On any manifold, homology groups are finitely generated abelian groups.  This means that they can always be written as the sum of copies of the integers and copies of the integers modulo powers of a prime number
\beq
\H_p(M)=\Z^{b_p}\oplus_i\Z_{p_i^{k_i}}.\label{JEINT}
\eeq
The number $b_p$ of powers of the integers that appears in the $p$th homology class is called the $p$th Betti number.  The second term, which is a sum of finite order cyclic groups, is said to be the torsion part of the homology group, while the $\Z^{b_p}$ term is called the free part.  In the case of the genus 2 Riemann surface we have seen that the 1st Betti number is equal to 4, meaning that D1-branes on this surface can carry four distinct kinds of conserved charges, all of which are quantized.

None of the homology groups of a Riemann surface contains a torsion part.  One example of a manifold with a torsion part is the three-dimensional real projective space $\rp^3$, which is the quotient of the 3-sphere $S^3$ by the antipodal map, which generates a $\Z_2$ symmetry.  The homology groups of $\rp^3$ are
\beq
\H_0(\rp^3)=\Z,\ \ \ \H_1(\rp^3)=\Z_2,\ \ \ \H_2(\rp^3)=0,\ \ \ \H_3(\rp^3)=\Z \label{JERP3OM}
\eeq
where $0$ is the trivial group which only contains the identity element.   Now we can use the homology classification to classify conserved charges of D-branes wrapping a $p$-cycle in $M$ where $M$ is the product of $\rp^3$ and an irrelevant 6-manifold.  As above we will restrict our attention to branes that are a single point on the 6-manifold.

The above homology classes (\ref{JERP3OM}) tell us that in type IIA, where there are only even-dimensional D-branes, the only available conserved charge is the D$0$-brane charge, which is an integer.  The fact that the second homology group is the trivial group means that all D$2$-branes on $\rp^3$ can decay.  On the other hand in type IIB there are odd-dimensional branes.  D$3$-branes can wrap the whole of $\rp^3$ and these carry a conserved charge which measures the number of times that the $\rp^3$ is wrapped.  The new ingredient in this example is $\H_1=\Z_2$, which contains the torsion term $\Z_2$.  This means that D$1$-branes on $\rp^3$ carry a $\Z_2$ torsion-valued conserved charge.  Physically, $\rp^3$ contains a single nontrivial loop which can be represented, for example, by a meridian that extended from the north pole to the south pole in the $S^3$ before it was quotiented by $\Z_2$.  After the quotient the north and south poles are identified and so the meridian becomes a loop.  A D$1$-brane wrapping this loop a single time is stable.  However if it wraps the loop twice, the brane is the image of a brane that extends from the north pole to the south pole and back on the original $S^3$.  Like any loop on a sphere, this loop can be contracted out of existence, and so the brane can decay.  A brane that wraps the loop twice is the same as, or can turn into, two branes that wrap the loop once.  Therefore a single brane wrapping the nontrivial loop in $\rp^3$ is stable, but if two come into contact then they can annihilate.

More generally, if the $p$-th homology group contains a $\Z_n$ factor then there is an interaction in which $n$ coincident D$p$-branes annihilate.  Such $\Z_n$-charged objects are familiar in gauge theories, for example in $U(n)$ pure Yang-Mills there are $\Z_n$-charged strings, and in $\N=1$ supersymmetric pure Yang-Mills there are $\Z_n$-charged Douglas-Shenker strings \cite{DS}.  In examples these strings attach to each other and form bound states with a binding energy that increases for each string that is bound.  When $n$ strings are bound then the binding energy is equal to the total energy and all of the strings decay into radiation.  For this reason torsion charged D-branes can never preserve any supersymmetry.  More generally, degree $n$ torsion cycles can never be calibrated because $n$ times a calibrating cycle is a boundary and so is homologous to the empty set, and so if a cycle minimizes the integral of the calibrating form then $n$ times the integral of the calibrating form is at most equal to the integral over the empty set which is equal to zero.  However the homology classification indicates that D-branes wrapping torsion cycles may nonetheless be stable.  Thus the homology classification is in a sense more powerful than a classification of conserved charges based on supersymmetry, because there exist conserved homology charges which are carried by objects which are stable but are never BPS.

\subsection{Generalizations: Other Coefficients and Cohomology} \label{JEGEN}

The homology that we have discussed so far is a particular kind of homology, known as homology with integer coefficients or simply as integral homology.  Representatives of integral homology classes are not precisely submanifolds for two reasons.  First, they are sometimes singular, as we will discuss further in subsection \ref{JEMMS}.  A homology class whose representatives are all singular is said to be nonrepresentable.  On a manifold of dimension nine or less every homology cycle is representable, and so in the classification of conserved D-brane charges we can think of homology classes as nonsingular submanifolds.  However in 10-dimensions there can be nonrepresentable 7-dimension homology classes, although some multiple of the cycle is always representable.  Thus later when we classify D-brane trajectories, in particular the 7-dimensional surfaces swept out by D6-branes in IIA, and also when we look at WZW models, which can have arbitrarily large dimensions, this distinction will play a role.  D-brane physics on nonrepresentable cycles is poorly understood, and we will see that the K-theory classification makes concrete predictions as to which singular submanifolds can be wrapped and which cannot.

The second difference between an integral homology class and a submanifold is that integral homology classes come with weights, which are integers.  When the weights are positive they can be thought of intuitively as winding numbers of the submanifold around some cycle, but the weights can also be negative.  These weights are the charges, thus negative weights can be thought of as winding numbers for anti D-branes.  So a homology class contains a little more information than just the submanifold which is wrapped, it also contains integers which are the charges of the objects doing the wrapping.

In quantum gauge theories the charges are integers because of the Dirac quantization condition.  However there are other physical theories in which the charges are not integers.  For example, in the classical Maxwell theory electric charges are real numbers.  Thus charges in classical electrodynamics are not of the form (\ref{JEINT}) and so are not classified by integral homology.  More generally, in the classical supergravity limit of a superstring theory there is no Dirac quantization condition and so charges are not classified by integral homology.  

There is another kind of homology, called real homology or homology with real coefficients, that does classify $p$-branes in supergravity.  The term ''real coefficients'' means that real homology classes are submanifolds weighted by real numbers.  In general one can define homology with coefficients valued in any abelian group $G$.  When we want the choice of abelian group to be explicit, we will write
\beq
\H_p(M;G)
\eeq
for the $p$th homology of $M$ with coefficients in $G$.  This is always the quotient of the kernel of the boundary map with coefficients in $G$ by its image with coefficients in $G$.  Thus it will always be a quotient of a subgroup of some copies of $G$ by another subgroup of some copies of $G$.

One can obtain real homology from integral homology by tensoring with the real numbers, which kills the torsion part but leaves the Betti numbers $b_p$ invariant
\beq
\H_p(M;\R)=\H_p(M;\Z)\otimes\R=(\Z^{b_p}\oplus_i\Z_{p_i^{k_i}})\otimes\R=\R^{b_p}. \label{JETENSORE}
\eeq
Thus torsion charged D-branes are unstable in the classical limit.  Physically their decay process is easy to understand.  If a D$p$-brane is of charge one under a group
\beq
\H_p(M;\Z)=\Z_n
\eeq
then in the quantum theory it is absolutely stable, as it wraps a noncontractible cycle and it cannot split into smaller pieces as its charge, one, is defined to be the lowest charge compatible with the Dirac quantization condition.  

However in the classical theory there is no Dirac quantization condition and so without even moving it can divide into $n$ branes of charge $1/n$ that wrap the same cycle.  These branes can then reattach in a different way, again without moving, to form one brane of charge $1/n$ that wraps the cycle $n$ times.  However the fact that the homology is $\Z_n$ means that a submanifold which wraps the generator $n$ times is the trivial element of homology and so can be deformed, maybe after nucleating some other branes as we saw on the Riemann surface, into nothingness.  Thus the long charge $1/n$ classical brane may decay.  The fact that all $p$-branes are unstable in this background reflects the triviality of the real homology group
\beq
\H_p(M;\R)=\H_p(M;\Z)\otimes\R=\Z_n\otimes\R=0
\eeq
where $0$ is the group consisting of only the identity element, which corresponds to zero charge.

Similarly in the classical theory every $p$-brane can be considered to wrap a representable cycle.  This is because for every nonrepresentable cycle there is some multiple $N$ of the cycle which is representable.  One can then divide the charge of the classical $p$-brane by $N$ and declare that it wraps the cycle $N$ times.  A small deformation then makes the brane honestly wrap a nonsingular cycle.

In general the relationship between homology groups with different coefficients is not so simple as in Eq.~(\ref{JETENSORE}).  Homology groups with different coefficients are related by a mathematical structure which is called an exact sequence of abelian groups, or simply an exact sequence.  An exact sequence consists of a series of abelian groups $G_i$ indexed by an integer $i\in\Z$ with homomorphisms
\beq
f_i:G_i\longrightarrow G_{i+1}
\eeq
such that the image of each homomorphism is the kernel of the next
\beq
\textup{Im}(f_i:G_i\longrightarrow G_{i+1})=\textup{Ker}(f_{i+1}:G_{i+1}\longrightarrow G_{i+2}).
\eeq
An exact sequence is usually represented by a list of groups separated by arrows with the functions over the arrows
\beq
...\stackrel{f_{i-1}}{\longrightarrow}G_i\stackrel{f_{i}}{\longrightarrow}G_{i+1}\stackrel{f_{i+1}}{\longrightarrow}G_{i+2}\stackrel{f_{i+2}}{\longrightarrow}... .
\eeq

If all of the abelian groups are the trivial group $0$ except for three consecutive groups
\beq
0\stackrel{0}{\longrightarrow}G_0\stackrel{f_{0}}{\longrightarrow}G_{1}\stackrel{f_{1}}{\longrightarrow}G_{2}\stackrel{0}{\longrightarrow}0 \label{JESES}
\eeq
then one calls the sequence a short exact sequence.  In this case $0$ is the zero map, whose image is $0$.  Exactness implies that $f_0$ is into and that $f_1$ is onto.  Thus elements of $G_1$ are roughly pairs of elements from $G_0$ and $G_2$, but the additive structures are mixed.  An example of a short exact sequence is 
\beq
0\stackrel{0}{\longrightarrow}\Z\stackrel{\times 2\pi}{\longrightarrow}\R\stackrel{exp}{\longrightarrow}U(1)\stackrel{0}{\longrightarrow}0
\eeq
where $\times 2\pi:\Z\longrightarrow\R$ is multiplication by $2\pi$ and $exp:\R\longrightarrow U(1)$ is the exponential map $r\mapsto e^{ir}$.  The kernel of the exponential map consists of all multiples of $2\pi$, which is the image of the previous map and so the sequence is exact.

Short exact sequences will be useful for us because given any short exact sequence of abelian groups (\ref{JESES}) one can construct a long exact sequence of homology groups which have those groups as coefficients.  In what follows we will be interested less in homology groups than in cohomology groups, which are like homology groups but one replaces the boundary map with its transpose which is called the coboundary map.  Weighted submanifolds which are closed under the coboundary are called cocycles while those in the image of the coboundary map are called cochains. The $p$th cohomology group $\H^p$ is then defined to be the quotient of the $p$-cocycles by the $p$-cochains.  When the $n$-dimensional space $M$ is orientable, as it always is in type II string theory compactifications, there is a theorem called Poincar\'e duality which states that the $p$th homology and $(n-p)$th cohomology groups are isomorphic
\beq
\H_p(M;G)=\H^{n-p}(M;G).
\eeq
A short exact sequence of abelian groups also induces a long exact sequence of cohomology groups, which we will use in Subsec.~\ref{JEFWS} when discussing the Freed-Witten anomaly.

Like homology groups, cohomology groups also may be defined using any abelian group of coefficients.  When one uses real coefficients the elements may be represented by differential forms and they generate a ring whose product is the wedge product $\wedge$ of differential forms.  

In a quantized quantum theory instead one uses integral cohomology classes, which in general cannot be represented by differential forms of a single degree.  Instead a degree $p$ class can be represented by a collection of differential forms, one in each degree between 0 and $p$, with a series of gauge equivalence relations.  In the mathematics literature this collection is called a Deligne cohomology class, named after a ULB graduate.  Physically the collection of forms corresponds to the tower of ghosts of ghosts familiar in the quantization of $p$-form electrodynamics.  Integral cohomology is also a ring and the product is known as the cup product $\cup$.  However for simplicity in the sequel we will often abuse the notation and pretend that integral classes are represented by differential forms and write the wedge product $\wedge$ instead of the cup product.

\section{Fluxes and Worldvolume Actions} \label{JESUGRA}

\subsection{The Failure of the Homology Classification}

The homology classification of D-branes has met with a great deal of success.  However in general it suffers from a number of shortcomings.  First, some homology cycles cannot be represented by any smooth submanifold and so any brane carrying such a homology charge would necessarily have a singularity, even if the spacetime is nonsingular.  D-branes wrapped on such cycles were first considered in Ref.~\cite{BHK} in a study of the Freed-Witten anomaly and later in Ref.~\cite{Szabo} in the context of the K-theory classification.  It was argued that sometimes, but not always, such wrappings are inconsistent in Ref.~\cite{ES}.  

When $M$ is 9-dimensional or less, as is the case in the classification of conserved charges in critical string theories considered above, all homology classes can be represented by smooth submanifolds.  But it may still be that a brane wrapping such a submanifold is necessarily anomalous.  Freed and Witten \cite{FW} have found a topological expression for this anomaly, which will be the subject of Sec.~\ref{JEFWSEZ}.  Examples of nontrivial homology classes such that any brane wrapping any representative of the class will be anomalous have appeared in, for example, Ref.~\cite{MMS}.  Thus homology is too big to be the class of conserved charges, because in general it contains charges that cannot be realized by any physical branes.  Such classes need to be removed.

In addition to containing unrealizable charges, homology also contains charges that are not conserved.  In Ref.~\cite{DMW} it was argued that sometimes a brane wrapping a nontrivial homology cycle can nonetheless decay.  Such classes of unstable branes need to be quotiented out of homology to arrive at a classification of conserved charges.  Thus the real group of conserved charges is equal to homology, minus the unphysical branes, quotiented by the unstable branes.  In Section \ref{JEKFW} we will review the argument of Ref.~\cite{MMS} that this prescription leads one precisely to twisted K-theory.  In Section~\ref{JEPROBLEMI} we will then argue that yet more unphysical branes need to be removed and more unstable branes need to be quotiented, which leaves one with an as of yet unidentified mathematical structure.

One may now wonder why the homology classification has been so successful.  Part of the reason is that, in the absence of the topologically nontrivial fluxes which lead to twisting, homology and K-theory and the unidentified structure are still unequal.  However they can all be expressed as the sum of a free part, which is a sum of copies of the integers, and a torsion part, which is a sum of finite cyclic groups.  Homology is so successful because the free parts of all of these structures are isomorphic, only the torsion parts differ.  Recall that the charges of BPS states are always elements of the free part of the charge group, this is true not only for homology but for K-theory as well.  Therefore in the absence of nontrivial fluxes the homology classification correctly identifies the free conserved charges, and so classifies all of the BPS states.

The first hint of the failure of the homology classification scheme came from the study of the charges of D-branes embedded in the worldvolumes of higher dimensional D-branes, which were calculated by demanding a cancellation of worldvolume anomalies on intersecting branes.  To explain this development, we will begin with a short review of D-brane charges.

\subsection{D-brane Charges and RR Fields}

D-brane charges, like the charges of electrons in QED, measure the coupling of a D-brane to a gauge potential.  Field configurations in QED are described by a 2-form field strength $F$, which in the absence of magnetic monopoles is closed.  The closure of $F$ ensures that locally one can introduce a one-form $A$, called the vector potential, such that $F$ is the exterior derivative of $A$.  The electric coupling $e$ of an electron to the field $F$ is defined by the topological term of the worldline action of an electron
\beq
S=e\int_\gamma A,\ \ \ F=dA
\eeq
where $\gamma$ is the electron's worldline.  This coupling leads to the celebrated Berry's phase in the electron's wave function $e^{iS/\hbar}$.

In the democratic formulation \cite{Townsend,GHT,VanProeyen} of type II supergravity, D$p$-branes couple not to two-form field strengths, but to $(p+2)$-form field strengths $G_{p+2}$ which are called Ramond-Ramond (RR) field strengths or sometimes improved RR field strengths.  In the case $p=0$ the D$p$-brane is a particle and, like an electron, it couples to a two-form $G_2$. When the NSNS 3-form field strength, which is called the $H$ flux, vanishes, the RR field strengths are closed.  Again this implies that locally one can introduce a gauge potential, this time a $(p+1)$-form $C_{p+1}$, whose exterior derivative is the field strength.  One can then define the D-brane's coupling to the RR field strength to be the coefficient $k$ in the following term in its worldvolume action
\beq
S\supset k\int_N C_{p+1},\ \ \ G_{p+2}=dC_{p+1} \label{JED}
\eeq
where $N$ is the $(p+1)$-dimensional worldvolume of the D$p$-brane.

While only the above term calculates the D$p$-brane's RR charge $k$, there are three physical effects in type II string theories with no analogue in QED which add extra couplings to the worldvolume action beyond those in Eq.~(\ref{JED}).  There may be a nontrivial $H$ field, D-branes support gauge theories whose gauge fields couple to the RR fields and also D-branes couple to gravity which couples to both the RR fields and the gauge fields.  We will consider each of these corrections in turn.

An $H$ flux is a 3-form NSNS field strength $H$.  In the absence of NSNS magnetic monopoles, which are called NS5-branes, $H$ is closed $dH=0$.  In the presence of a nontrivial $H$ flux, the RR field strengths are no longer closed, but instead they satisfy a twisted Bianchi identity
\beq 
dG_{p+2}+H\wedge G_p=0.
\eeq
Notice that the twisted Bianchi identity mixes RR field strengths which are differential forms of different degrees.  It will be convenient to combine all of the RR fields of different degrees into a single polyform.  In type IIA supergravity, the classical limit of IIA string theory, all of the RR field strengths are of even degree, whereas they are all of odd degree in type IIB.  Therefore in type IIA or IIB we can add all of the RR field strengths together in one even or odd polyform $G$, which satisfies the Bianchi identity
\beq
(d+H\wedge)G=0.
\eeq
As in QED, one may introduce magnetic monopoles that are defined to be sources for violations of the Bianchi identity, in type II supergravity these sources are $p$-branes.

The inclusion of an $H$ flux had the effect of changing the operator in the Bianchi identity from the exterior derivative $d$ to the twisted exterior derivative $(d+H\wedge)$.  The closedness of $H$ guarantees that the twisted exterior derivative is still nilpotent.  To see this, note that for any polyform $\omega$
\begin{eqnarray}
(d+H\wedge)(d+H\wedge)\omega&=&d^2\omega+d(H\wedge\omega)+H\wedge d\omega+H\wedge H\wedge\omega\\&=&0+(dH)\wedge\omega-H\wedge d\omega+H\wedge d\omega+0=0+0+0=0\nonumber
\end{eqnarray}
where $d^2\omega$ vanishes because $d$ is nilpotent, $dH$ vanishes because $H$ is closed and $H\wedge H$ vanishes because $H$ is an odd form.  The nilpotence of $(d+H\wedge)$ means that any $(d+H\wedge)$-closed form is locally $(d+H\wedge)$-exact.  In other words, we can introduce a polyform $C$ on each open patch such that
\beq
G=(d+H\wedge)C. \label{JEC}
\eeq
The terms of $C$ are the RR gauge potentials $C_{p+1}$, which are odd differential forms in IIA and even forms in IIB.  It is these gauge potentials which define the D$p$-brane charge Eq.~(\ref{JED}) in the presence of $H$ flux.

The condition Eq.~(\ref{JEC}) does not define the gauge potentials $C$ uniquely.  Instead, analogously to the condition $F=dA$ in QED, the gauge potentials are only determined up to gauge transformations.  In this case one may add any $(d+H\wedge)$-closed polyform $\Phi$ to $C$ and obtain another solution to (\ref{JEC})
\beq
C\longrightarrow C+\Phi,\ \ \ (d+H\wedge)\Phi=0. \label{JETRASFORMA}
\eeq
This means that the D-brane worldvolume action is in turn not well-defined.  Fortunately, the action of a configuration is not an observable, the quantity which needs to be well-defined is the path integral measure $e^{iS}$, which implies that the action must be well-defined up to a shift of an integral multiple of $2\pi$.  

The fact that $C$ is not uniquely defined is not so surprising.  After all, the closedness of $G$ only implied that $C$ exists on contractible patches.  Thus we may consider different solutions $C$ of (\ref{JEC}) to correspond to values of $C$ on different contractible patches.  Consider now a $(p+1)$-dimensional worldvolume $N$ that lies in a single contractible patch.  Then inside the patch there exists a $(p+2)$-manifold $X$ whose boundary is $N$.  We may then use Stoke's theorem to recast (\ref{JED}) as an integral on $X$
\beq
S\supset k\int_X dC_{p+1}. \label{JESTOKE}
\eeq
The polyform $dC$ is sometimes called the unimproved field strength.  

When the $H$ flux is nontrivial it may not seem like this expression is much of an improvement on (\ref{JED}), because $dC_{p+1}$ is still not gauge-invariant.  Using Eq.~(\ref{JETRASFORMA}) we see that it is subject to the gauge transformations
\beq
dC_{p+1}\longrightarrow dC_{p+1}+d\Phi_p=dC_{p+1}-H\wedge\Phi_{p-2} \label{JEDC}
\eeq
where we have used the $(d+H\wedge)$ closure of $\Phi$.  Notice that the shift in $dC_{p+1}$ is a closed form because
\beq
-d(H\wedge\Phi_{p-2})=H\wedge d\Phi_{p-2}=-H\wedge H\wedge\Phi_{p-4}=0\wedge\Phi_{p-4}=0
\eeq
so after the gauge transformation (\ref{JETRASFORMA}) $dC_{p+1}$ remains closed.

\subsection{The Dirac Quantization Condition}

Now we are ready to ask whether the path integral measure $e^{iS}$ is really invariant under the gauge transformations (\ref{JETRASFORMA}).  The answer is no.  Even if all of spacetime is contractible, so that there is only a single patch and even if $H=0$ then if the integral of $\Phi$ over $N$ is not a multiple of $2\pi$ it will change the phase of the measure.  This phase generalizes the Wilson loop in QED.  It is an observable in the sense that two different trajectories can interfere with each other, and the total wave function depends on the relative phases of the wave functions at coincident points.  This relative phase is the Berry's phase, and it is a new observable in the quantum theory that is not captured by $G$ and so it does not exist in the classical supergravity.  Thus the path integral measure need not be entirely determined by $G$ as there are other observables available.  

However all of the observables are invariant under integer shifts in the integral of $k\Phi$ over a closed surface, which are called large gauge transformations.  Thus we still need to check that $e^{iS}$ is invariant under large gauge transformations.   Large gauge transformations may be transition functions between patches and so it remains to check that the measure is the same when calculated on two different patches, in other words, it must be independent of the choice of $X$.

To this end, let us choose a different manifold, $Y$, such that the boundary of $Y$ is again $N$, but this time with the opposite orientation.  In particular let us assume that the union of $X$ and $Y$ is a smooth manifold without boundary.  Then we can use Stoke's theorem to rewrite the action (\ref{JED}) as
\beq
S\supset -k\int_Y dC_{p+1}
\eeq
and demand that this expression for $S$ give the same measure as (\ref{JESTOKE}).  The measures $e^{iS}$ will be equal if the actions differ by $2\pi$ times an integer.  Therefore we demand
\beq
k\int_X dC_{p+1}-(-k\int_Y dC_{p+1})=k\int_{X\cup Y} dC_{p+1}\in 2\pi\Z. \label{JEDIRACCOND}
\eeq
$X\cup Y$ can be any closed submanifold, and so we need to demand that the product of the D-brane charge $k$ and the flux $dC_{p+1}$ over any closed manifold be integral.  

One might worry that this condition is ill-defined, because $dC_{p+1}$ is only defined patchwise, as it enjoys the gauge transformations (\ref{JEDC}).  One would be right.  Fortunately branes may only wrap those cycles on which the integral of $H$ is zero, so that $H$ is exact
\beq
H=dB.
\eeq
On these cycles the transition functions are topologically trivial and so one may simultaneously define $dC$ on $X$ and $Y$ with no transition function and thus render (\ref{JEDIRACCOND}) well-defined.  Now that the term (\ref{JEDIRACCOND}) is well-defined, we need to understand when it is integral. 

There is only one way to ensure that $k\int dC_{p+1}/2\pi$ is always integral, one needs to quantize the RR charge $k$.  Let us define our units such that the smallest quantity of charge allowed is a single unit $k=1$.  Then the well-definedness of the path integral measure implies that
\beq
\frac{1}{2\pi}\int dC_{p+1}\in\Z \label{JEGAUSS}
\eeq
integrated over any cycle.  This is the Dirac quantization condition for RR fluxes, and we have seen that we also need to quantize the RR charges $k$ carried by D-branes.  Roughly speaking the quantization condition implies that RR field strengths, in particular the unimproved RR field strengths $dC$, are classified by integral cohomology.  However one needs to bear in mind the caveat that we have only been able to define $dC$ on those cycles on which the pullback of $H$ is exact.  This is a major shortcoming of the cohomology classification of RR fields, but is automatic in the K-theoretic interpretation.

The quantization condition was already implied in the previous section when we classified D-brane charges by finitely generated abelian groups.  In classical supergravity, where there is no quantization condition, RR charges instead are classified by $n$-tuples of real numbers called real homology classes.  In the quantum theory instead only integral classes are allowed.  However this does not mean that for each real-valued charge in the classical theory there is a single integer-valued charge in the quantum theory.  In fact, we have seen that the integral homology classes that classify D-branes also contain extra torsion charges, which were nontrivial in the example of string theory on $\rp^3$.

\subsection{Gauge and Gravitational Couplings}

Finally we are ready to discuss the worldvolume gauge bundle, which is often called the Chan-Paton gauge bundle.  A D-brane of charge $k$, which may alternately be thought of as a stack of $k$ coincident D-branes, is a place where open strings can end.  Quantizing these open strings one finds, among other massless modes, gluons that transform under a $U(k)$ gauge symmetry.  Notice that the quantization of D-brane charge in the quantum theory is critical, as the rank of the gauge group must be integral.  Configurations of the gluon field in a $U(k)$ gauge theory describe a geometrical object which is called a $U(k)$ gauge bundle $E$.  The gauge bundle $E$ is a vector bundle which consists of a copy of the complex vector space $\C^k$ at each point on the worldvolume $N$ of a stack of $k$ D$p$-branes.

While a D$p$-brane couples to the $(p+1)$-form RR gauge potential $C_{p+1}$, the gauge fields on the D-brane's worldvolume couple to the lower dimensional RR forms.  Given any gauge field configuration one can determine a gauge bundle $E$, and the information of the bundle is partially characterized by an even-dimensional polyform called the total Chern character $ch(E)$.  Different degree forms in the Chern character have different physical interpretations.  For example, the 0-form part, which is called the 0th Chern character $ch_0(E)$ is a constant integer which is just the rank of the gauge group $k$.  The next lowest dimensional component, the 2-form which is called the 1st Chern character $ch_1(E)$ is the trace of the gauge field strength $F$.  It measures, for example, the flux of a magnetic vortex in the gauge theory.  The next component is a 4-form called the 2nd Chern character $ch_2(E)$, which is the trace of the gauge field strength squared.  It roughly measures the instanton number of a gauge field configuration.  

The coupling to the RR fields is now easy to describe, the $j$th Chern character is a 2$j$-form characterizing the gauge field configuration and it couples to the RR gauge connection $C_{p+1-2j}$ via
\beq
S\supset \int_N ch(E) \wedge C \supset \int_N (k C_{p+1}+ \frac{\mathbf{Tr}(F)}{2\pi} \wedge C_{p-1} +  \frac{\mathbf{Tr}(F\wedge F)}{8\pi^2} \wedge C_{p-3}). \label{JEGAUGE}
\eeq 
Notice that the first term on the right hand side, the $ch_0(E)=Tr(1)=k$ term, is the term (\ref{JED}) which measures the D$p$-brane charge $k$, which is defined to be the strength of the coupling to $C_{p+1}$.  The second term $F \wedge C_{p-1}$ instead is a coupling to $C_{p-1}$ so it measures D$(p-2)$-brane charge.  Thus we conclude that a D$p$-brane may carry D$(p-2)$ brane charge, and that this charge is equal to the trace of its gauge field strength divided by $2\pi$.  Intuitively, magnetic flux tubes in a D$p$-brane carry D$(p-2)$-brane charge.  In fact if one dissolves a parallel D$(p-2)$-brane inside of a D$p$-brane one finds that the gauge field configuration of the worldvolume theory on the D$p$-brane changes, a magnetic flux tube appears where the D$(p-2)$ was, with a flux equal to the D$(p-2)$-brane's RR charge.  Similarly D$(p-4)$-branes carry instanton charge in the D$p$-brane's worldvolume theory.  

A simple dimension-counting argument using the gauge theory kinetic term $F\wedge\star F$ shows that while one gains energy by smearing out a D$(p-2)$ in a D$p$, and the size of a D$(p-4)$ is a modulus, it costs energy to dissolve a yet lower dimensional brane and so in general these tend to be ejected by the dynamics.  However in this review we will be interested in a classification of conserved charges and rarely discuss their dynamics.  The total charge of two D-branes is independent of whether they are separated or whether one is dissolved in the other, describing some nontrivial gauge configuration.

There are yet more couplings in the D-brane worldvolume action.  Usually one includes the coupling of NSNS fields.  Recall that the 3-form $H$ field is everywhere closed, and on a D-brane worldvolume it is exact.  Therefore on a worldvolume it can be written as the exterior derivative of a 2-form which is called the $B$ field.  The $B$ field, like the other gauge potentials in string theory, is not quantized.  One often includes the factor $e^B$ in the worldvolume action.  This may seem disastrous, as its couplings appear to lead to nonquantized lower dimensional D-brane charges \cite{BDS}.  However these charges, or at least the nonintegral parts, are canceled by contributions from the action of the bulk supergravity \cite{Wati} and so we will neglect them for now.  They will play a role later when we note that, in the case of a single D-brane $k=1$, the action is invariant under gauge transformations that leave $B+F$ constant even if $F$ is not constant.

The final couplings that we need to consider are those of the D-brane to gravity.  Gravity couples similarly to gauge fields.  Instead of using a gauge field configuration to determine a gauge bundle, one uses the configuration of gravitons, which determines the topology of the spacetime itself, to determine another vector bundle which is called the tangent bundle \textbf{TM}.  This vector bundle is the direct sum of two subbundles, the tangent bundle to the brane \textbf{TN} and the normal bundle to the brane inside of $M$, which we will denote \textbf{NN}.  Unlike the $U(k)$ gauge bundle, whose fibers were complex $k$-dimensional vector spaces, the fibers of the tangent bundle are 10-dimensional real vector spaces $\R^{10}$ while those of \textbf{TN} and \textbf{NN} are $\R^{p+1}$ and $\R^{9-p}$ respectively.  One rarely considers the Chern characters of a real vector bundle, because the odd Chern characters vanish being traces of odd numbers of the antisymmetric matrices that generate the Lie algebra of SO($N$).  Thus instead of defining a form of each even degree $2k$, real vector bundles determine a form in each degree of the form $4k$.  One common basis of these forms, which plays the role played by Chern characters in the case of complex vector bundles, is the set of Pontrjagin classes $p_k$ which are traces of degree $2k$ homogeneous polynomials in the curvature tensor.  Just as the Chern character differential forms can be assembled into distinct polyforms, the total Chern class and the total Chern character, by weighting them differently, the Pontrjagin classes can be assembled into a number of distinct polyforms.  Gravitational anomaly cancellation then determines how these polyforms couple to RR fields and to the worldvolume gauge field.  

It turns out that this coupling is most simply expressed in terms of a polyform called the A-roof genus $\hat{A}$.  The lowest terms in the A-roof genus are
\beq
\hat{A}=1-\frac{p_1}{24}+\frac{7 p_1^2-4p_2}{5760}+... 
\eeq
where the higher order terms are forms of degree at least 12, and so will vanish in the 10-dimensional world of critical superstrings.  In fact, crucially for the success of the K-theory interpretation, the RR fields and gauge fields will couple not to the A-roof genus but to its square root
\beq
\sqrt{\hat{A}}=1-\frac{p_1}{48}+\frac{9 p_1^2-8p_2}{23040}+... 
\eeq
whose wedge product with itself is the A-roof genus.  

With these ingredients, the unique coupling of RR fields and gauge fields to gravity which renders the measure of the path integral of the chiral fermions at the intersection of two branes well-defined was calculated in \cite{MM}, generalizing the result in \cite{GHM} which applied when \textbf{NN} is the trivial bundle.  At the level of differential forms they found
\beq
S\supset\int_N C\wedge ch(E)\wedge \frac{\sqrt{\hat{A}(\textup{TN})}}{\sqrt{\hat{A}(\textup{NN})}} \label{JESPIENO}
\eeq
where division is the inverse of the wedge product.  The integral of all forms of degree not equal to $p+1$ is defined to be zero.  This is our final answer for the worldvolume action of a D-brane.  In fact this is not the complete action, it is a collection of terms known as the Wess-Zumino terms, but it will suffice to determine the lower-dimensional D-brane charges as the other terms do not contain couplings to RR fields.

\section{K-Theory from the Sen Conjecture} \label{JESEN}

\subsection{Charges and K-Theory's Inner Product}

As we did with Eq.~(\ref{JEGAUGE}), one may analyze the action (\ref{JESPIENO}) and read the charges of the various D-branes off of the coefficients of each RR field.  This gives an expression for each lower-dimensional D-brane charge in terms of the gauge field of the original D-brane and the topology of its embedding.  However, to classify D-brane charges in all of spacetime one would like to discuss some object which lives not on the worldvolume of a particular brane like the gauge bundle \textrm{E} and the bundles \textbf{TN} and \textbf{NN}, but rather an object that lives in the bulk.  In Ref.~\cite{MM} the authors reexpress these lower D-brane charges in terms of the cohomology of the bulk spacetime, using the pushforward map $f_{!}$ which takes the worldvolume gauge bundle to a bundle on all of spacetime.  Here $f$ is an embedding of the D$p$-brane's worldvolume $N$ into the spacetime $M$.  

They found that the charges with respect to all RR fields may be summarized by the following simple expression
\beq
Q=ch(f_! E)\sqrt{\hat{A}(\textup{TM})}\in\H^*(M). \label{JEMMCARICA}
\eeq
E is a gauge bundle on the D-brane worldvolume $N$ and so its Chern character $ch(\textup{E})$ is a polyform which represents a class in the de Rham cohomology of $N$.  However $f_!$ E is a bundle on the entire spacetime $M$, and so its Chern character $ch(f_!\textup{E})$ represents a class in the cohomology of $M$. Also the Pontrjagin classes of the tangent bundle, which are the summands of the square root of the A-roof genus of \textbf{TM}, represent cohomology classes in $M$.  Therefore the D-brane charge $Q$ in (\ref{JEMMCARICA}) is an element of the cohomology of $M$.

To relate the charge (\ref{JEMMCARICA}) to the earlier worldvolume expression (\ref{JESPIENO}) Minasian and Moore used a version of the Grothendieck-Riemann-Roch theorem that was proven by Atiyah and Hirzebruch.  This theorem states that the Chern character of the bundle $f_!$E multiplied by the A-roof genus of the normal bundle is the class in $\H^*(M)$ which is Poincar\'e dual, in $M$, to the pushforward $f_*$ of the Poincar\'e dual in $N$ of the element $ch(\textup{E})$ of $\H^*(N)$
\beq
ch(f_!\textup{E})\hat{A}(NN)=\textup{PD}|_M(f_*(\textup{PD}|_N(ch(\textup{E}))))\hsp f_*:\H_*(N)\longrightarrow \H_*(M).
\eeq
These Poincar\'e dualities are necessary because, unlike cohomology classes, homology classes pushforward naturally.  The $\hat{A}(NN)$ correction is the price one must pay for the unnatural pushforward.  It kills the denominator of Eq.~(\ref{JESPIENO}) and combines with the worldvolume tangent bundle's A-roof genus to form the spacetime tangent bundle's A-roof genus in Eq.~(\ref{JEMMCARICA}).  

It may seem that this argument demonstrates that D-branes are classified by the de Rham cohomology of $M$.  However in \cite{MM}, Minasian and Moore argued that instead (\ref{JEMMCARICA}) suggests that D-brane charges are more naturally characterized by the K-theory classes $f_! $E.  Before explaining why this seemed natural to Minasian and Moore, we will digress to describe what K-theory is and why $f_! $E represents a K-theory class.

\subsection{What is K-Theory?}

Recall that a complex vector bundle E over $M$ consists of a complex vector space $\C^k$ fibered over every point in $M$ such that, on each contractible neighborhood $U\subset M$, the total space of the vector bundle is topologically $U\times \C^k$.  These neighborhoods are then glued together via transition functions in $U(k)$.  Given two vector bundles \textrm{E} and \textrm{F} of rank $j$ and $k$ over the same base $M$, there is an easy way to add them, called the direct sum \textrm{E}$\oplus$\textrm{F}.  One simply takes the direct sum fiber by fiber of the vector spaces, so that the total space above $U$ is $U\times\C^{j+k}$ and the transition function in $U(j+k)$ is block diagonal with the $j\times j$ and $k\times k$ blocks being the original transition functions in \textrm{E} and \textrm{F}.

While the direct sum provides an easy way to add vector bundles, subtraction is more difficult.  If \textrm{E} and \textrm{F} are vector bundles, one can define their difference \textrm{E-F} to be the pair (\textrm{E,F}).  Subtraction should in some sense be the inverse of addition, which motivates the following equivalence relation.  If \textrm{E}, \textrm{F} and \textrm{G} are vector bundles then one identifies
\beq
(\textup{E},\textup{F})=(\textup{E$\oplus $G},\textup{F$\oplus $G}). \label{JEEQUIVALENZA}
\eeq
This intuitively corresponds to subtraction because \textrm{E-E}$=0$
\beq
\textup{E}-\textup{E}=(\textup{E},\textup{E})=(0\oplus\textup{E},0\oplus\textup{E})=0-0=0
\eeq
where \textrm{0} is the trivial bundle of rank 0.  One can now define addition and subtraction for a pair of vector bundles
\beq
(\textup{E},\textup{F})+(\textup{E}\p,\textup{F}\p)=(\textup{E}\oplus\textup{E}\p,\textup{F}\oplus\textup{F}\p),\ \ \ (\textup{E},\textup{F})-(\textup{E}\p,\textup{F}\p)=(\textup{E}\oplus\textup{F}\p,\textup{F}\oplus\textup{E}\p).
\eeq
With this definition of addition and subtraction, the space of pairs of bundles (\textrm{E},\textrm{F}) is a group.  The inverse of any element is
\beq
-(\textup{E},\textup{F})=(\textup{F},\textup{E})
\eeq
and the identity is (\textrm{0,0}).  This group is called the \textit{K-theory} of $M$, and is denoted K${}^0(M)$.

Now we can identify $f_!\textup{E}$ with an element of the K-theory of $M$.  It is a complex vector bundle on $M$, so it can be identified with the pair ($f_!\textup{E},0$).  The Chern character of a bundle does not contain all of the information about a bundle, and so one may worry that assigning charges to K-group elements really uses more information than appears in the physical coupling.  However the Chern character is the coupling to the differential form representing the RR fields.  Moore and Witten have conjectured \cite{MW} that RR fields should also be classified by twisted K-theory, as we will review in Subsec.~\ref{JERR}.  Thus one may suspect that since the Chern character of $f_!E$ is a differential form approximation of a K-class, and since the RR fields are differential form approximations of a K-class, perhaps the coupling expressed in Ref.~\cite{MM} in terms of cohomology is just an approximation of a direct coupling of the K-theory.  In this interpretation, it appears that the charge $Q$ only knows about the Chern character, and not about the whole K-theory class, because of the approximation used in the calculation of D-brane charges which is inherent in the use of cohomology classes.  If one had a formulation which used only K-theoretic operations from the beginning perhaps the charge would depend upon all of the information in the K-theory class.

One may wonder about the presence of the A-roof genus term in the charge (\ref{JEMMCARICA}).  This term is independent of the choice of gauge bundle. Minasian and Moore explain that this term has a natural interpretation in K-theory.  It ensures that the assignment of a charge in cohomology to a K-class is an isometry.  More explicitly, as we will review momentarily, there is a natural inner product in both cohomology and in K-theory, and the A-roof genus term ensures that $Q$, which maps a K-theory class to a cohomology class, preserves these inner products.  

A natural inner product between any two elements $[\eta]$ and $[\omega]$ of de Rham cohomology is as follows.  Choose two differential forms $\eta$ and $\omega$ that represent the two classes, the choice of representative will not matter.  The inner product of the two classes is just the integral of the wedge product of the forms over the spacetime
\beq
([\eta],[\omega])=\int_M\eta\wedge\omega.
\eeq

The inner product of two K-theory classes E and F is slightly more complicated.  Given two vector bundles \textrm{E} and \textrm{F} of rank $j$ and $k$ we have seen that one may form their sum by taking the direct sum of each complex vector space fiber.  We will be interested in a different operation.  One may also form the product of \textrm{E} and \textrm{F} by taking the tensor product of each complex vector space fiber.  The product is a new vector bundle E$\otimes$F whose fibers are $jk$ dimensional.  This product makes K${}^0(M)$ into a ring.  Consider a Dirac operator $D\hspace{-.3cm}\slash$ acting on sections of the bundle E$\otimes$F, which intuitively are just $jk$-tuples of functions on each patch with transition functions between patches which are the same as the transition functions that define the bundle E$\otimes$F.  The Dirac operator is just a differential operator times a gamma matrix, so it takes vectors in $\C^{jk}$ to vectors in $\C^{jk}$ and sections of the $\C^{jk}$ bundle to other sections.  In particular, it has a kernel which consists of all of the sections that it takes to the zero section, which is the section that consists of the origin of $\C^{jk}$ above each point in $M$.  Also it has a cokernel, which is the kernel of its transpose.  

While neither the dimension of the kernel nor the dimension of the cokernel is a topological invariant of the bundle E$\otimes$F, the difference between these dimensions is an invariant.  This difference is called the index of $D\hspace{-.3cm}\slash$\ and its value is determined from the topology of the bundle via the Atiyah-Singer Index Theorem 
\beq
\textrm{ind}D\hspace{-.3cm}\slash_{E\otimes F}=\int_M ch(E)\wedge ch(F)\wedge\hat{A}(\textup{TM}).
\eeq
Now we can define the K-theoretic inner product $(\textrm{E},\textrm{F})_K$ of the K-classes corresponding to the vector bundles \textup{E} and \textup{F} to be the index of the Dirac operator on their tensor product
\beq
(\textup{E},\textup{F})_K=\textrm{ind}D\hspace{-.3cm}\slash_{\textup{E}\otimes \textup{F}}.
\eeq

Minasian and Moore noticed that this K-theoretic inner product is precisely equal to the de Rham inner product of the corresponding D-brane charges $Q$.  Concretely, given two D-branes with worldvolume gauge bundles \textrm{E} and \textrm{F} and embedding maps $e$ and $f$ of their worldvolumes into $M$, the de Rham inner product of their charges from Eq.~(\ref{JEMMCARICA}) equals the K-theory inner product of their gauge bundles pushed forward onto the full spacetime $M$ via the maps $e_!$ and $f_!$
\begin{eqnarray}
(Q(\textup{E}),Q(\textup{F}))&=&(ch(e_!(\textup{E}))\wedge\sqrt{\hat{A}(\textup{TM})},ch(f_!(\textup{F}))\wedge\sqrt{\hat{A}(\textup{TM})})\\
&=&\int_M ch(e_!(\textup{E}))\wedge ch(f_!(\textup{F}))\wedge \hat{A}(\textup{TM})
=\textrm{ind}D\hspace{-.3cm}\slash_{e_!\textup{E}\otimes f_!\textup{F}}=(e_!\textup{E},f_!\textup{F})_K.\nonumber
\end{eqnarray}
Therefore the square root of the A-roof genus in Eq.~(\ref{JEMMCARICA}) guarantees that the charge map $Q$ is an isometry of K-theory onto de Rham cohomology.

While this evidence for the K-theory classification of D-branes was quite circumstantial, resting on the compatibility of the inner products in the known cohomological expressions for D-brane charge and in a conjectured K-theory framework, it suggested that the older homological classification might be incorrect.  Witten later demonstrated in \cite{WittenK} that in the case of type I string theory the K-theory classification, or more precisely a variation of K-theory using real bundles, successfully predicts the existence of nontrivial $SO(32)$ gauge configurations missed by the homological classification.  In fact several months earlier he had already, in the context of the AdS/CFT correspondence, presented an example of a failure of the homological classification in type II with an orientifold 3-plane in \cite{BBA}.  Thus it was eventually irrefutable that the homology classification of branes was incorrect, it contains unphysical branes and also unstable branes, and it needed to be replaced.  

However the K-theory description of Minasian and Moore was mysterious as it described D-brane configurations in terms of a gauge bundle on the entire spacetime.  The original gauge bundle on a D-brane was easy to interpret physically.  Open strings end on the D-brane, when quantized some of their modes correspond to gluons and nontrivial configurations of the gluon field are described by gauge bundles on the D-brane.  But there were no gluons in the bulk, open strings must end on branes, and so the presence of a gauge field in the bulk was a mystery.

\subsection{The Sen Conjecture}

An interpretation for this apparent bulk gauge bundle that classifies all D-brane configurations came from a parallel development in string field theory.  In quantum field theories, a particle with a negative mass squared, called a tachyon, is a sign of instability.  These particles will be spontaneously created, reducing the energy of the system and changing the vacuum itself via a process called tachyon condensation.  Sometimes, as in the case of the standard model's Brout-Englert-Higgs mechanism, this process will end when the system arrives in a new vacuum with no tachyonic excitations available.  One example of such a process in string theory is the closed string tachyon condensation when spacetime is a discrete quotient of the complex plane \cite{APS} with a conical singularity, which decays to a stable vacuum in which spacetime is the full complex plane.  Sometimes the field theory is hopelessly unstable and the decay will never end or will end once all of spacetime has been destroyed, as has been conjectured to occur in the wrong sign heterotic theory of Ref.~\cite{FH}.  More often, no one knows the end point of this process, as is the case for closed string tachyon condensation in critical bosonic string theory.

Consider a D-brane and an anti D-brane in type II string theory.  Open strings may extend from one to the other, and quantizing these strings one finds a tachyonic mode.  Tachyonic modes correspond to instabilities, and in this case Ashoke Sen argued in 1998 \cite{Sen} that the condensation of the tachyon corresponds to a dynamical process in which the brane and antibrane annihilate.  We have seen that D$p$-branes can carry the charges of lower dimensional D-branes, encoded in the topology of their worldvolume gauge bundles.  Therefore if the D$p$ and anti D$p$ carried topologically distinct gauge bundles, while the total D$p$-brane RR charge is zero there is still a net RR charge of lower-dimensional D-branes.  As net RR charge is conserved, this means that lower dimensional D-branes will be present after the annihilation of the D$p$-branes is complete.  In the case $p=9$ the D$9$-brane and anti D$9$-brane fill all of space, and so one might suspect that any configuration of lower dimensional D$p$-branes could be encoded in the gauge bundle on this pair.  

This is not possible, as these branes each carry a single $U(1)$ gauge bundle and abelian gauge theories have, for example, no instantons and so there would be no D$5$-branes.  However if one allows the gauge group to be large enough then there is no such obstruction.  This led Sen to conjecture that any configuration of D-brane charges can be realized as a gauge field configuration on a stack of sufficiently many D$9$-branes and anti D$9$-branes.  Then, after these D$9$-branes and anti D$9$-branes have decayed, or equivalently after the open string tachyons have condensed, the arbitrary configuration of lower dimensional D$p$-branes will remain.  For example if the D9-branes contained a magnetic flux tube with $k$ units of magnetic flux then after the annihilation $k$ D$7$-branes will remain.  If instead the anti D$9$-branes carry this flux tube then $k$ units of anti D$7$ charge will remain.  If the D9 worldvolume contained an instanton then a D$5$-brane will remain wrapped around the same cycle as the instanton, and so on.

Now we can interpret the classification of D-brane charges (\ref{JEMMCARICA}) by bundles $f_!$E on the bulk, these are the gauge bundles on the D$9$-branes.  We can also interpret the K-theory equivalence relation (\ref{JEEQUIVALENZA}).  Consider a brane anti-brane system that represents the element $(\textbf{E}\oplus \textbf{G},\textbf{F}\oplus \textbf{G})$ in K-theory.  If a brane and antibrane with the same gauge bundle, which pushes forward to \textbf{G}, annihilate, then the equivalence condition states that the conserved charges (\textbf{E,F}) are conserved, as they must be.  Thus Witten has argued \cite{WittenK} that Sen's conjecture implies that D-branes are classified by K-theory.

These vector bundles over all of space time are not conserved charges in the usual sense.  Usually, when one speaks of the conserved charges of a configuration, one integrates a conserved current over a timeslice and then the conservation implies that the choice of timeslice was irrelevant.  If instead the charge is different at two timeslices then a process has occurred which can destroy this charge.  Here instead one is considering a D-brane charge on all of the spacetime, not just on a timeslice.  The process of tachyon condensation is not being treated as a process relating configurations at two moments in time, but rather it relates two different trajectories on the entire spacetime.  This is not to say that tachyon condensation is not a real dynamical process that can occur over time, but it is merely to say that in the viewpoint taken in this subsection, tachyon condensation is something which relates two entire trajectories.  

Instead of a dynamical process which proceeds in the time direction, closed string tachyon condensation is treated in Ref.~\cite{APS} as an RG flow that proceeds along an internal direction of configuration space.  Similarly the deformations in Ref.~\cite{WittenK} relating different representatives of the same K-theory class identify configurations related by deformations in the configuration space.  Thus the K-theory of the entire spacetime classifies not charges that are invariant under time-dependent physical processes, but rather charges associated with entire trajectories which are invariant under deformations such as renormalization group flow.  

While the K-theoretic classification of D-brane charges using the Sen conjecture was revolutionary, it was somewhat limited.  First, it did not apply to type IIA string theory which has no spacefilling D-branes.  Ho\v{r}ava later claimed \cite{Petr} that one can make the same arguments in IIA string theory using unstable D9 sphalerons, and one arrives not at K${}^0$ but at a similar object named K${}^1$.  Also Kapustin has argued \cite{Kapustin} that D9-branes are inconsistent in the presence of $H$-flux, and so the K-theory classification does not apply in these cases.  As we will discuss presently, in these cases one needs to use a variation of K-theory known as twisted K-theory.  Finally, as was emphasized in the last two paragraphs, this classification scheme does not classify charges that are preserved under dynamical processes, but rather a kind of property of trajectories that is preserved under deformations.  But the question of what structure classifies conserved RR charges was left open.  These shortcomings were all addressed by Maldacena, Moore and Seiberg \cite{MMS} using a new approach to understand K-theory based not on tachyon condensation, but based on the Freed-Witten anomaly.  This approach will be the subject of Section~\ref{JEKFW}.

\section{The Freed-Witten Anomaly} \label{JEFWSEZ}

\subsection{How Are Homology and K-theory Different?} \label{JEMMS}

The above K-theory classification of D-brane charges in the absence of $H$ flux is not equivalent to the homology classification.  There are three differences:
\begin{itemize}
\item  Some homology classes do not correspond to any K-theory class.
\item  Some K-theory classes correspond to multiple homology classes.
\item  While homology and K-theory are both abelian groups, their addition rules are not compatible.  
\end{itemize}
Each of these differences has an interpretation in terms of D-brane physics, and in each case K-theory gets it right while homology gets it wrong.  As we will argue shortly, the homology classes that do not correspond to K-theory classes correspond to inconsistent D-branes that lead to worldsheet global anomalies on the strings that end on them.  The multiple homology classes that correspond to the same K-theory class correspond to distinct brane configurations such that processes exist which transform one configuration into another, so that the conserved charges corresponding to all of these configurations must be the same.  The transformations of D-branes that create homology charge can be thought of as a violation of the conservation of a RR current due to the aforementioned anomalies \cite{BRST}.  Thus the first two differences are the consequences of the same anomalies. 

The difference in the addition rules of homology and K-theory corresponds to the fact that a bound state of two D-branes should carry the sum of the conserved charges of the constituents.  In particular the homology classification would imply that the sum of two D$p$-branes is again a D$p$-brane, whereas K-theory successfully predicts that a D$p$-brane can, for example, carry a half unit of D$(p-2)$-brane charge so that two D$p$-branes can combine to yield a D$(p-2)$.  For brevity we will have little to say about this additive structure, but the literature is filled with examples.  A particularly rich example involving D-branes of several dimensions is described in Ref.~\cite{aussy2005}.

Despite all of these differences, homology and K-theory, in the absence of $H$ flux, are remarkably similar.  Classically they are identical.  Mathematically one says that they are isomorphic when tensored with the real numbers
\beq
\H^{even}\otimes\R\cong\textup{K}^0\otimes\R,\ \ \ \H^{odd}\otimes\R\cong\textup{K}^1\otimes\R
\eeq
which yields homology and  K-theory with real coefficients.  In physics we saw in Subsec.~\ref{JEGEN} that this tensoring eliminates the Dirac quantization condition and it is interpreted as a classical limit.  One finds that branes in supergravity are classified by homology with real coefficients which is isomorphic to K-theory with real coefficients.  Intuitively, the inconsistencies which lead some D-branes to be anomalous all disappear when there are enough D-branes, or more precisely when the number of D-branes has the right prime factors, as occurs in the classical limit.

There appear to be two things that can go wrong with a D-brane wrapped on a homology cycle that imply that the homology cycle does not correspond to any K-theory cycle.  One is that the D-brane suffers from the Freed-Witten anomaly \cite{FW}, which implies that worldsheets of open strings ending on the D-brane are afflicted with a global anomaly that does not allow one to globally define the phases of the measures of their path integrals.  No D-branes with Freed-Witten anomalies carry K-theory charges, and we will argue that they are all inconsistent.  If a D-brane wraps a nonsingular submanifold and is not Freed-Witten anomalous then it automatically carries a K-theory charge.  This does not mean that all homology classes that can be wrapped by Freed-Witten anomaly-free branes correspond to K-classes, because some homology classes can not be represented by nonsingular submanifolds \cite{Thom,Sull}.  Thus D-branes can fail to carry K-theory charge if they are Freed-Witten anomalous or if they wrap a homology cycle which cannot be represented by any nonsingular submanifold \cite{ES}.

It may seem odd that there are homology classes that are not representable by any nonsingular submanifold.  In fact, homology classes in a space $M$ of dimension 9 or less can always be represented by nonsingular submanifolds, but precisely in dimension 10 there appear 7-dimensional homology cycles which cannot be represented by any nonsingular submanifold, meaning that the representability of cycles is only an issue for D6-branes in IIA string theory and generically for branes in noncritical string theories like WZW models.  When one says that a class is not representable by a nonsingular submanifold, this means that it can still be represented by a subset, but that this subset is not a nonsingular manifold because it contains at least one point, called a singularity, where the tangent space is not of the form $\R^n$.  For example, the real cone over the complex projective space $\cp^3$ has a singularity at the tip of the cone.  

The singularities of these cycles are different from the singularities usually encountered in string theory because the ambient spacetime is nonsingular, it is only the wrapped cycle which is singular.  Not only is the cycle singular, but every cycle in its homology class is singular and so there is no way that the singularity may be blown up or deformed away.  The nonsingularity of the spacetime means that there are no strong gravitational effects that can change the geometry, the physics of the singularity is governed by open strings.  However some nonrepresentable homology cycles nonetheless lift to K-theory, and so the corresponding branes carry K-theory charges.  In Ref.~\cite{ES} the authors gave an example of a nonrepresentable homology cycle that lifts to a K-theory class, and one that does not.  If one believes the type IIA version of the Sen conjecture, and therefore that type IIA branes are classified by K-theory, then the first representable cycle can be consistently wrapped by a D6-brane and the second cannot.  However, unlike the Freed-Witten case, the global anomalies of strings ending on these two D-branes have not yet been analyzed.  Such an analysis would be a strong test of the K-theory classification hypothesis.

Summarizing, a D-brane wrapping a homology cycle is inconsistent if it suffers from a Freed-Witten anomaly, and is sometimes inconsistent if the homology cycle cannot be represented by any nonsingular submanifold.  This brings us to the question, what is a Freed-Witten anomaly?  To answer this question we will first need some mathematical background on $spin$ structures and $spin^c$ structures.

\subsection{$Spin$ Structures}

The remainder of this section is technically more difficult than the rest of the lectures.  The uninterested reader can skip to Sec.~\ref{JEKFW}.  To understand the following sections he will only need the conclusion that on a D-brane worldvolume there is an anomaly called the Freed-Witten anomaly which is equal to $W_3+H$.  Here $W_3$ is a class in the third integral cohomology of the worldvolume called the third Stiefel-Whitney class.  It is determined by the topology of the D-brane's worldvolume.  $H$ is also an element of the third integral cohomology of the worldvolume, it is equal to the pullback of the spacetime NSNS field strength.  A D-brane can only wrap cycles $N$ such that
\beq
W_3+H=0\in\H^3(N;\Z).
\eeq

The Freed-Witten anomaly is an obstruction to the existence of certain representations of the Lorentz group in the spectrum of a theory on spacetimes with some topological properties.  To understand it, we will first need to review the action of the Lorentz group on fields defined on a topologically nontrivial space.  

The wavefunctions of fermions do not transform under any representation of the Lorentz group.  To see this, consider a 360 degree rotation on your favorite plane.  This is the identity rotation of the Lorentz group.  Representations are, by definition, homomorphisms of groups to matrices which means among other things that they map the identity group element to the identity matrix.  However a 360 degree rotation is the identity element of the Lorentz group but it changes the sign of the wavefunction of a fermion, and so is not represented by the identity matrix.  If $U$ is the map from the Lorentz group to the group action on the spinor $\psi$ then
\beq
\psi=U(1)\psi\neq U(e^{2\pi i})\psi=-\psi
\eeq
implies that $U$ is not a representation.  However $U$ is a representation up to a phase, which in this case is a choice of sign.  Maps like $U$ that are representations up to a choice of phase are called projective representations, of which representations are examples in which the phase factor is one.  Symmetries in quantum theories always act via projective representations.

In a topologically trivial spacetime one can always define an action on states via a projective representation.  This is because a physical state does not correspond to a vector in a Hilbert space, but to a ray in a Hilbert space, which is the set consisting of a vector multiplied by any phase.  In particular, the multiplication of a fermion wave function by minus one negates the corresponding vector in the Hilbert space, but it is still described by the same ray and so it still corresponds to the same physical state.  Thus the action of the Lorentz group on rays, that is to say on physical states, is well-defined.  

One might then be tempted to do away with the Hilbert space and just deal with the rays directly, but perhaps unfortunately the known formalisms for doing calculations in quantum field theory require a choice of a representative vector in each ray.  In the case of the Lorentz group, this means that given any action of the Lorentz group, one must also determine a phase.  A choice of an element of the Lorentz group plus a phase is the same as a choice of an element of the nontrivial central extension of the Lorentz group which is called the $Spin$ group.  An element of the $Spin$ group is an element of the corresponding Lorentz group plus a $\Z_2$ choice of sign, which topologically means that the group manifold of a $Spin$ group is a certain $\Z_2$ bundle over the group manifold of a Lorentz group $SO(N)$
\beq
\begin{array}{ccc}\Z_2 & \longrightarrow & Spin(N)\\
&&\downarrow\pi\\
&&SO(N)\end{array}
\eeq
where $\pi$ is the projection map from the $Spin$ group to the Lorentz group.

The simplest example of the relationship between the Lorentz and Spin groups occurs in the 2-dimensional Euclidean space $\R^2$.  Here the Lorentz group is $U(1)$, and the $Spin$ group is another copy of $U(1)$ which is twice as big.  $\pi$ is the identification of antipodal points on the big $U(1)$.  In particular a rotation by $\alpha$ in the Lorentz group is only a rotation by $\alpha/2$ in the $Spin$ group.  Vectors, like the gauge connection $A$ in electrodynamics, are sections of the tangent bundle of 2-dimensional Euclidean space which is a trivial vector bundle whose fibers are $\R^2$ and transform in the 2-dimensional $SO(2)$ representation of the Lie algebra $u(1)$, which topologically means that the structure group of the bundle is the Lorentz group $SO(2)$.  Spinors are also sections of a 2-dimensional trivial bundle, which again is subject to transition functions in $SO(2)$, which is the $Spin$ group of $SO(2)$.  These are the same transition functions as were used for the tangent bundle, but with a projective representation in which all rotations are divided by two.  However, as the bundle is trivial, these transition functions can be taken to be trivial and so the tangent and spin bundles are equivalent in this case.  

If instead one considers QED on a topologically nontrivial 2-dimensional Euclidean space $\Sigma$ then the two bundles are different.  Now the photon wavefunction is a section of the tangent bundle $\mathbf{T\Sigma}$ while the electron wavefunction is a section of the square root of the tangent bundle $\sqrt{\mathbf{T\Sigma}}$, which is called the $Spin$ bundle.  Both bundles may be constructed from local trivializations and transition functions.  The transition functions that create both bundles are the same, but the transition functions act on the square root bundle using a different representation, in which the rotation angles are all halved.  If one considers the fiber to be $\C$ and the transition functions to be 1-dimensional complex representations of $U(1)$ then the $Spin$ bundle transition functions act on the $Spin$ bundle via the projective representation
\beq
U(e^{i\alpha})=e^{i\alpha/2} \label{JEPREP}
\eeq
where $e^{i\alpha}$ is the action on the tangent bundle. 

Notice that the map (\ref{JEPREP}) is not well-defined, as, given a transition function $e^{i\alpha}$ one may interpret $\alpha$ as $\alpha+2\pi$ by choosing a different fundamental domain for the angles and one would find a different sign for the right hand side, which translates into a minus sign in the transition functions for the spinors.  To define the spin bundle one needs to choose all of these signs in the transition functions.  The choice of these signs is called the {\it{spin structure}}.  

The choice of spin structure is a physical choice, it can determine, for example, whether a compactification is supersymmetric or not.  The choice of spin structure on the worldsheet of a closed string determines whether one is describing NS or R modes, which in turn determines the spin of the spacetime field corresponding to an excitation of the worldsheet spinor.  Sometimes different choices of spin structure are physically equivalent.  Two choices are only inequivalent if, after a spinor moves around some closed loop, the two spin structures disagree on the spinor's sign.  Thus there is a $\Z_2$ choice of $spin$ structure for each inequivalent closed loop of the theory.  For example, the fundamental group of the genus $g$ Riemann surface $\Sigma_g$ has $2g$ generators and its group of spin structures is
\beq
\H^1(\Sigma_g,\Z_2)=\Z_2^{2g}. \label{JE2DSS}
\eeq
In this example one can see that a choice of $spin$ structure is just a choice of whether a spinor will have periodic or antiperiodic boundary conditions on each of the $2g$ inequivalent loops.

We have seen that a choice of $spin$ structure is just a choice of an element of $\Z_2$ at each transition function between patches.  This information defines a $\Z_2$ bundle.  Actually, not every choice of $\Z_2$ on overlaps gives a bundle.  For example, consider three patches $U_i$, $U_j$ and $U_k$ which are discs and whose double and triple overlaps are all discs.  When all possible intersections of a set of patches covering a space are topologically trivial, one says that the patches form a {\it{good cover}}.  We first trivialize the $Spin$ bundle on each patch.  To define the $Spin$ bundle globally we must define transition functions which are $+1$ or $-1$ on each of the three overlaps $U_i\cap U_j$, $U_j\cap U_k$ and $U_k\cap U_i$.  Name these transition functions $x_{ij}$, $x_{jk}$ and $x_{ki}$ respectively.  The $\Z_2$ bundle is only well-defined if these patches satisfy the triple overlap condition on the triple overlap $U_i\cap U_j \cap U_k$
\beq
x_{ij}x_{jk}x_{ki}=1. \label{JETRIPOLA}
\eeq
In fact, the $spinor$ is also only well-defined when we impose this condition.  This is because, given the value of a spinor on the overlap written using the trivialization of $U_i$, one may switch to the $U_j$ trivialization and then to the $U_k$ trivialization and finally back to the $U_i$ trivialization without ever moving, therefore the value of the spinor should not change.  This means that we must impose the condition (\ref{JETRIPOLA}) at every triple overlap in our definition of $spin$ structures.  In the example (\ref{JE2DSS}) we have already done this.

Mathematically the group of choices of elements of a group $G$ on 2-way overlaps of an atlas of $M$ subject to the 3-way overlap condition (\ref{JETRIPOLA}) has a name.  It is called the first \v{C}ech cohomology group with $G$ coefficients, and is denoted
\beq
\check{\H}^1(M;G).
\eeq
When one uses a good cover \v{C}ech cohomology is isomorphic to the cohomology $\H^*(M;G)$ that we have been using all along.  From now on we will restrict our attention to good covers.  Therefore the group of spin structures on $M$ is $\H^1(M;\Z_2)$.

The classification of $\Z_2$ bundles by characteristic classes in the first cohomology group with $\Z_2$ coefficients is an example of a more general feature in the classification of bundles.  As we will later be interested in a different example, $Spin^c$ bundles, we will now summarize the main features of this structure.  Consider a bundle whose fiber $F$ is $(p-1)$-connected, meaning that it $k<p$ then the $k$th homotopy group of $F$ vanishes.  Imagine that the $p$th homotopy group is nontrivial, and is equal to the group $G$
\beq
\pi_{k<p}(F)=0\hsp \pi_p(F)=G.
\eeq
Then $F$ bundles will be characterized entirely by a degree $p+1$ characteristic class in the cohomology with $G$ coefficients
\beq
\omega_{p+1}\in \H^{p+1}(M;G) \label{JECHAR}
\eeq
and also some characteristic classes of higher degree.  If furthermore all of the homotopy classes of $F$ of degree higher than $p$ vanish then the bundles are entirely characterized by just $\omega_{p+1}$.  

In our case we are interested in $\Z_2$ bundles, so $p=0$ and $G=\Z_2$.  The characteristic class $\omega_1\in\H^1(M;\Z_2)$ is the $spin$ structure.  We will also be interested in circle bundles, for which $p=1$ and $G=\Z$.  Circle bundles are completely characterized by a single characteristic class
\beq
c_1=\omega_2\in\H^2(M;\Z)
\eeq
which is called the first Chern class.

\subsection{When is a Manifold $Spin$?}

Beyond two dimensions a new complication will arise, due to the fact that we are not really classifying $\Z_2$ bundles over spacetime, but rather $Spin$ bundles over spacetime which are $\Z_2$ bundles over Lorentz bundles over spacetime.  $\Z_2$ bundles over any space always exist, for example the trivial bundle always exists.  However, given a particular Lorentz bundle over spacetime, which is the principal bundle associated to the tangent bundle \textbf{TM}, it is not always possible to define a $Spin$ bundle whose transition functions are those of the tangent bundle but in a $Spin$ representation.  The problem is that when the Lorentz bundle is nontrivial, sometimes no choice of signs for the $Spin$ lift satisfies the triple overlap condition (\ref{JETRIPOLA}).  

The triple overlaps are determined entirely by the topology of the tangent bundle, and so provide a characteristic class of the tangent bundle.  While transition functions of our $\Z_2$ bundle are choices of elements of $\Z_2$ on each 2-way overlap and define elements of the first \v{C}ech cohomology group $\check{H}^1(M;\Z_2)$, the obstruction to (\ref{JETRIPOLA}) is a choice of element $x_{ij}x_{jk}x_{ki}$ in $\Z_2$ on each 3-way overlap and defines an element of the second \v{C}ech cohomology group $\check{H}^2(M;\Z_2)$.  

This is part of the above general structure.  $F$ bundles are classified by elements $\omega_{p+1}$ of $\H^{p+1}(M;G)$ but if $G$ is the extension of a group $H$ by $F$ then, given an $H$ bundle, one can define another characteristic class
\beq
\omega_{p+2}\in \H^{p+2}(M;G)
\eeq
which is the obstruction to lifting the $H$ bundle $P$ over $M$ to a $G$ bundle $P\p$ over $M$
\beq
%\begin{array}{ccc}F=\Z_2 & \rightarrow & G=Spin(N)\\
%&&\downarrow\pi\\
%&&H=SO(N)\end{array}
\begin{array}{ccc}F & \rightarrow & G\ \ \\
&&\downarrow\pi\\
&&H\ \ \end{array}
\hsp\hspace{-.1cm}
\begin{array}{ccc}H & \rightarrow & P\ \\
&&\downarrow p\\
&&M\ \end{array}
\hspace{.3cm}\stackrel{\textup{Lift}}{\Longrightarrow}\hspace{.3cm}
\begin{array}{ccc}F & \rightarrow & P\p\ \ \\
&&\downarrow\pi\\
&&P\ \ \end{array}
\hsp
\begin{array}{ccc}G & \rightarrow & P\p \ \ \\
&&\downarrow p\p\\
&&M\ \ .\end{array}
%\begin{array}{ccc}G & \rightarrow & P\p\ \ \\
%\ \downarrow\pi&&\downarrow\pi\ \\
%H & \rightarrow & P\ \ \\
%&&\downarrow p \\
%&&M\ \ .\end{array}
\eeq
%Note that the arrows of the rightmost cartoon do not form an exact sequence, but they commute.  
The $H$ bundle $P$ can be lifted to a $G$ bundle $P\p$ only when $\omega_{p+2}=0$.  In our case $F$ is $\Z_2$, $H$ is the Lorentz group, $G$ is the $Spin$ group and $\omega_{p+2}$ is called the second Stiefel-Whitney class and is denoted $w_2$.  The bundles $P$ and $P\p$ are the tangent and $Spin$ bundles respectively, or more precisely the associated principle bundles.  When $w_2$ is nonzero the triple overlap condition is never satisfied and spinors cannot be globally defined.  If
\beq
w_2(\mathbf{TM})=0
\eeq
then the spacetime $M$ is said to be $spin$.  More generally $w_2$ can be defined for any $SO$ bundle $P$, not just the tangent bundle, and it always measures the obstruction to lifting the $SO$ bundle $P$ to a $Spin$ bundle $P\p$.  

This discussion has been somewhat abstract.  We will now describe a concrete example of the obstruction to the lift of an $SO$ bundle to a $Spin$ bundle.  We will consider a higher rank bundle, although already in two dimensions the Hopf bundle has no $spin$ lift.  In general the topologies of the Lorentz group and the corresponding $Spin$ group are different.  The simplest case in which the topologies are different is the 3-dimensional Euclidean space $\R^3$.  Now the Lorentz group is $SO(3)$ and its $spin$ cover is $Spin(3)\cong SU(2)$.  If spacetime is a Euclidean 3-manifold $M$, then vector fields like the photon are sections of the tangent bundle \textbf{TM} and spinor fields are sections of its square root, which is the spin bundle $\sqrt{\mathbf{TM}}$.  Again, both of these bundles are built using the same trivializations and transition functions, but while one uses the 3 real dimensional $SO(3)$ representation of the Lie algebra $su(2)$ in the definition of the action of the transition functions of the rank 3 tangent bundle \textbf{TM}, one uses the 2 complex dimensional representation $SU(2)$ for the transition functions of the rank 2 spin bundle $\sqrt{\mathbf{TM}}$.

We now ask the question, given an $SO(3)$ bundle $P$ on a manifold $M$, is it always possible to lift $P$ to an $SU(2)$ bundle $P\p$\ on $M$?  Recall that the group manifold $SU(2)$ is a $\Z_2$ bundle over the group manifold $SO(3)$ and so there is a projection map
\beq
\pi:SU(2)\longrightarrow SO(3)
\eeq
with kernel $\Z_2$ that maps $SU(2)$ to $SO(3)$.  The $SU(2)$ bundle $P\p$ is then said to be the lift of the $SO(3)$ bundle $P$ if $\pi$ acting fiberwise on $P\p$ takes it to $P$.  If $P$ is the tangent bundle \textbf{TM} then $P\p$ exists if and only if $M$ is $spin$.  All manifolds of dimensional less than 4 are $spin$, an example of a 4-dimensional manifold that is not $spin$ is the complex projective space $\cp^2$.  However manifolds of dimension four or more are difficult to picture, so we will consider a 2-dimensional example.  As all 2-dimensional manifolds are $spin$, the tangent bundle of a 2-manifold always lifts to a $Spin$ bundle.  Thus to find an example of a bundle that does not lift, we will have to consider a bundle which is not the tangent bundle.

When discussing the worldvolume gauge theories on D-branes one is often interested not in the tangent bundle but in the normal bundle \textbf{NN} of the submanifold $N\subset M$ wrapped by a D-brane.  This is because worldvolume scalars are often sections of the normal bundle and worldvolume fermions are often sections of the $spin$ lift of the normal bundle.  

When classifying topologically nontrivial configurations in gauge theories one is often interested instead in the topology of the gauge bundle itself.  For example, consider an $SU(2)$ gauge theory with no matter or with only adjoint matter.  Gluons always transform in the adjoint representation of the gauge group, and by hypothesis any matter in this theory also transforms in the adjoint.  The only generators of the gauge group that appear in this theory are then those of the adjoint representation, which are three-dimensional and generate the Lie group $SO(3)$.  Therefore this theory really only exhibits an $SO(3)$ gauge symmetry.  The difference between $SU(2)$ and $SO(3)$ is that $SU(2)$ contains an element which is minus the identity, which gets identified with the zero element in the quotient $\pi$ to $SO(3)$.  However conjugating any field by minus the identity leaves the field invariant and so this element acts trivially on our field configurations, only $SO(3)$ can act faithfully.

The fact that this gauge theory is really an $SO(3)$ gauge theory means that gluon configurations are classified by $SO(3)$ bundles and not by $SU(2)$ bundles.  This is a physically important distinction because there are more $SO(3)$ bundles than there are $SU(2)$ bundles.  This is true in part because only $SU(2)$ is simply-connected
\beq
\pi_1(SU(2))=0\hsp \pi_1(SO(3))=\Z_2. \label{JEPGRECA}
\eeq
The nontriviality of the fundamental group of $SO(3)$ implies that, unlike $SU(2)$ gauge theories, $SO(3)$ gauge theories admit $\Z_2$-charged magnetic monopoles and Dirac strings, which in the Higgs phase become physical $\Z_2$ vortices.  

For example, consider an $SO(3)$ gauge theory on $S^2$.  We can trivialize the $SO(3)$ gauge bundle on the northern and southern hemispheres.  The bundle is then topologically classified by the transition function from the equator to $SO(3)$, which is a map in $\pi_1(SO(3))=\Z_2$.  The nontrivial map leads to the $\Z_2$-charged vortex.  If two of these vortices collide they may annihilate, but a lone vortex is topologically stable.  The Douglas-Shenker strings \cite{DS} are examples of these stable vortices in supersymmetric gauge theories with only adjoint matter.

The key observation is that there are no vortices in the $SU(2)$ gauge theory, and so this $SO(3)$ bundle has no $spin$ lift.  In fact, using the general characterization of bundles in Eq.~(\ref{JECHAR}), the first nontrivial homotopy class of $SO(3)$ is at $p=1$ and is given by (\ref{JEPGRECA}).  Therefore $SO(3)$ bundles are partially classified by a characteristic class
\beq
\omega_2\in \H^2(S^2;\Z_2)=\Z_2.
\eeq
This characteristic class is precisely the second Stiefel-Whitney class $w_2$.  Therefore the nontrivial vortex configuration of the $SO(3)$ bundle has nonvanishing $w_2$ and so does not have a $spin$ lift to an $SU(2)$ bundle.

The absence of a $spin$ lift means that one cannot define spinors.  One spinor representation of $SO(3)$ is the two-dimensional fundamental representation of $SU(2)$.  Therefore $SU(2)$ fundamental matter cannot be defined in the presence of a $\Z_2$ vortex, as its wavefunction would be ill-defined.  Instead, in the presence of $SU(2)$ fundamental matter one may only consider $SU(2)$ gauge bundles, which contain no vortices.

Similarly, when spacetime is not $spin$ or equivalently when $w_2$ of the tangent bundle is nonzero, the wavefunctions of Lorentz spinors are ill-defined and so spinors cannot be consistently incorporated in the theory.  In the next subsection we will see that sometimes spinors can be included if they are charged under a $U(1)$ gauge symmetry.

\subsection{$Spin^c$ Structures} 

%However in the presence of a $U(1)$ gauge field, sometimes charged fermions can exist even on manifolds which are not $spin$.  To see how this works, we will examine in more detail how a manifold can fail to be $spin$.  If a manifold is not $spin$, then the $Spin$ bundle, of which the fermion's wavefunction is a section, does not exist.  The problem is that it does not satisfy the triple overlap condition, which must be satisfied by all bundles.  It can be locally trivialized on each path $U_i$.  And there are transition functions relating each pair of patches
%\beq
%f_{ij}:U_i\rightarrow U_j
%\eeq
%which would determine the $spin$ structure had one existed.  But the problem is that at least one of the triple overlaps is negative
%\beq
%f_{ij}f_{jk}f_{ki}=-1
%\eeq
%whereas in the case of a bundle all triple overlaps must be positive.  The set of triple overlaps defines, using what is called \v{C}ech cohomology, a cohomology 2-class which is called the second Stiefel-Whitney class $w_2$.  This class is always $\Z_2$-torsion, so it is a purely quantum phenomenon.  It is zero precisely when all of the triple overlaps are equal to one which occurs precisely when our spacetime is $spin$.

We have seen that when spacetime is not a $spin$ manifold there is a nontrivial class $w_2$ and that this implies that the transition functions $x_{ij}=\pm 1$ of the sign of a spinor do not satisfy the triple overlap condition (\ref{JETRIPOLA}).  Instead at some triple overlaps the product of the transition functions is
\beq
x_{ij}x_{jk}x_{ki}=-1.
\eeq
The fact that the product is not equal to one means that the $Spin$ bundle is not really a bundle.  Therefore spinors cannot be sections of the $Spin$ bundle and therefore cannot appear in the spectrum.  In general a particle can only appear in a theory if its wavefunction is a section of a legitimate bundle that satisfies the triple overlap condition.

While the $Spin$ bundle does not exist in this case, it is sometimes possible to construct a different bundle that does exist from the inconsistent $Spin$ bundle.  The strategy will be to consider a $U(1)$ bundle $Q$ whose second Stiefel-Whitney class is equal to that of $P$.   The tensor product bundle $P\otimes Q$ will then have a $spin$ lift and the sections of its associated vector bundle will be the wavefunctions of $U(1)$-charged spinors.  So while uncharged spinors are still inconsistent, charged spinors may exist if we can define $Q$. 

If there is a $U(1)$ gauge group, then there is also a $U(1)$ gauge bundle $Q$.  $U(1)$ bundles are entirely characterized by a degree two integral cohomology class called the Chern class
\beq
c_1(Q)\in\H^2(M;\Z).
\eeq
Any element of $\H^2(M;\Z)$ defines a $U(1)$ bundle $Q$, and therefore one can construct trivializations and $U(1)$ transition functions and the transition functions always satisfy the triple overlap condition (\ref{JETRIPOLA}).  

Recall that $U(1)$ is isomorphic to $SO(2)$, and so one may also try to define a $spin$ lift $\sqrt{Q}$ of $Q$ whose transition functions are the square roots of those of $Q$.  The Chern class of the square root bundle $\sqrt{Q}$ will satisfy
\beq
2c_1(\sqrt{Q})=c_1(Q). \label{JEC1}
\eeq
This implies that $\sqrt{Q}$ will only be a bundle if $c_1(Q)$ is divisible by two.  We have another criterion for when $\sqrt{Q}$ exists.  The lift $\sqrt{Q}$ of $Q$ is a bundle if and only if its transition functions satisfy the triple overlap condition (\ref{JETRIPOLA}).  These triple overlaps are measured by the \v{C}ech class
\beq
w_2(Q)\in\check{H}^2(M;\Z_2)
\eeq
which consists of the triple products $x_{ij}x_{jk}x_{ki}$ of the transition functions.  The triple products vanish, or more precisely are equal to the identity, precisely when $w_2(Q)$ is equal to zero, but we have seen that the bundle also exists precisely when $c_1(Q)$ is even.   This suggests that $w_2(Q)$ is the mod 2 reduction of $c_1(Q)$
\beq
w_2(Q)=c_1(Q)\textup{\ mod\ 2}
\eeq
and in fact it is.  

Our strategy to make a bundle from the $Spin$ nonbundle $\sqrt{P}$ is to construct another nonbundle $\sqrt{Q}$ that fails the triple overlap conditions at just the same triple overlaps as $\sqrt{P}$.  The tensor product of these two nonbundles $\sqrt{P}\otimes\sqrt{Q}$ will then satisfy the triple overlap condition because the $U(1)$ commutes with everything, so the products of the triple overlaps just multiply.  At each triple overlap the products of the transition functions for the two bundles are either both $+1$ or both $-1$, in either case the products of the transition functions of the tensor bundle is $+1^2=1$ or $-1^2=1$ and so the tensor bundle satisfies the triple overlap condition and is a legitimate bundle.  

To complete the construction, we need only choose $Q$ such that
\beq
c_1(Q)\textup{\ mod\ 2}=w_2(Q)=w_2(P)\hsp c_1(Q)\in\H^2(M;\Z). \label{JEQDEF}
\eeq
While there exists a $U(1)$ bundle for any class $c_1(Q)$ in the integral cohomology group $\H^2(\Z)$, not every class $w_2\in\H^2(M;\Z_2)$ may be obtained as the mod 2 reduction of an integral class.  As we will see in the next subsection, such an integral class exists if and only if a certain 2-torsion integral 3-class called the third Stiefel-Whitney class vanishes
\beq
W_3(P)=0\in\H^3(M;\Z).
\eeq
For the moment we will assume that $W_3$ vanishes and so a U(1) bundle $Q$ exists satisfying (\ref{JEQDEF}).  In this case the tensor product bundle $P\otimes Q$ has a $spin$ lift because
\beq
w_2(P\otimes Q)=w_2(P)+w_2(Q)=2w_2(P)=0
\eeq
where we have used the fact that $w_2$ is $\Z_2$ torsion.  Now we can define our charged spinor wavefunction.  If the spinor has an odd $U(1)$ charge $2q+1$ then its wavefunction is a section of the associated vector bundle to the principal bundle $\sqrt{P}\otimes\sqrt{Q}\otimes Q^q$.

The fiber of the tensor bundle is the product of the groups $Spin$ and $U(1)$.  Both of these groups share the element which is equal to $-1$ times the identity, so to avoid double-counting the fiber of the tensor bundle is the group
\beq
Spin^c(N)=\frac{Spin(N)\times U(1)}{\Z_2}.
\eeq
The tensor bundle is called a $Spin^c$ bundle.  For example if $N=2$ then
\beq
Spin^c(3)=U(2)=\frac{SU(2)\times U(1)}{\Z_2}=\frac{Spin(3)\times U(1)}{\Z_2}.
\eeq
$U(2)$ gauge theories admit Dirac strings, which are the $spin^c$ lifts of the $\Z_2$ vortices in the $SO(3)$ gauge theory.

Using the above general classification of bundles combined with the fact that
\beq
\pi_1(Spin^c)=\Z\oplus\Z_2
\eeq
we find that $Spin^c$ bundles are partially characterized by two degree two cohomology classes.  There is an element of integral cohomology which is just the magnetic flux of the $U(1)$ gauge theory, and there is an element of cohomology with $\Z_2$ coefficients $\H^2(M;\Z_2)$ which is called the $spin^c$ structure.  

Recall that for every degree $p+1$ characteristic class there is a degree $p+2$ obstruction to a lift.  In this case the obstruction to the existence of a $spin^c$ structure is the third Stiefel-Whitney class
\beq
w_3\in\H^3(M;\Z_2).
\eeq
While $w_3$ is always $\Z_2$ torsion, unlike $w_2$ it always has a lift to cohomology with integral coefficients, which will be denoted
\beq
W_3\in\H^3(M;\Z)
\eeq
and is also always $\Z_2$ torsion.  When the third Stiefel-Whitney class of the tangent bundle \textbf{TM} is equal to zero, $M$ is said to be a $spin^c$ manifold and a $spin^c$ lift of the tangent bundle exists.  On such manifolds one can define $U(1)$-charged spinors.  However recall that the $U(1)$ bundles $\sqrt{Q}$ themselves may fail the triple overlap condition by a sign, and in fact they must fail if $M$ is not $spin$ in order to cancel the failure of the $spin$ bundle.  In this case the field strength $c_1(\sqrt{Q})$ defined in (\ref{JEC1}) fails the Dirac quantization condition by a half-integer.  These shifted quantization conditions are often responsible for fractional brane charges in string theory.

\subsection{The Freed-Witten Anomaly} \label{JEFWS}

We will be interested in fermions not in the bulk spacetime, but in the spectra of the open strings that end on a particular D$p$-brane.  While Freed and Witten have demonstrated the existence of their anomaly with and also without the need of the following assumption, we will assume, in line with Sen's conjecture, that it is always possible to nucleate a spacefilling D9 and anti D9 pair from nothing.  Then we may consider the open strings that stretch from any fixed D$p$-brane to the D9.  The ground state Ramond sector open strings that stretch from the D$p$ to the D9 are fermions whose wavefunctions $\psi$ are sections of the tensor product of the spin bundle over the D$p$-brane worldvolume of $N$ and the $U(1)$ line bundle which is the gauge bundle of the $U(1)$ worldvolume gauge field on the D$p$-brane.  This bundle exists precisely when the D$p$-brane's worldvolume is $spin^c$, and so the wavefunctions of the strings between the D$p$ and the D9 only exist if the D$p$ wraps a $spin^c$ submanifold $N$ of the spacetime $M$.  If it does not, the D$p$-brane is said to suffer from a Freed-Witten anomaly which is equal to $W_3$, the failure of $N$ to be $spin^c$.

Fundamental strings couple electrically to the NSNS $B$ field.  Therefore if one introduces a nontrivial $B$ field, the path integral measure of a fundamental string trajectory will change.  In particular, the wave functions of the fermions found in the Ramond groundstate of our $p$-9 strings will be shifted by a $B$ field.  The phase of the wavefunction will increase by the integral of the $B$ field over the worldsheet.  However, if the D$p$-brane was $spin^c$ then the wavefunction was well-defined, and so after multiplying it by a constant shift it will continue to be well defined.  Conversely, if the D$p$-brane wrapped a non-$spin^c$ submanifold then the constant shift provided by the $B$ field will not render the wave function well-defined, in this case the wavefunction is a section of a $spin^c$ bundle which cannot be globally extended over the entire D$p$ worldvolume.  Below we will argue that a nontrivial $H$ flux, which corresponds to a $B$ flux that cannot be globally defined in exactly the same way as the fermion wavefunction, can render the wave function well-defined if and only if
\beq
W_3+H=0. \label{JEFW}
\eeq
To arrive at (\ref{JEFW}) we will need to understand more precisely just what $W_3$ measures.  We will now introduce an exact sequence based characterization of $W_3$ that generalizes to the case with $H$ flux.

Remember that when a $Spin$ bundle does not exist, as in the case of $\cp^2$ which is $spin^c$ but is not a $spin$ manifold, one may not define uncharged fermions but one may define $U(1)$ charged fermions if the gauge bundle fails to be a gauge bundle in just the right places.  %This is because charged fermion wavefunctions are sections of the tensor product $spin\otimes U(1)\cong spin^c$ bundle which does exist on a $spin^c$ manifold.  A shift in the holonomy of a legitimate $U(1)$ bundle is not sufficient for this cancellation, one needs to shift the curvature, in fact one needs to make the Chern class half-integral such that the mod two reduction of twice the Chern class exactly cancels the second Stiefel-Whitney class.  This cancellation ensures that the tensor product of the $Spin$ bundle and the $U(1)$ gauge bundle is an honest bundle which satisfies the triple overlap condition.

If $W_3\neq0$, so that the D$p$-brane's worldvolume is not $spin^c$, then no choice of gauge bundle $Q$, even one that fails the triple overlap condition, can render the $spin^c$ bundle well-defined.  We have already noted that this is a consequence of the condition that $w_2(\textup{TN})$ be the mod 2 reduction of the integral class $c_1(Q)$, as no such integral class exists when $W_3(\textup{TN})$ is nonzero.  Now we will see this more explicitly.  First we will need to understand how the gauge bundle's field strength $F=c_1(\sqrt{Q})$, the second Stiefel-Whitney class $w_2$ and the third Stiefel-Whitney class $W_3$ are related.  They are related by a long exact sequence which arises from the short exact sequence
\beq
0\longrightarrow\Z\stackrel{\times 2}{\longrightarrow}\Z\stackrel{mod\ 2}{\longrightarrow}\Z_2\longrightarrow 0 \label{JESESB}
\eeq
where the first map is multiplication by two and the second is reduction modulo two.  This short exact sequence leads to a long exact sequence in cohomology with $\Z$ and $\Z_2$ coefficients.  

Given a short exact sequence (\ref{JESESB}) one can always form a long exact sequence of cohomology groups with coefficients given by the elements of the short sequence.  In this case one finds
\beq
...\stackrel{\beta}{\longrightarrow}\H^2(M;\Z)\stackrel{\times 2}{\longrightarrow}\H^2(M;\Z)\stackrel{mod\ 2}{\longrightarrow}\H^2(M;\Z_2)\stackrel{\beta}{\longrightarrow}\H^3(M;\Z)\stackrel{\times 2}{\longrightarrow}...
\eeq
where $\beta$ is a map called the Bockstein homomorphism, which is a close relative of the exterior derivative of differential forms.  All of the elements that we have been discussing, the $U(1)$ field strength $F=c_1(\sqrt{Q})\in\H^2(M;\Z)$, the second Stiefel-Whitney class $w_2\in\H^2(M;\Z_2)$ and the third Stiefel-Whitney class $W_3\in\H^3(M;\Z)$ fit into this long exact sequence
\beq
...\stackrel{\beta}{\longrightarrow}c_1(\sqrt{Q})\stackrel{\times 2}{\longrightarrow}c_1(Q)\stackrel{mod\ 2}{\longrightarrow}w_2\stackrel{\beta}{\longrightarrow}W_3\stackrel{\times 2}{\longrightarrow}...\label{JEESATTA}
\eeq
In particular we see that $W_3$ is defined to be the Bockstein of $w_2$
\beq
W_3=\beta w_2.
\eeq

If and only if the D$p$-brane wraps a $spin^c$ submanifold, $W_3$ will be zero.  $W_3$ is the image of $w_2$ under the Bockstein, and so $W_3$ is zero if and only if $w_2$ is in the kernel of the Bockstein homomorphism.  The above sequence is exact, which means that the kernel of the Bockstein map $\beta$ is the image of the preceding mod two reduction map.  Thus $w_2$ is in the kernel of the Bockstein map if and only if there exists an element $c_1(Q)$ of integral cohomology such that the mod two reduction of $c_1(Q)$ is equal to $w_2$.  Remember that this was the condition such that a $U(1)$ bundle $Q$ exists satisfying
\beq
w_2(Q)=w_2(\textup{TM}).
\eeq
Thus the existence of $c_1(Q)$ which reduces modulo two to $w_2$ is equivalent to the condition that the D-brane wrap a $spin^c$ cycle, as desired.

If the D-brane wraps a $spin^c$ cycle which is not $spin$ then $w_2$ is not equal to zero and so $c_1(Q)$ is not in the kernel of mod two reduction.  By exactness of the sequence, this means that $c_1(Q)$ is not in the image of multiplication by two, and so $c_1(\sqrt{Q})$ does not really exist as an integral class.  This corresponds to the fact that the compensating $U(1)$ bundle $\sqrt{Q}$ fails the triple overlap condition.  Circle bundles that fail triple overlap by a sign do not have integral Chern classes, instead they are classified by the integral class $c_1(Q)$ which is intuitively twice their Chern class.  

$c_1(Q)$ is the degree 2 characteristic class which partially characterizes the topology of the $Spin^c$ bundle.  When it fails to exist, because $w_2$ is not in the kernel of $\beta$, then the $Spin^c$ lift of the tangent bundle does not exist.  The fermion wavefunction is a section of the $Spin^c$ bundle, so it does not exist either.  

What changes when we add a nontrivial $H$ field?  The fermion wavefunction no longer is a section of just the $Spin^c$ bundle, but of the tensor product of the $Spin^c$ bundle with another bundle that has characteristic class $B$.  Recall that the sum of the $U(1)$ gauge field $F$ and the $B$ is gauge invariant, and so $F$ and $B$ always appear together in gauge-invariant expressions.  The $B$ field, like $w_2$, is naturally an element of cohomology with torsion coefficients.  This is because it enters in the string partition function via the factor $e^{\int_\Sigma B}$, and so adding an integer to the $B$ field has no observable effect.  

Thus one may add $B$ to $w_2$ in the long exact sequence (\ref{JEESATTA}).  In fact it is the sum that appears in the string partition function.  The failure of $B$ to lift to an integer class in $\H^2(M;\Z)$ is, by exactness, measured by its image under the Bockstein homomorphism.  We refer to its image as the $H$ flux, even though in general it may contain torsion components.  Thus we have the modified long exact sequence 
\beq
...\stackrel{\times 2}{\longrightarrow}c_1(Q)+B\stackrel{mod\ 2}{\longrightarrow}w_2+B\stackrel{\beta}{\longrightarrow}W_3+H\stackrel{\times 2}{\longrightarrow}...
\eeq
where by abuse of notation we refer to both the $B$ field and its integer lift as $B$.

Now $c_1(Q)+B$ is the characteristic class for the combined bundle, whose sections are the fermion wavefunction.  The class $c_1(Q)+B$ is defined to be any class such that its mod two reduction is $w_2+B$, thus the combined bundle exists if and only if $w_2+B$ is in the image of the mod 2 reduction map.  By exactness, this is the case if and only if $w_2+B$ is in the kernel of the Bockstein, which implies (\ref{JEFW})
\beq
0=\beta(w_2+B)=W_3+H.
\eeq
Therefore a D-brane can sometimes consistently wrap a non-$spin^c$ cycle, but this can happen precisely when the $H$ flux cancels $W_3$, the obstruction to the cycle being $spin^c$.  Summarizing, the Freed-Witten anomaly condition states that a D-brane may wrap a cycle if and only if $W_3+H$ of that cycle vanishes as an integral cohomology class.

\section{Twisted K-Theory from the Freed-Witten Anomaly} \label{JEKFW}

\subsection{Brane Insertions and MMS Instantons} \label{JEMMSSEC}

So what happens if you ignore the Freed-Witten anomaly and wrap a D$p$-brane on a cycle with $W_3+H$ flux anyway?  For simplicity, we will just consider the free part of $W_3+H$, the $\Z^k$ terms that survive the supergravity limit, but the same conclusions apply to the torsion part.  We will now provide a heuristic argument for the claim of Ref.~\cite{MMS} that this anomaly may be canceled by a D($p-2$)-brane insertion.  

Recall that there are large gauge transformations that change the $B$ field and the $U(1)$ gauge field strength $F$ but preserve their sum.  Thus one may intuitively try to gauge transform $H=dB$ into $dF$.  Concretely this means the following.  Consider a small, contractible sphere inside of the D$p$-brane.  The integral of the $B$ field over the sphere is not well-defined, as the $B$ field is not gauge invariant.  One may define it via Stoke's theorem as the integral of $H$, either over the inside or the outside of the sphere.  In the first case, one finds that the integral of $B$ is about zero because the sphere is small and $H$ is well-defined.  In the second case, one finds that the integral of $B$ is about equal to the integral of the $H$ flux on some three-cycle.

These two cases are related by a large gauge transformation, and so $F$ must also differ in the two cases by the same amount, the integral of $H$ over a three-cycle.  By Stoke's theorem, we can then conclude, using either gauge, that the integral of $dF$ over the union of these two three-cycles is equal to the integral of $H$.  $dF$, by the Bianchi identity, is equal to the magnetic monopole charge that intersects the 3-cycle.  Thus for every unit of $H$ flux on a given cycle there is a single unit of magnetic monopole charge on that cycle.  Considering again the full integral cohomology we conclude that there is a magnetic monopole which is Poincar\'e dual to $W_3+H$
\beq
Q_{\mathbf{monopole}}=\textup{PD}(W_3+H).
\eeq

Recall that magnetic flux $F$ on a D$p$-brane worldvolume couples to the RR field $C_{p-1}$ and so carries D$(p-2)$-brane charge.  A magnetic monopole is a codimension three surface that is a source for magnetic flux, so it is the endpoint of a D$(p-2)$-brane.  Thus a D$p$-brane is always free to wrap a cycle on which $W_3+H$ is nonvanishing, but the price is that there must be a D$(p-2)$ brane which ends on the D$p$-brane on a cycle dual to the offending $W_3+H$.  The D$(p-2)$ extends away from the brane until it finds another place to end, or, if there is no other place to end, it is semi-infinite.  Such configurations of branes ending on branes were referred to as baryons in Ref.~\cite{BBA} and, for reasons that will be apparent momentarily, as (MMS) instantons in Ref.~\cite{MMS}.

Depending on the compactification in question, baryons may have infinite masses because the D$(p-2)$-branes may be forced to extend to infinity.  If the transverse dimensions are compact then the D$(p-2)$-branes may have no place for the other endpoint, and so the baryons configurations may be inconsistent.  Of course, the D$(p-2)$-branes may end on an anti D$p$-brane wrapping the same cycle, but the D$p$ and anti D$p$ may decay so this system carries no conserved charges.  Thus baryonic configurations are often dropped in classifications of D-brane charges.  As we have mentioned, while baryonic configurations carry homology charges, they do not carry K-theory charges.  Thus K-theory does \textit{not} classify baryons.

The Freed-Witten anomaly not only eliminates certain configurations from our classification, but also it leads to the nonconservation of a charge.  To see this, consider again a baryonic configuration in which a D$p$-brane wraps a cycle with nonvanishing $W_3+H$ and a D$(p-2)$ ends on the D$p$ and extends away in another direction.  If this other direction is a spatial direction, then one arrives at the high-mass spider-like configuration described above.  But the other direction could also be timelike.  This corresponds to D$(p-2)$-branes being created or destroyed.  Thus the nonconserved charge is a D$(p-2)$-brane charge, it is the charge of the homology $(p-2)$-cycle wrapped by the created or destroyed branes.  This charge maps to the trivial K-theoretic charge, as K-theory is the quotient of the legal homology charges by a set of nonconserved charges.  These timelike configurations are called MMS instantons because the D$p$-brane may live for only a short period of time as it sweeps out a $(p+1)$-cycle and its trajectory may be a solution to the Euclidean equations of motion.

When $g_s\neq 0$ the intersection of a D$(p-2)$-brane and a D$p$-brane is not a sharp corner.  Rather there is a continuous transition, in which one sees, if one looks closely at the D$(p-2)$, that it is always a thin tube of D$p$-brane with $S^2$ cross-sections that carry D$(p-2)$-charge because the integral of the $U(1)$ field strength over these spheres is nonzero.  The spheres may be thought of as D$p$ dipoles, as antipodal points carry opposite D$p$ charge, thus they carry no net D$p$ charge and appear to be D$(p-2)$-branes from far away.  The actual monopole is at the end of the tube, at infinity, so in a sense these branes avoid the Freed-Witten anomaly by putting it all at a point and cutting out that point.  If instead of extending away in a spatial direction the D$(p-2)$ extends in the time direction then the configuration is an MMS instanton.  Smoothing out the corners at nonzero string coupling one finds that this corresponds to, for example, a homologically nontrivial stack of D$(p-2)$ branes that grows, via the Myers dielectric effect \cite{Myers} for example, into a D$p$-brane and sweeps out a $(p+1)$-cycle before eventually collapsing into heat as it finishes and closes up.

In conclusion, the Freed-Witten anomaly is responsible for two physical effects.  First, D-branes carrying $W_3+H$ not equal to zero are, without extra insertions, inconsistent.  Secondly D-branes wrapping homology cycles which are Poincar\'e dual to $W_3+H$ in some bigger cycle are unstable, and their corresponding homology charges are not preserved.  The inconsistent charges do not lift to K-theory classes, while the unstable charges correspond to the zero K-theory class.

Now that we have turned on fluxes, the tachyon condensation argument for the K-theory classification no longer applies.  In fact D-branes are not quite classified by K-theory when there is a topologically nontrivial $H$ flux, instead they are classified by a variation of K-theory known as twisted K-theory \cite{BM} where the twist is given by $H$.  Some details on the construction of twisted K-theory from cohomology will appear in the next subsection.  This construction, known as the Atiyah-Hirzebruch spectral sequence, will make it apparent that the Freed-Witten anomaly indeed removes and quotients out the correct homology charges to arrive at twisted K-theory.

\subsection{Atiyah-Hirzebruch Spectral Sequence}

D$p$-branes in type IIA string theory exist for every even $p$.  The branes carrying conserved RR charges are extended in one time direction and $p$ spatial dimensions, and so are classified by the even homology groups $\H_p(M)$.  By Poincar\'e duality these are isomorphic to the odd cohomology groups
\beq
\H^{9-p}(M)=\H_p(M).
\eeq
While D$p$-brane charges for a fixed $p$ are classified by the single odd cohomology group $\H^{9-p}$, the collection of all D$p$-brane charges in type IIA string theory is the direct sum of these groups for all even values of $p$.  We refer to this direct sum as the odd cohomology and denote it $\HO$.  Similarly D-branes in IIB are classified by the even cohomology $\HE$.  Summarizing, all D-branes in type II string theories carry charges in either $\HE$ or $\HO$ which are defined by
\beq
\HE(M)=\oplus_k \H^{2k}(M)\hsp \HO(M)=\oplus_k \H^{2k+1}(M).
\eeq
However we know that these charge groups are too big, we really want the twisted K-groups $\K^0_H$ and $\K^1_H$ which will be quotients of subsets of $\HE$ and $\HO$ respectively.  

The Atiyah-Hirzebruch spectral sequence (AHSS) was introduced by Atiyah and Hirzebruch to compute untwisted K-groups, although it has since been generalized to the twisted case.  It is an algorithm which begins with $\HE$ and $\HO$.  It arrives at twisted K-theory after a series of approximations, the $n$th of which are denoted $E^0_n$ and $E^1_n$.  At each step some elements are removed and some are quotiented out.  After a finite number $N$ of steps the algorithm arrives at groups which as sets are equivalent to the twisted K-groups $\K^0_H$ and $\K^1_H$
\beq
E^0_1=\HE(M)\hsp E^1_1=\HO(M)\hsp |E^i_N|=|\K^i_H(M)|
\eeq
which classify D-branes in IIB and IIA string theories respectively.

To calculate the $n$th approximation $E^i_n$ of $\K^i_H$, one needs both of the $(n-1)$st approximations $E^0_{n-1}$ and $E^1_{n-1}$ and also a differential operator $d_{2n-1}$ which maps one to the other
\beq
d_{2n-1}:E^0_{n-1}\leftrightarrow E^1_{n-1}\hsp d_{2n-1}d_{2n-1}=0.
\eeq
The subscript $2n+1$ on the differential reflects Atiyah and Hirzebruch's observation that each element of $E^i_{n-1}$ can be represented by a class in the degree $2k+i$ cohomology group $\H^{2k+i}$ for some $k$, and $d_{2n-1}$ increases the degree of this class by $2n-1$.  However, in the cases $n\ge 2$, the action of $d_{2n-1}$ on a given cohomology class is only well defined up to the quotients that occurred at previous steps in the spectral sequence, and so the action is only well defined on $E^i_{n-1}$ itself.  

Given $E^i_{n-1}$ and $d_{2n-1}$ one can determine the $E^i_n$.  They are simply the cohomology of the $E^i_{n-1}$ with respect to the differentials $d_{2n-1}$  
\beq
E^i_{n}=\frac{\Ker(d_{2n-1}:E^i_{n-1}\rightarrow E^{i+1}_{n-1})}{\Im(d_{2n-1}:E^{i+1}_{n-1}\rightarrow E^{i}_{n-1})}.    
\eeq

Only the first differential, $d_3$, has been explicitly computed in the literature.  It is
\beq
d_3x=Sq^3x+H\cup x \label{JED3}
\eeq
where $Sq^3$ is a cohomology operation known as a Steenrod square which takes an integral class in the degree $k$ cohomology to a class in the degree $k+3$ integral cohomology, as does the cup product with $H$.  The image of $Sq^3$ is always a $\Z_2$ torsion component of $\H^{k+3}(M;\Z)$.  Recall that in the supergravity limit the torsion components disappear, as the finite cyclic group $\Z_2$ tensored with the real numbers is zero.  Therefore in the classical theory the $Sq^3$ term vanishes and $d_3$ reduces to the wedge product with the $H$ flux.  

While limited results about the higher differentials have appeared, they have been computed in general in the case of real coefficients \cite{AS}, in which they compute the tensor products of the K groups with the real numbers.  The $n$th differential on a differential form $x$ is
\beq
d_{2n+1}x=[H,...,H,x] \label{JEREALE}
\eeq
which is the Massey product of $n$ copies of a differential form representing $H$ with $x$.  Massey products are intuitively products of differential forms multiplied by inverses of the exterior derivative.  These inverses in general are not well-defined on differential forms, or even on the de Rham cohomology of differential forms.  However they are well-defined on $E^i_{n-1}$, as required.  For this reason the Massey product is called a secondary cohomology operation.  Note in particular that the real differential (\ref{JEREALE}) disappears in the case $H=0$.  Recalling that the real case corresponds to the supergravity limit, we find that $p$-branes in supergravity are classified by de Rham cohomology if $H=0$.  

However the nontriviality of (\ref{JEREALE}) when $H$ is nontrivial implies that even in the supergravity limit the homology or equivalently the cohomology classification of D-branes fails.  There is a an important exception to this last statement.  A manifold is called {\it{formal}} if all Massey products with more than two terms vanish.  Many of the kinds of manifolds that are of interest in compactifications are formal.  For example, all K\"{a}hler manifolds are formal, as are all simply-connected six-manifolds.  The differentials (\ref{JEREALE}) at $n>1$ vanish on a formal manifold and so branes in supergravity on a formal background are just classified by the $H\wedge$ cohomology of the de Rham cohomology, which is isomorphic to the $d+H\wedge$ twisted cohomology.

\subsection{Freed-Witten from the AHSS}

One may now ask what the AHSS has to do with the Freed-Witten anomaly.  Recall that the Freed-Witten anomaly provided a necessary condition for the consistency of a homology-valued D-brane charge, and also yielded a set of D-brane charges that are not conserved.  It did not provide a sufficient condition for consistency, because some nonrepresentable cycles that do not lead to Freed-Witten anomalies cannot be wrapped.  On the other hand, the AHSS computes the twisted K-theory precisely, at least as a set, from the cohomology.  Thus it must eliminate all unphysical branes, those afflicted with the Freed-Witten anomaly and those wrapping unwrappable singularities.  It must also quotient by all nonconserved charges, those corresponding to branes that may decay via MMS instantons and those that may decay via trajectories in which the brane grows, wraps a nonrepresentable cycle and then is sucked into a worldvolume singularity that forms, eats its hosts and finally eats itself.

Thus the consistency of the AHSS and the Freed-Witten anomaly, which is critical for the K-theory classification, requires that the AHSS indeed eliminates all Freed-Witten anomalous D-branes, although it may eliminate others.  We will now argue that this is the case.  In particular, we will show that cohomology classes which are not in the kernel of $d_3$ are Poincar\'e dual to D-brane worldvolumes that suffer from the Freed-Witten anomaly, while its image consists of cohomology classes which are dual to the worldvolumes of unstable D-branes.  However some Freed-Witten anomalous D-branes and Freed-Witten unstable D-branes will be missing from this classification, we will note that in the only known example of this phenomenon \cite{MMS} these anomalies and instabilities are captured by $d_5$ and correspond to MMS instantons in which a D$(p-4)$-brane ends on a D$p$-brane.

We will need the definition of the Steenrod square $Sq^3$, which appears in the differential $d_3$ in (\ref{JED3}).  Consider an integral cohomology class $x^p$ of degree $p$.  This class is Poincar\'e dual, in the 9-dimensional timeslice $M^9$, to a homology class $x_{9-p}$ of degree $9-p$
\beq
\textup{PD}(x^p)=x_{9-p}\in\H_{9-p}(M;\Z)
\eeq
which corresponds to a spatial slice of the worldvolume of a D$(9-p)$-brane.  In particular, a spatial slice of the worldvolume is a $(9-p)$-cycle $N\subset M$ that represents the homology class dual to $x^p$.  The submanifold $N\subset M$ has a normal bundle \textbf{NN}, which is a $p$-dimensional real vector bundle on $N$.  

%asses, there is a more general usage of the third Stiefel-Whitney class $W_3$.  Given any real vector bundle \textrm{E} on $N$, one can define an $n$th Stiefel-Whitney class $w_n(\textup{E})$.  In the case in which $E$ is the tangent bundle \textbf{TN}, the Stiefel-Whitney classes $w_n(\textup{E})=w_n(\mathbf{TM})$ are often simply denoted $w_n$ and named the $n$th Stiefel-Whitney class.  Thus the Stiefel-Whitney class $W_3$, that determines whether a submanifold $N$ is $spin^c$, is the third Stiefel-Whitney class of the tangent bundle $W_3(\mathbf{TN})$.

%Now we will be interested in the Stiefel-Whitney classes of the normal bundle \textbf{NN}.  

Recall that $W_3(\mathbf{NN})$ is the integral cohomology 3-class which measures the obstruction to the existence of a $spin^c$ lift of the normal bundle \textbf{NN}.  The Steenrod squares are defined to be the pushforwards $i_*$ of the Stiefel-Whitney classes of the normal bundle of $N$ onto $M$ using the inclusion map $i:N\hookrightarrow M$
\beq
sq^n x=i_* w_n(\textup{NN})\hsp i_*:\H^n(N)\longrightarrow \H^{p+n}(M). \label{JESQ}
\eeq
Recall that homology classes pushforward naturally and cohomology classes pullback naturally.  The pushforward of a cohomology class is defined to be the Poincar\'e dual in $M$ of the pushforward of its Poincar\'e dual in $N$
\beq
sq^n x=i_* w_n(\textup{NN})=\textup{PD}|_M(i_*\textup{PD}|_N(w_n(\textup{NN}))).
\eeq
The unnaturalness of the pushforward of a cohomology class is responsible for the fact that we need an infinite tower of differentials on the cohomology of $M$ to capture a single anomaly on the worldvolume of $N$, as well as the fact that the pushforward of a cohomology class is a class of a different degree.  We use the notation $Sq^3$ to denote the integral lift of $sq^3$, which always exists.  Equation (\ref{JESQ}) combined with the linearity of the pushforward map implies that $Sq^3$ will be trivial when $W_3(\mathbf{NN})$ is trivial, but it does not imply the inverse.  

This is not quite the Freed-Witten condition for two reasons.  First $W_3(\mathbf{NN})$ is not quite the Freed-Witten anomaly, the Freed-Witten anomaly was the the third Stiefel-Whitney class of the tangent bundle.  Second, $i_*$ can have a nontrivial kernel, so some D-branes afflicted with the Freed-Witten anomaly and so having a nontrivial $W_3$ can nonetheless lead to a trivial $d_3$.  We will now resolve the first issue and speculate about the second, then we will extend to the case with nontrivial $H$ flux.

To relate the differential defined in (\ref{JESQ}) to the Stiefel-Whitney class of the tangent bundle, we first observe that the direct sum of the tangent and normal bundles to the submanifold $N\subset M$ is the tangent bundle of $M$, restricted to $N$
\beq
\textup{TN}\oplus \textup{NN} = \textup{TM}|_N.
\eeq
We can now insert this relation into the addition formula for Stiefel-Whitney classes
\beq
w(A\oplus B)=w(A)\wedge w(B) \label{JEWADD}
\eeq
where $w(A)$ is the sum of the Stiefel-Whitney classes of the bundle $A$.  As usual a lower case $w$ denotes a Stiefel-Whitney class in the cohomology with $\Z_2$ coefficients.  Only the odd degree Stiefel-Whitney classes $w_{2k+1}$ can be lifted to integral cohomology classes $W_{2k+1}$, and even these will continue to be $\Z_2$ torsion
\beq
2W_{2k+1}=0. \label{JENIL}
\eeq

Expanding the addition law (\ref{JEWADD}) in components, in the case $A$=\textbf{TN}, $B$=\textbf{NN}, $A\oplus B$=\textbf{TM}$|_N$, and using the identity $w_0=1$ we find at degree one
\beq
w_1(\textup{TM}|_N)=w_0(\textup{TN})\wedge w_1(\textup{NN})+w_1(\textup{TN})\wedge w_0(\textup{NN})=w_1(\textup{NN})+w_1(\textup{TN}). \label{JEPRIMA}
\eeq
We recall that the first Stiefel-Whitney class of a manifold is 0 if and only if the manifold is orientable.  D-branes in type II string theory are always orientable, as are compactification manifolds.  Therefore 
\beq
w_1(\textup{TM})=w_1(\textup{TN})=0. \label{JEORI}
\eeq
Let $i$ be the inclusion map of the D-brane worldvolume $N$ into $M$, then the Stiefel-Whitney classes of the tangent space \textbf{TM} of $M$ restricted to $N$ are just the pullbacks of those of \textbf{TM} using $i^*$
\beq
i:N\hookrightarrow M\hsp w_k(\textup{TM}|_N)=i^*w_k(\textup{TM}).
\eeq
In particular, the fact that the first Stiefel-Whitney class of the tangent space of $M$ vanishes (\ref{JEORI}) implies that its image $w_1(\mathbf{TM}|_N)$ under the homomorphism $i^*$ also vanishes.  Thus the left hand side and the last term on the right hand side of (\ref{JEPRIMA}) both vanish.  This means that the other term on the right hand side of (\ref{JEPRIMA}) must also vanish
\beq
w_1(\textup{NN})=0.
\eeq
We will need this fact in the following argument.

Our goal is to relate the two third Stiefel-Whitney classes $W_3(\mathbf{TN})$ and $W_3(\mathbf{NN})$, the first of which appears in the Freed-Witten anomaly and the second of which appears in the AHSS.  These classes both appear in the degree three part of (\ref{JEWADD})
\begin{eqnarray}
w_3(\textup{TM}|_N)&=&w_0(\textup{TN})\wedge w_3(\textup{NN})+w_1(\textup{TN})\wedge w_2(\textup{NN})\\
&&+w_2(\textup{TN})\wedge w_1(\textup{NN})+w_3(\textup{TN})\wedge w_0(\textup{NN})\nonumber\\
&=&1\wedge w_3(\textup{NN})+0+0+w_3(\textup{TN})\wedge 1=w_3(\textup{NN})+w_3(\textup{TN}).\nonumber\label{JETRE}
\end{eqnarray}
Compactification manifolds $M$ in type II string theory are always $spin^c$.  Therefore
\beq
w_3(\textup{TM}|_N)=i^*w_3(\textup{TM})=i^*0=0.
\eeq
Thus (\ref{JETRE}) implies that the third Stiefel-Whitney classes of the normal and tangent bundles to $N$ add to zero.  Using the fact (\ref{JENIL}) that Stiefel-Whitney classes are mod 2 torsion, either of these classes may be moved to the other side of the equality to yield
\beq
w_3(\textup{TN})=w_3(\textup{NN}).
\eeq
This equality is preserved by their integral lifts.

We can now reformulate the Freed-Witten anomaly as follows.  In the absence of $H$ flux, D$p$-branes must wrap cycles $N$ such that the third Stiefel-Whitney class of the normal bundle of $N$ is trivial.  If $W_3(\mathbf{NN})$ is zero then we saw in (\ref{JESQ}) that square three of the Poincare dual of $N$ vanishes.  Thus the cohomology class $x$ dual to a D$p$-brane's worldvolume is annihilated by the first AHSS differential
\beq
d_3x=Sq^3 x=i_* W_3(\textup{NN})=0 \label{JEWCON}
\eeq
if $W_3(\mathbf{NN})=0$ which in turn occurs if and only if $W_3(\mathbf{TN})=0$, which occurs if the D$p$-brane wrapping is Freed-Witten anomaly free.  Therefore consistent D$p$-brane wrappings are in the kernel of $d_3$, as they must be in order to lift to K-theory classes.  

This argument is not invertible.  In fact the invertibility fails at two points.  First, some inconsistent D$p$-brane wrappings have trivial $W_3(\mathbf{TN})$, although only when $N$ is nonrepresentable so this will not occur when $M$ is 9-dimensional or less.  It may also fail because sometimes $W_3(\mathbf{NN})\neq 0$ but $Sq^3 x^p=0$ because $W_3$ is in the kernel of the pushforward map $i_*$ in (\ref{JESQ}).  In Ref.~\cite{MMS} it was shown that such a situation arises in the identification of the twisted K-theory of $SU(3)$ with D-brane charges in the SU(3) WZW model.  As will be seen in Subsec.~\ref{JEWZW}, in this case the unphysical branes are in the kernel of $d_3$ but are not in the kernel of $d_5$, and so they do not carry K-theory charges.  Generalizing this result, in Ref.~\cite{Uday} a general form of the Freed-Witten part of the $d_5$ term was proposed.

The main advantage of the Freed-Witten perspective of the K-theory classification over the Sen conjecture perspective is that one can easily include a nontrivial $H$ flux.  We will now argue that, even in the presence of $H$ flux, D$p$-branes that are Freed-Witten anomaly free are Poincar\'e dual to cohomology classes that are in the kernel of the AHSS differential $d_3$, as they must be if their charges are to lift to K-theory classes.

The Freed-Witten anomaly on a D$p$-brane's worldvolume is equal to the sum of the third Stiefel-Whitney class of its normal bundle plus the integral of the pullback of the $H$ flux onto its worldvolume.  More precisely, now using the language of integral cohomology and not that of differential forms, the $H$ flux in the bulk defines an integral 3-class in $\H^3(M;\Z)$ and its contribution to the anomaly is
\beq
H\cap N\hsp \cap:\H^j\times \H_k\longrightarrow \H_{k-j}
\eeq
where $\cap$ is an operation called the cap product.  Roughly speaking the cap product takes a $p$-manifold $N$ and a 3-class $H$ and produces a $(p-3)$-manifold which is the Poincar\'e dual of $H$ in $N$, in other words, it produces the magnetic monopole in the worldvolume gauge theory of the D$p$-brane.

The cap product satisfies a useful relation.  If $x^{9-p}$ is the degree $9-p$ cohomology class which is Poincar\'e dual to a $p$-dimensional timeslice $N$ of the worldvolume of our D$p$-brane, then the magnetic monopole is the Poincare dual (PD) of $H\cup x^{9-p}$
\beq
H\cup x^{9-p}=H\cup \textup{PD}(N)=\textup{PD}(H\cap N). \label{JEHCON}
\eeq
where we recall that the cup product $\cup$ is the integral cohomology version of the wedge product of differential forms, which we have been using implicitly for most of this note.  For example the $H$ flux on a D$p$-brane's worldvolume is topologically trivial if and only if $H\cup x^{9-p}$ is topologically trivial.  

Adding the contribution of the $H$ flux (\ref{JEHCON}) to the Freed-Witten anomaly to the contribution of $W_3$ from (\ref{JEWCON}) we find that the vanishing of the entire Freed-Witten anomaly $W_3+H$ implies the vanishing of
\beq
d_3 x=Sq^3 x + H\cup x = i_*(W_3(\textup{NN})+H)=0.
\eeq
Thus, as desired, Freed-Witten anomaly free configurations are in the kernel of $d_3$.  The higher differentials have never been explicitly calculated, but it is known that Freed-Witten anomaly free D$p$-branes which wrap representable cycles are in fact in the kernel of all of the differentials $d_{2k+1}$ at least in the untwisted case, and so these D-branes all carry K-theory charges. Furthermore, the AHSS guarantees that these charges are trivial precisely when the D-brane wraps a cycle whose Poincar\'e dual is in the image of one of the differentials $d_{2k+1}$.

\subsection{The Supergravity Limit and Twisted Homology}

In the supergravity limit everything is tensored by the real numbers $\R$.  This kills the $W_3$ term, as torsion groups tensored with the group of real numbers are trivial and $W_3$ is always $\Z_2$ torsion by (\ref{JENIL}).  Similarly, Steenrod squares are all identically equal to zero after tensoring by the reals, as their images are also torsion.  Thus in the supergravity limit the AHSS depends entirely on $H$.  The first nontrivial differential is given by the wedge product with $H$, and more generally all of the differentials are given by the Massey products in (\ref{JEREALE}).  

Working with differential forms, which capture all of the information in the classical theory, one may begin the AHSS with the set of not just closed differential forms but all differential forms.  Then one arrives at the set of closed differential forms by introducing the differential $d_1$ equal to the exterior derivative and taking its cohomology, which is de Rham cohomology.  The forms in the image of the exterior derivative are exact.  These do not correspond to trivial configurations, but rather to configurations that contain nontrivial RR fluxes and no D-branes.  %As a result of the absence of D-branes the RR fluxes, even the non-improved RR fluxes $dC$, are globally defined.

Poincar\'e duality only applies to closed differential forms, and so general differential forms are difficult to relate to D-branes.  However one may interpret every step in terms of D-branes if, instead of passing through differential forms, one works with submanifolds, or more precisely simplicial complexes, from the beginning.  Then the first differential $d_1$ is the boundary operator, and the cohomology of the boundary operator consists of all D-branes which have no boundaries quotiented by those that wrap boundaries and so can decay, in other words, the homology group.  The second differential $d_3$ can be defined to be the cap product with $H$, whose kernel consists of branes that contain no magnetic monopoles
\beq
d_1=\partial\hsp d_3=H\cap. \label{JETO}
\eeq
In the full quantum theory in which one uses integral homology, one needs to include a term in $d_3$ which captures $W_3$.  However in the classical theory, and also in the quantum theory in certain cases, $W_3$ vanishes.  In particular if $M$ is any orientable 6-manifold then $W_3$ vanishes, and so in orientifold-free flux compactifications of type II which yield 4-dimensional physics one can drop the $W_3$ term and use $d_3=H\cap$.  

In general one needs the entire series of differentials from (\ref{JEREALE}), but when the compactification manifold possesses a special property called formality we have noted that they all vanish, at least rationally.  Simply connected 6-manifolds are always formal, for example.  Also all manifolds which are diffeomorphic to Kahler manifolds are formal.  Generalized K\"{a}hler manifolds may also be formal, at least they share many of the nice properties of formal manifolds \cite{CavalcantiFormale}.  Thus, in flux compactifications one can generally compute the twisted K-theory using the simple differentials (\ref{JETO}).  

In fact, on formal manifolds the cohomology with respect to these two differentials is isomorphic to the cohomology with respect to the single differential \cite{Cavalcanti}
\beq
\partial_H=\partial+H\cap. \label{JEBORDO}
\eeq
The cohomology of $\partial_H$ is called twisted homology, in the real case it was applied to the classification of D-branes on calibrated cycles in Refs.~\cite{Luca1,Luca2}.  Notice that the two summands are of different dimensions, the first reduces the dimension of a submanifold by 1 and the second by 3.  Thus twisted homology classes contain networks of branes whose dimensions are all equal modulo 2.  These are all baryons. 

More generally, and perhaps more usefully for phenomenology, on all simply connected 6-manifolds twisted homology and twisted K-theory are equivalent as sets.  This is a consequence of the fact that the obstructions to this equivalence, $d_5$ in twisted K-theory and the next element of the spectral sequence in the identification of $d+H$ with $d$ and $H$ cohomologies are both degree 5 operations.  Degree 5 operations on a 6-manifold either have images in $H^5$ or operate on elements of $H^1$, both of which are trivial in the simply connected case, and so these two obstructions vanish.

The cap product with a differential form is only well defined on homology classes, but in (\ref{JEBORDO}) it is applied to a general submanifold which may have a boundary.  To render it well defined, one needs to include two pieces of extra data on the submanifold.  One needs a nonquantized 2-form called the $B$-field whose exterior derivative is the pullback of the $H$ field, and one needs an integral two-class $F$, that turns out to be the field strength of the worldvolume gauge bundle.  Then the boundary map yields
\beq
\partial:N\mapsto N\p\hsp \int_{N\p} B+F=\int_N H.
\eeq
This only defines the sum of $B+F$.  There are large gauge transformations which change $B$ and $F$, but these must always preserve both the sum and the quantization condition on $F$.

The extra data needed to define a twisted homology class not only matches with the physical expectations of a D-brane, but it also precisely matches the extra data needed to define a twisted generalized complex submanifold in Ref.~\cite{Gualtieri}.  This is because cycles in both cases can be thought of as sections of a  bundle whose fibers are the Eilenberg-MacLane space $\K(\Z,2)$ with 3-class $H$ or alternately of its $S^2$ subbundle with Euler class $H$.  We recall that an Eilenberg-MacLane space $\K(G,n)$ is defined to be any space whose only nontrivial homotopy group is
\beq
\pi_n(\K(G,n))=G
\eeq
so that, in particular, $\K(G,n)$ bundles are characterized completely by the characteristic class $\omega_{n+1}$ in $\H^{n+1}(M;G)$, which in the case of $\K(\Z,2)$ is the $H$ flux
\beq
H=\omega_3\in\H^3(M;\Z).
\eeq
The extra data is the 2-form connection $B$ on this bundle and also its choice of trivialization $F$.  The subbundle realization is particularly useful because the homology of the total space of the sphere subbundle is just an extension of the ordinary homology of $M$ by its twisted homology, thus it provides a tool for the calculation and also the visualization of twisted homology classes.

Unlike K-theory, which topologically can only be twisted by a 3-form, homology can be twisted by any $p$-form $G_p$.  One then needs a worldvolume $(p-1)$-form connection and trivialization, which describe the fibration of a $K(\Z,p-1)$ bundle or its $S^{p-1}$ subbundle with characteristic class or Euler class $G_p$.  Thus while D-branes are sections of $K(\Z,2)$ bundles, M-branes are sections of $K(\Z,3)$ bundles, or equivalently are subsets of the total space of the bundle.  The Freed-Witten anomaly is then interpreted as the topological obstruction to the existence of a section of the bundle restricted to the D-branes worldvolume.  If one uses the approximation $E_8$ for $K(\Z,3)$ and the based loopgroup of $E_8$ for $K(\Z,2)$ one arrives at the characterization of D-branes in $E_8$ gauge theories in Refs.~\cite{WitFluxQuant,MEK,MEMONO}, which was based on the $E_8$ gauge theory of Ref.~\cite{DMW,Allan}.  While the existence of this bundle away from the boundaries is not evident in perturbative string theories, $PU(\infty)$, which is the quotient of $U(\infty)$ by its $U(1)$ center, is also a model for $K(\Z,2)$ and the existence of a $PU(\infty)$ bundle in the bulk in string field theory is well-known, see for example Refs.~\cite{Harvey,Carlo}.

\subsection{RR Magnetic Field Strengths} \label{JERR}

Ramond-Ramond fluxes are sourced by D-branes.  Thus one might expect that inequivalent RR field strengths are in one to one correspondence with inequivalent D-branes and so should be classified by the same twisted K-group.  Moore and Witten proposed just this in Ref.~\cite{MW}.  If one believes that any acceptable RR field can be sourced by some D-brane, in other words, that if a nontrivial flux is supported on some cycle then that cycle is free to degenerate and at the point where it degenerates one will find a D-brane sourcing the flux, then one version of the K-theory classification of RR field strengths is as follows.

Consider type II string theory on a spacetime which is topologically $\R\times M$ where $\R$ may be thought of as the time direction and $M$ is compact.  Recall that conserved D$p$-brane charges are characterized by D$p$-branes that extend along the time direction and also wrap a $p$-cycle $N\subset M$.  If the spacetime metric were really the Cartesian product of $\R\times M$ then such an eternal configuration would be inconsistent because the sourced flux would have nowhere to go, however no such assumption is made on the metric.  Thus, as in the familiar cases of D-branes on $AdS^p\times M$, configurations in which cycles of $M$ are wrapped are consistent because $M$ gets large sufficiently quickly far from the origin of $AdS^p$.  So we will not worry about the problem of fluxes having no place to go, essentially because $\R\times M$ is noncompact.  We have argued that in type IIA $p$ is even and conserved D-brane charges are classified by $\K^1_H(M)$, whereas in type IIB $p$ is odd and conserved D-brane charges are classified by $\K^0_H(M)$.

To extend this classification to fluxes one needs the above conjecture, that fluxes are in one to one correspondence with the branes that source them.  More concretely, consider time-dependent D$p$-brane configurations in which a D$p$-brane nucleates, wrapping a contractible $p$-cycle, sweeps out a $(p+1)$-cycle $N$ and then shrinks into oblivion on the other side of $N$.  If $N$ fails to lift to a K-theory class, for example if there is a Freed-Witten anomaly or if it is nonrepresentable, then when the D$p$-brane disappears a lower-dimensional D-brane will remain.  However if $N$ does lift to a K-theory class then this process, while perhaps not energetically favorable, is possible.

Recall from Section~\ref{JESUGRA} that D$p$-branes are violations of the Bianchi identity
\beq
ddC_{7-p}=\textup{PD}(\textup{D$p$-brane})=\textup{PD}(N). \label{JEBIANCHI2}
\eeq
Therefore one can use Stoke's theorem to write Gauss' Law (\ref{JEGAUSS}) for the flux over any $(8-p)$-cycle $N\p$ that links the D$p$-brane worldvolume $N$.    As our D$p$-brane only existed during a finite lifetime, we can choose the union of a timeslice before and after the D-brane came into existence.  The difference between the unimproved field strength $dC_{7-p}$ on the timeslice before and after will, using (\ref{JEBIANCHI2}) and an integral cohomology version of Stoke's theorem, be the Poincar\'e dual of $N$.  As $N$ sweeps out a $(p+1)$-dimensional cycle in the 9-manifold $M$, the Poincar\'e dual will be a degree $(8-p)$ integral cohomology class of $M$
\beq
x=\textup{PD}(N)\in\H^{8-p}(M).
\eeq

It was crucial that the slices which link the brane are compact.  Technically this is necessary because otherwise the linking number would vanish and Stoke's theorem would not apply.  Also noncompact slices would be disastrous as Poincar\'e duality is an isomorphism of the cohomology groups of compact spaces.  In fact, the fluxes on a noncompact cycle are not quantized, as Dirac's argument requires a nontrivial linking number and so fluxes on noncompact cycles cannot be quantized by either integral cohomology or integral K-theory.  Thus only the fluxes along $M$, corresponding in the language of differential forms to forms with all legs along $M$, will be quantized.  In the language of electrodynamics these are the magnetic fluxes, which are related by Hodge duality to the electric fluxes, which by definition have one leg along the time direction.  This distinction is important, as RR field strengths in the democratic formulation \cite{GHT} of type II supergravity, which we are using, are self-dual
\beq
G_p=\star G_{10-p}
\eeq
where $\star$ is the Hodge star.  The self-duality implies that the magnetic fields completely determine the electric fields, and vice versa.  Thus a classification of the magnetic fields is a complete classification of all RR fields.

In particular, if we start with the trivial flux $dC=0$ then the conjectured relation between D-branes and RR field strengths implies that we can turn on any flux $x$ by nucleating a D-brane which then sweeps out the dual cycle $N$.  Conversely, given any consistent nucleation, sweeping and annihilation of a D-brane one creates some flux.  Thus RR field strengths $dC_{7-p}$ on $M$ are conjectured to be classified by the group that classifies not D$p$-brane charges but the surfaces that they sweep out in $M$.  In the case of IIA this is the twisted K-group $\K^0_H(M)$ and in the case of type IIB it is the twisted K-group $\K^1_H(M)$.

Recall that all D-branes carry homology charges.  Similarly all fluxes carry cohomology charges.  The AHSS eliminates some of these charges and quotients others to arrive at the twisted K-theory classification.  We identified the eliminated D-brane charges as those charges which are carried only by anomalous branes, and we noted that the anomalies may be canceled by inserting branes that end on certain submanifolds of the anomalous branes.  We also identified quotiented D-brane charges as nonconserved, and saw that they are carried by the branes whose insertions cancel anomalies.  Thus we provided a physical interpretation for the mismatch between the homology and the K-theory classification of branes.

A similar physical interpretation exists for the mismatch between the cohomology and K-theory classification of fluxes.  We have seen that fluxes that do not lift to K-theory are created by Freed-Witten anomalous branes.  These D$p$-branes fail to decay, leaving behind lower dimensional D-branes.  Thus RR field strengths whose cohomology classes fail to lift to K-theory can only exist in the presence of these lower-dimensional D-branes.  This is similar to the observation that D-branes whose charges do not lift to K-theory are only consistent in the presence of lower dimensional D-brane insertions.  

\begin{table}[h]
\begin{center}
\begin{tabular}{|l|l|l|}
\hline \textbf{} & \textbf{RR Charges} & \textbf{RR Field Strengths}\\
\hline In Image of $d_p$: &Charge is not conserved&Pure gauge\\
\hline Not In Kernel of $d_p$: & Baryons and MMS instantons & Contains D-branes\\
\hline
\end{tabular}
\end{center}
\caption{Interpretation of those RR charges and field strengths that are in the image of the AHSS differentials and those that are not in the kernel}
\end{table}

When discussing the worldvolume theories of branes in supergravity we argued that the field strengths $dC$ are only well-defined up to large gauge transformations, which at the level of differential forms are the wedge products of $H$ with another integral class.  By comparison, RR field strengths that lift to the zero twisted K-theory class are those that are in the image of the AHSS differentials, which include the image of $H\wedge$ as well as torsion and Massey product corrections.  This leads one to the conjecture that while the gauge symmetries of the classical supergravity (\ref{JEDC}) shift $dC$ by forms in the image of $H\wedge$, the gauge symmetries of the full quantum string theory shift $dC$ by the classes in the image of the AHSS differentials.  Thus RR field strengths whose cohomology classes lift to the trivial K-class are conjectured to be pure gauge.

Summarizing, magnetic RR field strengths $dC$ appear to be classified by twisted K-theory.  While naively they are classified by cohomology, those whose cohomology class does not lift to K-theory can only exist in the presence of unstable D-branes and those whose cohomology class lifts to the trivial K-theory class are pure gauge.

One can simultaneously classify D-branes and RR fluxes in configurations that contain both, as the RR fluxes that lift to twisted K-theory do not require the insertion of any D-branes.  However one must remember in such cases that the true RR flux is not just the corresponding K-theory element, but one must also add the RR flux sourced by the D-brane.  If the D-brane is unstable then it is a trivial element in K-theory, and one must add an RR flux which is a cohomology class that does not lift to K-theory such that an AHSS differential of the cohomology class of the RR flux yields the cohomology class which is Poincar\'e dual to the unstable D-brane.

Notice that we have classified RR field strengths $dC$, but have not classified RR potentials $C$.  Gauge potentials are continuously valued and so are not classified by ordinary twisted K-theory.  It has been conjectured that they are classified by a generalization of twisted K-theory known as differential K-theory \cite{Freed}, which characterizes not just the topologies of bundles but also their Wilson loops.

\section{Examples} \label{JEWZW}

The literature contains a modest number of examples of compactifications in which the twisted K-theory of the spacetime or of a spatial slice has been computed and has been successfully tested against an independent calculation of the D-brane spectrum.  This is because the full spectrum of stable D-branes can only be computed in a few special classes of compactifications.  More generally applicable D-brane classifying schemes, such as derived categories \cite{AD,Aspinwall}, only classify BPS D-branes.  D-branes which wrap torsion homology classes, for example, may be stable but are never BPS nor even calibrated and so are missed by schemes which are based on supersymmetry or calibrations.

However in some special classes of simple string models one can calculate the full D-brane spectrum.  The simplest class of examples consists of two-dimensional topological field theories in which the target space is a finite set of points.  In Ref.~\cite{MS} it was proven, using sewing conditions for the field theories, that boundary conditions are indeed classified by K-theory.  While this analysis relied on the fact that the spacetimes considered are $spec$ of semisimple Frobenius algebras, the authors hint that the K-theory classification of D-branes may extend to general topological string theories.

More traditional tests of the K-theory classification are in the context of exactly solvable conformal field theories.  One set of such exactly solvable conformal CFTs is the WZW model corresponding to a Lie group $G$, which describes strings propagating on the group manifold $G$.  When $G$ is semisimple and compact one can compute the D-brane spectrum directly from the admissible boundary conditions of the worldsheet \cite{FS}, and one invariably finds twisted K-theory with a twist determined by the level of the affine Lie algebra corresponding to $G$.  In this section we will review the first two cases in which the spectra derived from CFT boundary conditions and from twisted K-theory have been compared, the $SU(2)$ and $SU(3)$ WZW models.

\subsection{The SU(2) WZW Model}

The group manifold of $SU(2)$ is the three-sphere.  Three-spheres appear in spacetime, for example, in type II string theory on a topologically trivial space with $k$ flat, coincident NS5-branes.  The NS5-branes are linked by a three-sphere and, although the dilaton diverges close to the NS5-brane, fundamental strings away from the brane are described by an exactly solvable CFT.  The radial coordinate of the embedding of a fundamental string is described by a nonlinear dilaton model and the embedding on the linking three-sphere is described by an SU(2) WZW model at level $k-2$, which is unitary when $k>2$.  Using Gauss' Law, the integral of the $H$ flux on the linking 3-sphere is equal to $k$
\beq
\int_{S^3} H=k
\eeq
and so we will expect our twisted K-theory to be twisted by the class
\beq
k\in\H^3(S^3)=\Z.
\eeq

In the sequel we will not restrict our attention to this embedding of the CFT in type II superstring theory, but rather we will consider the WZW model on its own.  This CFT has a 3-dimensional target space and so is not critical when coupled to 2-dimensional gravity, in particular it has a central charge of 
\beq
c_{SU(2),k-2}=\frac{3k-6}{k-1}.
\eeq
Topologically inequivalent boundary conditions of the CFT wrap conjugacy classes of the three-sphere, which are points or two-spheres.  

Surprisingly only a discrete set of conjugacy classes is consistent \cite{AS}.  This is because a boundary state on a two-sphere is a wavefunction on each choice of loop in the two-sphere on which a fundamental string may end.  The phase of this wavefunction must be well-defined on the loopspace, or else the boundary state is ill-defined.  In particular if one takes the loop and moves if off of the 2-sphere to the north and then pulls it back on from the south, dragging it to its original position, one needs to arrive at the same state. In general, however, the state shifts by the integral of the worldvolume field strength $F$ over the 2-sphere, reflecting the fact that the loop on which the string ends is the trajectory of an electrically charged particle in the D-brane's worldvolume gauge theory.  This shift is also easily calculable in the CFT.  The authors of Ref.~\cite{AS2} found that, imposing a particular boundary condition on the string worldsheets, there are only $k$ possible wrappings in which this shift is a multiple of $2\pi$, and so only $k$ conjugacy classes may be wrapped.  

Today we know that any 2-dimensional submanifold may be wrapped, but in Ref.~\cite{BDS} it was shown that the the D-branes will feel a Myers dielectric force \cite{Myers} that integrates to a potential energy function which has one of $k$ minima, corresponding to the $k$ special conjugacy classes found in Ref.~\cite{AS2}.  The minimum applicable to a given brane depends on the integral of $F$ on its worldvolume.  Of these $k$ conjugacy classes, $k-2$ are two-spheres at constant, evenly spaced latitudes and two are points, the north and south poles.  The locations of the poles may be changed by applying an inner automorphism of $SU(2)$ to the ansatz of boundary conditions investigated in \cite{AS2}.  The existence of yet more general boundary conditions remains an open question.

Now we are ready to compute the twisted K-theory of the 3-sphere with $k$ units of $H$ flux and compare it to the conformal field theory result that there should be $k$ stable even-dimensional D-brane configurations and no odd-dimensional D-brane configurations.  The homology and cohomology of the three-sphere, like that of any sphere, is torsion-free and is generated by a single generator at the lowest and highest dimension
\begin{eqnarray}
&&\H_0(S^3)\cong H^3(S^3)=\Z\hsp \H_1(S^3)\cong H^2(S^3)=0\nonumber\\
&&\H_2(S^3)\cong H^1(S^3)=0\hsp 
\H_3(S^3)\cong H^0(S^3)=\Z
\end{eqnarray}
where $\cong$ is Poincar\'e duality.  Recall that the $H$ flux $H$ is the element $k$ of $\H^3$.

As the cohomology ring is trivial beyond dimension three, all cohomology operations of dimension greater than three are trivial.  Therefore only the dimension three AHSS differential $d_3$ may be nonzero.  In addition the image of $Sq^3$ is always torsion but there are no torsion classes and so the $Sq^3$ term in $d_3$ is trivial.  It is also trivial for dimensional reasons.  Therefore only the $H\cup$ term in $d_3$ may be nonvanishing.  This augments the degree of a cohomology class by 3, and so it kills all of the 3-classes since its image would be a six class but there are no nonzero six classes.

If we let $a$ and $b$ be the generators of $\H^0$ and $\H^3$ respectively, then we have found
\beq
d_3:\H^0(S^3)\longrightarrow\H^3(S^3):a\mapsto kb\hsp d_3:\H^3(S^3)\longrightarrow\H^6(S^3):b\mapsto 0.
\eeq
Therefore the kernel of $d_3$ consists of all elements in $\H^3(S^3)$ while the image consists of those elements of $\H^3(S^3)$ which are divisible by $k$.  The fact that the image is a subset of the kernel was guaranteed by the nilpotency of $d_3$.  

The twisted K-theory of $S^3$, as a set, is just the quotient of the kernel of $d_3$ by its image.  In particular there are no even classes in the kernel and so $\K_H^0(S^3)$ is trivial
\beq
\K^0_H(S^3)=\frac{\Ker(d_{3}:\HE\longrightarrow\HO)}{\Im(d_3:\HO\longrightarrow\HE)}=\frac{0}{0}=0. 
\eeq
On the other hand $\K^1_H$ is nontrivial.  The kernel of $d_3$ acting on the odd cohomology, in particular acting on $\H^3(S^3)$, is all of $\H^3(S^3)$ and so is a copy of the integers $\Z$ which contains the image $k\Z$ as a proper subgroup
\beq
\K^1_H(S^3)=\frac{\Ker(d_{3}:\HO\longrightarrow\HE)}{\Im(d_3:\HE\longrightarrow\HO)}=\frac{\Ker(d_{3}:\H^3\longrightarrow\H^6)}{\Im(d_3:\H^0\longrightarrow\H^3)}=\frac{\Z}{k\Z}=\Z_k. 
\eeq
These may be identified with the K-homology groups that classify D-branes via Poincar\'e duality
\beq
\K_0^H(S^3)\cong \K^1_H(S^3)=\Z_k\hsp\K_1^H(S^3)\cong \K^0_H(S^3)=0.
\eeq
The triviality of $\K_1^H$ agrees with the CFT expectation that there are no odd-dimensional branes.  $\K_0^H$ contains $k$ elements which again is in agreement with the CFT result that the level $k-2$ $SU(2)$ WZW model contains $k$ inequivalent D-brane embeddings.

Physically the quotient by $k\Z$ corresponds to the fact that a D2-brane may sweep out the 3-sphere, and the cancellation of its Freed-Witten anomaly causes it to change the D0-brane charge by $k$ units during this process.  As a result D0 charge is only conserved modulo $k$.  These D2-branes are the MMS instantons described in Subsec.~\ref{JEMMSSEC}.  The fact that one restricts to the kernel of $d_3$, thus eliminating all of $\H^0$, corresponds to the fact that all space-filling D-branes are FW anomalous since the integral of $H$ over their worldvolumes is equal to $k$ and not to zero.

\subsection{The SU(3) WZW Model}

The twisted K-theory of the group manifold $SU(3)$ was calculated in Ref.~\cite{MMS}.  The authors used their result to classify conserved D-brane charges in type II string theory on $\R^{1,1}\times SU(3)$, where branes do not wrap the spatial $\R$, so as to avoid the tadpoles resulting from D9-brane charge.  Again their results were successfully compared against the CFT expectations, and against the expectation that consistent branes on a manifold of dimension 9 or less are precisely those whose Freed-Witten anomaly vanishes.  Here we will sketch their calculation of the K-groups.

We begin by reviewing the nontrivial homology and cohomology classes of the group manifold $SU(3)$, which are identical to that of the similar manifold $S^3\times S^5$
\begin{eqnarray}
&&\H_0(SU(3))\cong H^8(SU(3))=\Z\hsp \H_3(SU(3))\cong H^5(SU(3))=\Z\nonumber\\
&&\H_5(SU(3))\cong H^3(SU(3))=\Z\hsp 
\H_8(SU(3))\cong H^0(SU(3))=\Z.
\end{eqnarray}
Notice that again there are no torsion groups and so $Sq^3$ is trivial.  Therefore $d_3$ is just the cup product with the $H$ flux, which is $k$ times the generator of $\H^3$ when the WZW model is at level $k-3$.

As $d_3$ increases the degree by 3, it may only be nontrivial on $\H^0$ and $\H^5$, and in fact it is.  Let $x^i$ be the generator of $\H^i$.  Then the nontrivial actions of $d_3$ are just 
\beq
d_3:\H^0(S^3)\longrightarrow\H^3(S^3):x^0\mapsto kx^3\hsp d_3:\H^5(S^3)\longrightarrow\H^8(S^3):x^5\mapsto kx^8
\eeq
while $d_3$ annihilates $x^3$ and $x^8$.  The second step in the AHSS is then the quotient of the kernel of $d_3$, consisting of $\H^3\oplus\H^8$ by its image, which is the sublattice of the kernel which is divisible by $k$
\begin{eqnarray}
E^0_{2}&=&\frac{\Ker(d_{3}:\HE\rightarrow \HO)}{\Im(d_{3}:\HO\rightarrow \HE)}=\frac{\H^8}{k\H^8}=\frac{\Z}{k\Z}=\Z_k\nonumber\\ 
E^1_{2}&=&\frac{\Ker(d_{3}:\HO\rightarrow \HE)}{\Im(d_{3}:\HE\rightarrow \HO)}=\frac{\H^3}{k\H^3}=\frac{\Z}{k\Z}=\Z_k. \label{JEPREK}
\end{eqnarray}
The D-branes that have survived up to this point are Poincar\'e dual to the third and eighth cohomology classes, identifying them as D0 and D5-branes wrapping a point and a 5-cycle in SU(3) respectively.  The nontrivial image of $d_3$ is the result of two MMS instantons, a D2-brane which violates D0 charge and a baryonic D7-brane which violates D5 charge.  The branes excluded because they are not in the kernel of $d_3$ are the D3-brane wrapping the $S^3$ with D1 insertions and the D8-brane wrapping all of $SU(3)$ with $k$ D6-brane insertions.  The D6 insertions, for example, wrap the cycle $N$ which is Poincar\'e dual to $x^3$ and also extend along one time direction.

Had we been considering the direct product of spheres $S^3\times S^5$, Eq.~(\ref{JEPREK}) would have already been the twisted K-theory.  But SU(3) is somewhat different.  The 5-manifold $N$ is not $spin^c$, and so a brane wrapping this 5-manifold may suffer from a Freed-Witten anomaly.  The nontrivial cohomology of the non-$spin^c$ 5-cycle $N$ is
\beq
\H^0(N)=\Z\hsp \H^3(N)=\Z_2\hsp \H^5(N)=\Z.
\eeq
As $N$ is not $spin^c$, $W_3$ is the nontrivial class in $\H^3(N)$.  The pushforward of this class onto the cohomology of $SU(3)$ is $\Z_2$-torsion, but there are no torsion classes in the cohomology of $SU(3)$, which as we noted implies that $Sq^3$ is zero.  Therefore $W_3$ is in the kernel of the pushforward that defines $Sq^3$ and so does not affect $d_3$.  However if $W_3+H$ is nonzero then the brane is nonetheless anomalous and so does not carry a K-theory charge. Therefore it must fail to be in the kernel of a higher AHSS differential.  As the only other odd spacing between cohomology classes is 5, anomalous branes must not be in the kernel of $d_5$.

We still have not yet commented on when $W_3+H$ is nonzero.  In Ref.~\cite{MMS} the authors found that
\beq
W_3+H=1+k\in\Z_2=\H^3(N)
\eeq
and so the D5 is anomalous if and only if $k$ is even.  Therefore $d_5$ is 2-torsion, and is nontrivial when $k$ is even, which identifies it as
\beq
d_5 x^3=\left\{
\begin{array}{cl}
\frac{k}{2}x^8&\textup{if $k$ is even}\\
0&\textup{if $k$ is odd}\\
\end{array}\right.\hsp d_5 x^8=0.
\eeq
While nontrivial actions of $d_3$ describe D$p$-branes on which D$(p-2)$-branes end, those of $d_5$ describe D$p$-branes on which D$(p-4)$-branes end.  In particular, when $k$ is even, a baryonic D5-brane wrapping $N$ has a FW anomaly which is canceled by $k/2$ D$1$-brane insertions which each end at a point on each timeslice.  Similarly, a D4-brane which sweeps out the 5-cycle absorbs $k/2$ D$0$-branes, and so D$0$-brane charge is conserved modulo $k/2$ and not modulo $k$ as one would have suspected using only (\ref{JEPREK}).

One may now calculate the twisted K-theory of $SU(3)$ with $k$ units of $H$ flux.  When $k$ is odd, $d_5$ is trivial and so as a set the K-groups are isomorphic to $E_2$
\beq
\K_0^{H=k}\cong\K^0_{H=k}=\Z_k\hsp \K_1^{H=k}\cong\K^1_{H=k}=\Z_k.
\eeq
On the other hand when $k$ is even one needs to take the quotient of the kernel of $d_5$ in $E_2$ by its image.  $d_5$ kills every class in $E_2^0$  and so the kernel includes all of $E_2^0=\Z_k$.  On the other hand it only kills the odd classes in $E^1_2$, and so its odd kernel is $2E_2^1=\Z_{k/2}$.  The image does not contain any nonzero classes in $E^1_2$, and contains, besides zero, only the single class
\beq
\frac{k}{2}\in\Z_k=E^0_2
\eeq
in $E^0_2$, therefore one needs to quotient only by the $\Z_2$ subgroup of $E^0_2$ generated by the element $k/2$.  In all we have then found \cite{MMS}
\begin{eqnarray}
\K_0^{H=k}\cong\K^0_{H=k}&=&\frac{\Ker(d_5:E^0_2\rightarrow E^1_2)}{\Im(d_5:E^1_2\rightarrow E^0_2)}=\frac{\Z_k}{\Z_2}=\Z_{k/2}\nonumber\\    
\K_1^{H=k}\cong\K^1_{H=k}&=&\frac{\Ker(d_5:E^1_2\rightarrow E^0_2)}{\Im(d_5:E^0_2\rightarrow E^1_2)}=\frac{\Z_{k/2}}{0}=\Z_{k/2}.   
\end{eqnarray}
In particular there are $k/2$ topologically distinct even and also odd brane charges when $k$ is even, whereas there are $k$ even and $k$ odd distinct charges when $k$ is odd.

\section{Problems} \label{JEPROBLEMI}

While the K-theory classifications of RR fields and D-branes has been reasonably successful, it suffers from several fundamental problems.  

\subsection{K-theory vs Homology Revisited}

One of the easiest objections to make is that D-branes can wrap any homology cycle, and so D-branes can be classified by homology.  D-branes wrapping homology cycles which are not in the kernels of some of the AHSS differentials will not carry K-theory charges, and so they will necessarily have anomalies which are canceled by the insertions of lower dimensional D-branes.  These configurations, as we have frequently reiterated, are the baryons introduced by Witten in Ref.~\cite{BBA}.  They are not inconsistent if the inserted branes are not themselves anomalous \cite{aussy2005} and if the other end of the inserted branes can go to infinity or to a horizon or boundary.  

However for some practical purposes one may wish not to include baryonic branes, as they often have infinite energies.  Upon Kaluza-Klein reduction they tend to correspond to particles which are confined, for example to 't Hooft-Polyakov monopoles in an $N=2$ gauge theory whose supersymmetry is softly broken to $N=1$.  Such monopoles may be confined by some number of vortices depending on the monopole charges \cite{ABE}, and each of these vortices is the dimensional reduction of one of the inserted branes.  An example in which the pattern of the confinement in $N=1$ gauge theories can be read from relative homology has appeared in Ref.~\cite{WittenMQCD2} and this discussion is generalized in Ref.~\cite{Stefano}.  Thus the homology classification of branes captures all brane charges, while the K-theory classification only captures the charges of nonbaryonic branes.  Therefore the choice of classification depends on the physical question being asked.

Not only are the branes which are not closed under the AHSS differentials legitimate states, but the brane charges in the image are in a sense conserved.  When a brane that carries a nontrivial homology charge but a trivial K-theory charge is destroyed by an AHSS instanton, the instanton leaves behind a RR field strength.  This RR field strength in turn can be used to reconstruct the original brane charge.  Thus, while the D-brane charge itself is not conserved, a combination of the brane charge and the flux is conserved.  Combining this with some information from the integrals of the RR connections one reproduces the fact that improved field strengths are precisely gauge invariant and so there are conserved charges valued in de Rham cohomology.  Thus, for some questions of conserved charges, the answer is not twisted K-theory but rather de Rham cohomology.  One arrives at twisted K-theory when one drops all information about RR fluxes and tries to only classify the D-branes.

Similarly one may argue that RR field strengths should be classified by cohomology.  Improved fields strengths are gauge invariant and are classified by twisted $d+H\wedge$ de Rham cohomology.  However one may also argue that unimproved field strengths should be classified by integral cohomology.  Some integral cohomology classes do not lift to twisted K-theory classes, but the corresponding field strengths are not inconsistent, they merely correspond to configurations with RR charge.  Thus, if one is willing to consider configurations with RR charges, then unimproved RR field strengths may assume any value in integral cohomology and not just those which lift to K-theory classes.

Again one may ask whether RR field strengths which are exact under the AHSS differentials should be identified with zero.  Such RR field strengths carry a K-theory charge which is equal to zero, as they can be shifted away by a large gauge transformation.  These large gauge transformations also add an integral cohomology class to an RR gauge connection.  Thus, just as the decay of K-theoretically trivial D-branes leads to an integral RR field strength, the decay of a K-theoretically trivial RR field strength leads to an integral RR gauge connection, for example
\beq
dC_p\longrightarrow dC_p-H\cup\Phi\hsp C_{p-2}\longrightarrow C_{p-2}+\Phi
\eeq
as in Eq.~(\ref{JEDC}).  While in the bulk such a shift, which is merely an integral shift of a Wilson loop, is unobservable, on the worldvolume of a D-brane it has a physical effect.  The RR gauge connections couple to the worldvolume field strengths of D-branes, for example as a theta angle when the flux is codimension four.  An integral shift of the theta angle leaves the gauge theory invariant, but may shift the charges assigned to some of the objects in a given solution.  Thus in some applications one may wish to distinguish between the different integral classes of connections and so between the different cohomology classes of RR field strengths which are identical as K-theory classes.

Summarizing, D-branes are classified by both K-theory and by homology, and fluxes are classified by both K-theory and cohomology.  However for certain physical applications one classification scheme may be preferable to the other.  For example, as we saw in Sec.~\ref{JEWZW}, twisted K-theory reproduces the spectrum of boundary states in some CFTs.  We will see in Subsec.~\ref{JEKS} that homology classes of D-branes classify the ranks of gauge groups in Klebanov-Strassler gauge theories whereas twisted K-theory classes classify the universality classes of the gauge theories.

\subsection{The Star Problem}

The RR field strengths of the democratic formulation of the type II supergravity are not independent.  Instead they double count the degrees of freedom.  This redundancy is killed by the star condition
\beq
G=\star G \label{JESTELLA}
\eeq
which relates the improved $p$-form field strength to the improved $(10-p)$-form field strength.  The operator $\star$ is the Hodge star, which is a continuous function of the metric.  In particular, the ratio between two dual components of $G$ varies continuously with the metric and so is in general irrational.  One may think that this poses no problem, as $G$ is a differential form which is not closed and so cannot be quantized in the usual way.  

To see that this is a problem, consider the special case in which the NS 3-form $H$ vanishes
\beq
H=0\hsp G_{p+1}=dC_p.
\eeq
The $(p+1)$-forms $dC_p$ are quantized, and so in this case $G_{p+1}$ is quantized.  This quantization is required for the D$(p-1)$-brane partition function to be well-defined, and in the K-theory perspective it is required because the $dC$'s are Chern characters, which are always rational.  However the ratio of two components depends on the star and so is generically irrational, which is a contradiction.

Physically this problem is solved by choosing a polarization, as is done in a similar context in Ref.~\cite{WitTop}.  That is, one must choose one independent set of half of the $dC$'s, called a polarization, and quantize them, this is the largest set that can be simultaneously quantized.  Then in the partition function one only sums over this half.  Finally, one must check that the partition function would be the same given any other choice of polarization.  For example, in QED on a compact spacetime one must choose between quantizing the electric and the magnetic field strengths, or perhaps some combination on various cycles, but it is impossible to quantize all of the fluxes on all of the cycles for a generic metric.

While physically this appears to be the correct way to make sense of the theory, it is difficult to interpret in the K-theory context.  If one interprets the $dC$'s as Chern characters, then only half of the Chern characters are well-defined.  There appear to be three proposals for when such a choice can make sense.

First, one may need to replace K-theory with some sort of quantized version of K-theory.  The K-groups $\K^0(M)$ and $\K^1(M)$ are isomorphic to the set of maps $[M\longrightarrow X]$ from $M$ to a space $X$ called the classifying space of $\K^0$ or $\K^1$
\beq
\K^0(M)=[M\longrightarrow BU(\infty)]\hsp \K^1(M)=[M\longrightarrow U(\infty)] \label{JECLASS}.
\eeq
Chern characters are pullbacks of cohomology classes corresponding to the homotopy classes of the classifying space $X$ using the map (\ref{JECLASS}) from spacetime to the classifying space of K-theory.  This is consistent with the fact that $BU(\infty)$ has nontrivial homotopy at all of the even degrees and $U(\infty)$ has nontrivial homotopy at all of the odd degrees
\beq
\pi_{2k}(BU(\infty))=\pi_{2k+1}(U(\infty))=\Z\hsp \pi_{2k+1}(BU(\infty))=\pi_k(U(\infty))=0
\eeq
and correspondingly $\K^0$ is characterized by even-degree Chern characters and $\K^1$ is characterized by odd degree Chern characters.

We want to choose a polarization, which means that only some of the Chern characters and so only some of the pullbacks may be well-defined.  Only some of these pullbacks will be well-defined if, for example, one replaces these maps by sections of bundles in which the classifying space is nontrivially fibered over the spacetime.  As no global section exists for a nontrivial fibration, one can only determine local sections over a submanifold of the base.  The choice of submanifold may give a choice of polarization.  However unfortunately while nontrivial $U(\infty)$ bundles over 10-dimensional manifolds always exist, $BU(\infty)$ bundles are not characterized by a 10-dimensional class, like $ch_5$ for a $U(\infty)$ bundle, and so it is not clear what fibration to use.  Thus it is easier to quantize $\K^1$ than $\K^0$.

A second way out is to consider a manifold in which there is a natural polarization which is the cohomology of a submanifold, and then to only consider the K-theory of the submanifold.  In fact, we have already done this when we applied K-theory to classify conserved D-brane charges.  In this case our spacetime was $\R\times M^9$.  When one direction is $\R$ it need not be quantized, but if we only insist that a K-class is defined on $M^9$ we may even choose to replace $\R$ with $S^1$ so that spacetime is compact.  In either case, we only quantized the fluxes on the compact $M^9$, which were classified by the twisted K-theory of $M^9$.  This is a polarization of the cohomology and of the K-theory by the K\"{u}nneth theorem
\begin{eqnarray}
\H^p(M^9\times S^1)&=&\H^p(M^9)\otimes\H^0(S^1)\oplus\H^{p-1}(M^9)\otimes\H^1(S^1)\nonumber\\
&=&\H^p(M^9)\otimes\Z\oplus\H^{p-1}(M^9)\otimes\Z=\H^p(M^9)\oplus\H^{p-1}(M^9)\nonumber\\
\K^p(M^9\times S^1)&=&\K^p(M^9)\otimes\K^0(S^1)\oplus\K^{p-1}(M^9)\otimes\K^1(S^1)\nonumber\\
&=&\K^p(M^9)\otimes\Z\oplus\K^{p-1}(M^9)\otimes\Z=\K^p(M^9)\oplus\K^{p-1}(M^9)
\end{eqnarray}
which demonstrates that the cohomology (K-theory) of $M^9\times S^1$ is just two copies of the cohomology (K-theory) of $M^9$, one in which the forms have a leg around the circle and one in which they do not.  

If the circle direction is considered to be time then the half of the classes with a leg on a circle are the electric fluxes and the other half are the magnetic fluxes.  This analysis works equally well with the circle replaced by $\R$ as it was above, but then there is no Dirac quantization argument for the quantization of electric fluxes.  In either case, however, it is consistent to quantize only the magnetic fluxes as we have done in the classification of conserved RR charges.  Thus, when spacetime is of the form $M^9\times\R^1$ or $M^9\times S^1$ the star problem has a natural solution which allows one to continue to classify brane charges and magnetic RR field strengths as elements of the K-theory of $M^9$.  It is still unknown whether the topology of our universe is of such a form, if it began at a fixed time in a big bang then it is not.

A third resolution to the star problem is to eliminate half of the Chern characters by replacing the classifying space of K-theory by something with less cohomology, in particular, which does not include the cohomology classes of the Chern characters to be eliminated.  While the first resolution only worked for $\K^1$, this resolution only works for $\K^0$.  The problem is that $\K^1$ is classified by 5-classes $ch_{5/2}$ which are interrelated by the Hodge duality, and so one cannot simply eliminate the cohomology class in the classifying space $U(\infty)$ of $\K^1$ whose pullback is $ch_{5/2}$, as one would lose all of the 5-classes.  Physically, such a solution would either annihilate all of the components of IIB supergravity's self-dual 5-form $G_5$ or it would eliminate none of them, but it would not eliminate the one-half required for a polarization.

For example, we may choose a polarization of type IIA supergravity in which $G_0$, $G_2$ and $G_4$ are quantized and independent and $G_6$, $G_8$ and $G_{10}$ are found by solving the star condition (\ref{JESTELLA}).  Recall that $G_{2p}$ is the Chern character $ch_p$ which is the pullback of the cohomology class of the generator
\beq
\pi_{2p}(BU(\infty))=\Z.  
\eeq
Therefore we need to replace $BU(\infty)$, the classifying space of $\K^0$, with a new space whose sixth, eighth and tenth homotopy groups are trivial.  

Rather than working with $BU(\infty)$, recall the fact that, by definition, $BU(\infty)$ is the classifying space not only of $\K^0$ but also of $U(\infty)$ bundles.  The homotopy groups of the classifying space $BP$ of $P$-principle bundles are related to the homotopy groups of the fiber $P$ by the theorem
\beq
\pi_k(BP)=\pi_{k-1}(P)
\eeq
therefore the classes of $\pi_{2p}(BU(\infty))=\Z$ correspond to the classes $\pi_{2p-1}(U(\infty))=\Z$.  Thus to eliminate the sixth, eighth and tenth homotopy groups of $BU(\infty)$ it suffices to remove the fifth, seventh and ninth homotopy groups of $U(\infty)$.  This obviously drastically changes the Lie group $U(\infty)$, and the resulting Lie group is not uniquely defined as, for example, the higher homotopy groups are invisible to the 10-dimensional physics and so can apparently be chosen at will.

One particularly simple choice of truncation of $U(\infty)$ is $LE_8$, the centrally extended loop group of $E_8$, this is affine $E_8$ at a level which we will identify with $G_0$.  The identification of the level with $G_0$ reproduces several Freed-Witten anomalies as topological obstructions to the existence of this bundle \cite{MEK}.  Considering massless IIA, the Romans mass $G_0$ is equal to zero and so the $LE_8$ fiber is just the Cartesian product of the free loop group of $E_8$ with $U(1)$.  The only nontrivial homotopy groups, up to dimension 14, of this fiber are
\beq
\pi_1(LE_8)=\pi_2(LE_8)=\pi_3(LE_8)=\Z. \label{JEOMLE8}
\eeq

At first this may seem too big.  We wanted the homotopy groups to be subgroups of those of $U(\infty)$, but $\pi_2(U(\infty))=0$ and so we have also added a class.  This new class will be characterized by a characteristic 3-class which in the mathematics literature is called the Dixmier-Douady class.  In Ref.~\cite{Allan} it was argued that this new class be identified with the $H$ flux.  Evidence came from the fact that, again, topological obstructions to the existence of the bundle reproduce known examples of Freed-Witten anomalies with this interpretation.  Another suggestion that this new 3-class is the $H$ field comes from the fact that an $LE_8$ bundle over a ten-manifold $M$ carries the same data as an $E_8$ bundle over a circle bundle over the 10-manifold, that is to say, over the M-theory spacetime.  An $E_8$ bundle is characterized by a four class.  If one considers the $E_8$ bundle of \cite{WitFluxQuant,DMW} then this four class is the M-theory 4-form and its dimensional reductions give the IIA 4-form RR field strength and 3-form NS field strength.  In \cite{Allan} it was argued that the characteristic 3-class of the $LE_8$ bundle is precisely the dimensional reduction of this M-theory 4-form, and so is indeed the NS 3-form.

While the replacement of the classifying space of $\K^0$ with $LE_8$ may appear far-fetched, it simultaneously achieves two objectives.  First, it naturally chooses a polarization and so solves the star problem in any background.  This is an improvement over the second solution which required that every spatial slice of spacetime have the same topology.  Secondly it naturally integrates the NS 3-form flux into the bundle framework, as it must be as IIA string theory admits 9-11 flips which interchange the $H$ flux and the RR 4-form field strength $dC_3$.  As an added bonus, if one correctly identifies a gravitational correction to the relation between $dC_3$ and the Pontrjagin class of the M-theory $E_8$ bundle, it produces the correct topology of the $E_8$ gauge bundles of the end of the world, as found from gravitational anomaly cancellation in Ref.~\cite{HW}.

\subsection{S-duality Covariance}

We have seen that the star problem does not appear in several variations of the K-theory classification.  However all mild variations of the K-theory classification suffer from a more fundamental problem.  All schemes treat the NSNS $H$ field as part of the background data in which one seeks to classify RR fields or charges.  Indeed, the $H$ flux in many ways resembles the topological data more than it resembles the RR fields.  For example, the topology of spacetime is determined by the metric, which is also a NSNS field and it is mixed with the $H$ field, for example, in the Seiberg-Witten map \cite{SW} and also in T-duality \cite{BEM} as we will review in Subsec.~\ref{JET}.  By contrast, RR fields mix among themselves under T-duality.

While one can argue that morally the $H$ field is or is not a part of the topological background data, in type IIB string theory the $H$ flux and the RR field strength $dC_2$ are related by S-duality and this S-duality leaves the topology of the spacetime invariant.  Therefore, in a given background, $dC_2$ and $H$ must be classified identically.  In Subsec.~\ref{JEKS} we will argue that this means that one can, for example, fix $dC_2$ as part of the background data and then classify the S-duals of the D-branes by twisted K-theory with twist equal to $dC_2$.  However none of these S-duals can be the correct classification, as every one of them gives a different consistency condition for the fields $dC_2$ and $H$.

The inconsistency of the K-theory classification and S-duality was demonstrated concretely in Ref.~\cite{DMW}.  Consider the twisted K-theory classification of RR field strengths.  An RR field strength lifts to K-theory if it is annihilated by $d_3$, for example the 3-form field strength $dC_2$ lifts if it satisfies
\beq
0=d_3 (dC_2)=(Sq^3+H\cup)dC_2=dC_2\cup dC_2+H\cup dC_2. \label{JENONCOV}
\eeq
Here we have used the fact that $Sq^3$ squares three-classes.  If $dC_2$ does not satisfy Eq.~(\ref{JENONCOV}), it is still a consistent field configuration but the configuration will carry a D3-brane charge which is Poincar\'e dual to $d_3(dC_2)$, in other words
\beq
Q_{D3}=\textup{PD}(dC_2\cup dC_2+H\cup dC_2) \label{JED3CARICA}
\eeq
where $\textup{PD}$ is the Poincar\'e dual.

D3-brane charge is believed to be S-duality invariant.  We will assume that even torsion D3-brane charges, like those in the image of $Sq^3$, are S-duality invariant.  The authors of Ref.~\cite{DMW} then performed the S-duality transformation
\beq
dC_2\longrightarrow dC_2\hsp H\longrightarrow H+dC_2
\eeq
under which the D3-charge (\ref{JED3CARICA}) becomes
\beq
Q_{D3}=\textup{PD}(dC_2\cup dC_2+H\cup dC_2+dC_2\cup dC_2)=\textup{PD}(H\cup dC_2) \label{JED3CARICAB}
\eeq
where we have used the fact that $dC_2\cup dC_2$ is $\Z_2$ torsion, as is the cup product of any odd degree class with itself.  Notice that the charges (\ref{JED3CARICA}) and (\ref{JED3CARICAB}) in general do not agree, unless one imposes that $dC_2\cup dC_2=0$ which would correspond to imposing that $W_3$ and $H$ must vanish independently on any D-brane worldvolume, a conjecture which has not been supported by the analysis of worldsheet anomalies in \cite{FW}.  Therefore the K-theory classification, which rests upon (\ref{JED3CARICA}), appears to be inconsistent with S-duality.

We can see this inconsistency in the language of MMS instantons by noting that NS5-branes may also form, sweep out a cycle and decay.  In type IIB S-duality guarantees that NS5-branes will also be subject to a Freed-Witten anomaly
\beq
Q_{D3}=W_3+dC_2 
\eeq
which will be canceled by the insertion of $Q_{D3}$ D3-branes.  However in the derivation of the K-theory classification from the AHSS we have only considered D-brane MMS instantons.  If one also includes NS5-brane MMS instantons one needs to further quotient the group of D3-brane charges by those charges which may be created or destroyed by NS5-brane MMS instantons. 

Given that the twisted K-theory classification of RR fields and S-duality appear to conflict, there are several directions to proceed.  Perhaps the easiest is to claim that S-duality simply is not understood for torsion fields and so we should not worry about it.  A second approach is to abandon the K-theory classification and replace K-theory with a different generalized cohomology theory.  This avenue has been explored in Refs.~\cite{Hisham,Hisham2,Hisham3,Hisham4,Hisham5,Hisham6} in which the authors instead use elliptic cohomology.

However it may be that even this is not radical enough.  When we consider both the $H$ flux and the RR fields to be independent fields, the equations of motion that describe them, even in the supergravity limit, are nonlinear.  This is in contrast to the case in which the $H$ flux is fixed in which case the equations for the RR field are linear and so the solutions form a linear space and so naturally have the structure of an abelian group.  The solutions of a nonlinear set of equations do not obviously form a group.  In the case of type IIB string theory on the conifold for example, one finds that pairs $(H,dC_2)$ are classified, when they are both nonzero, by the semigroup of natural numbers which are the greatest common divisors of the three-classes, although perhaps if one motivates a choice of sign for these elements one may conclude that pairs are classified by the integers, which do form an abelian group.  In general no argument has been presented that the combined set of NSNS and RR fields, subjected to the equations of motion for all of the fields and quotiented by all of the gauge symmetries, should have any additive structure.

The $E_8$ bundle framework discussed in the previous subsection provides a natural approach to finding topological classification schemes which naturally include both NSNS and RR fields and, unlike generalized cohomology theories do not impose that there exist a group structure.  As we will see in Subsec.~\ref{JET}, the $E_8$ formalism already knows about T-duality, the dual circle is in the fiber itself.  Thus it may be applied to type IIB at least in the case of backgrounds which are T-dual to type IIA backgrounds, although the bundle one arrives at in type IIB depends not just on the background, as in IIA, but also on the choice of circle to be T-dualized.

While no one knows what mathematical object classifies all NSNS and RR fields that solve all equations of motion quotiented by gauge invariances, the S-duality covariant generalization of Eq.~(\ref{JED3CARICA}) was guessed in Ref.~\cite{DMW} and proven in Ref.~\cite{aussy2005} using the Freed-Witten anomaly.  The D3-brane charge is in general given by
\beq
Q_{D3}=\textup{PD}(dC_2\cup dC_2+H\cup dC_2+H\cup H+P) \label{JED3CARICAC}
\eeq
where $P$ is independent of $dC_2$ and $H$ and is S-duality invariant.  Therefore if one finds $P$ for any value of $(dC_2, H)$ then one can use Eq.~(\ref{JED3CARICAC}) to find the D3-brane charge for every value of $(dC_2,H)$.  Setting the D3-charge to zero one finds the S-duality covariant twisted K-theory condition on the 3-classes.  Partial results on the value of the 6-class $P$ were presented in Ref.~\cite{DFM}.  Analogous conditions for the other RR field strengths are still lacking.

\section{Applications} \label{JEAPPLICAZIONI}

Despite its shortcomings, the K-theory classification has had several applications in different areas of string theory.  For example it has been used to identify inconsistencies in string models that satisfy the usual tadpole cancellation conditions but in which certain probes suffer from global anomalies in their worldvolume gauge theories \cite{Uranga}.  In these lectures we will mention two applications of the K-theory classification scheme.  In Subsec.~\ref{JET} we will describe how an isomorphism of twisted K-groups known as Takai duality led to a conjectured solution to the age old problem of finding a formula for the topology of a T-dual manifold given the original topology and $H$ flux.  Then in Subsec.~\ref{JEKS} we will describe how an S-dual of the twisted K-theory classification classifies universality classes of cascading gauge theories.  We will see that the homology classes of branes which lift to the same K-theory class correspond to different steps in the same cascade.

\subsection{T-duality} \label{JET}

When the $H$ flux is topologically trivial, in other words when it represents the zero cohomology class, one can globally define the NS $B$ field.  If the compactification 10-manifold is topologically the product of a circle with a 9-manifold then, given the metric and $B$ field, one can use the Buscher rules \cite{Buscher} to calculate the metric and $B$ field of the compactification manifold which is T-dual with respect to the aforementioned circle.  Using this prescription the topologies of the original compactification and the dual compactification are always identical.

It has been known for more than a decade \cite{AABL} that if instead the compactification 10-manifold is a nontrivial circle fibration over a 9-manifold then in general the topology of the dual manifold is different from that of the original manifold.  Over the years many examples of topology changing T-dualities have been found, see for example Ref.~\cite{KSTT}.  However the dual topology was often found through a somewhat laborious technique involving the exact metric and sometimes consistency conditions which may admit more than one unique solution.  In the case of Calabi-Yau compactifications, for example, the metric is often unavailable and so no method of computing the topology of the T-dual manifold was known.

The twisted K-theory classification led to a conjectured formula for the topology and also the $H$ flux of a T-dual manifold which, computationally, is far easier than early approaches even in the examples in which those approaches are feasible.  Recall that magnetic fluxes on a timeslice are classified by $\K^1$ in IIB, while in IIA they are classified by $\K^0$.  Similarly in IIB D-brane charges on a timeslice and also D-brane trajectories in spacetime are classified by $\K^0$ while they are classified by $\K^1$ in IIA.  Type IIA and IIB compactifications are T-dual, and T-duality exchanges branes in IIA with branes in IIB and fluxes in IIA with fluxes in IIB.  Therefore one may hope that T-duality exchanges $\K^0$ and $\K^1$.  

More concretely, if one begins with a compactification on $M$ with NS 3-form flux $H$ and the T-dual manifold is $\hat{M}$ with NS 3-form flux $\hat{H}$ then one expects
\beq
\K^0_H(M)=\K^1_{\hat{H}}(\hat{M})\hsp \K^1_H(M)=\K^0_{\hat{H}}(\hat{M}). \label{JEKISO}
\eeq
Thus the question of finding the T-dual compactification data $(\hat{M},\hat{H})$ is reduced to the problem of finding a solution to Eq.~(\ref{JEKISO}).

In the language of algebraic K-theory a solution to Eq.~(\ref{JEKISO}), in the case in which $M$ is a circle bundle over $X$, was provided nearly 20 years ago in Ref.~\cite{RR}.  Recall that the topology of the total space $M$ of a circle bundle $M\rightarrow X$ is completely classified by the topology of $X$ and a 2-class
\beq
c_1(M)\in\H^2(X;\Z)
\eeq
called the Chern class of the circle bundle.  Recasting the results of Ref.~\cite{RR} in the language of topology, one finds that if $M$ is a circle bundle over $X$ with Chern class $c_1(M)$ then Eq.~(\ref{JEKISO}) is solved by an $\hat{M}$ which is a circle bundle over the same base $X$ but with Chern class $c_1(\hat{M})$ and NS 3-form flux $\hat{H}$ satisfying
\beq
c_1(M)=\int_{\hat{S^1}}\hat{H}\hsp c_1(\hat{M})=\int_{S^1}H. \label{JEBEM}
\eeq
Here $S^1$ and $\hat{S^1}$ are the circle fibers of $M$ and $\hat{M}$ respectively.  

We have used the language of differential forms to write the relation between the NS fluxes and Chern characters as an integral, but in terms of integral cohomology the map from $\hat{H}$ to $c_1(M)$ and $H$ to $c_1(\hat{M})$ is the pushforward $\pi_*$ using the projection map $\pi$ of the respective circle bundle.  To completely specify the dual $H$ flux one must also specify that T-duality leaves invariant any $H$ flux which is in the image of the pullback $\pi^*$ with respect to $\pi$.  Intuitively, this means that $H$ flux which is completely supported on $X$ is unchanged by T-duality, instead T-duality merely exchanges the component of the $H$ flux which has two legs along $X$ and one along with fiber with a Chern class which has those same two legs along $X$.

In Ref.~\cite{BEM} the authors conjectured that the solution $(\hat{M},\hat{H})$  of Eq.~(\ref{JEBEM}) is the T-dual compactification to $(M,H)$.  They showed that this conjecture reproduces all of the topology changing T-dualities that they knew of in the literature and locally reproduces the Buscher rules.

While the formula (\ref{JEBEM}) is a success of the K-theory classification, it also follows from the competing $E_8$ classification.  In this case the solution is manifestly unique, whereas in the K-theory case it is not known whether there are in some cases distinct solutions to Eq.~(\ref{JEKISO}).  Recall that in the $E_8$ scheme one interprets the NS $H$ flux as the degree 3 characteristic class of an $LE_8$ fibration.  Using a construction reviewed in Ref.~\cite{MEK}, the data of a $G$ bundle over a circle bundle over $X$ is equivalent to the data of an $LG$ bundle over $X$, where $LG$ is the trivially centrally extended based loop group of $G$.  In what follows we will need to use the theorem
\beq
\pi_p(LG)=\left\{
\begin{array}{cl}
\pi_p(G)\oplus\pi_{p+1}(G) &\textup{if $p\geq 2$}\\
\pi_p(G)\oplus\pi_{p+1}(G) \oplus\Z&\textup{if $p=1$}\\
\end{array}\right. \label{JEOMTEOREMA}
\eeq
which allows one to construct the homotopy groups of $LG$ from those of $G$.

Consider type IIA on the 10-manifold $M^{10}$ which is a circle bundle over $X^9$.  Then the data of the $LE_8$ bundle over $M^{10}$ is captured precisely by the date of an $LLE_8$ bundle over $X^9$.  Using Eqs.~(\ref{JEOMLE8}) and (\ref{JEOMTEOREMA}) one can now find the homotopy groups of $LLE_8$
\beq
\pi_1(LLE_8)=\Z^3\hsp \pi_2(LLE_8)=\Z^2\hsp \pi_3(LLE_8)=\Z.
\eeq
The fact that the fundamental group of $LLE_8$ is dimension three means that the fiber contains three circles.  The first two are the M-theory and type IIA circles which have been compactified.  The third descends from $\pi_3(E_8)=\Z$ in the $E_8$ fibration over the M-theory spacetime.  

In Ref.~\cite{MEMONO} it was conjectured that this third circle is the T-dual type IIB circle, and that the first two form the F-theory torus.  This third circle is the dimensional reduction of the class $\pi_2(LE_8)=\Z$, whose characteristic class was the NS 3-form $H$.  Therefore the characteristic class of the third circle, which is the Chern class $c_1(\hat{M})$ of the IIB circle bundle, is just the dimensional reduction of $H$ on the IIA circle
\beq
c_1(\hat{M})=\int_{S^1_{IIA}}H
\eeq
in accordance with the conjecture (\ref{JEBEM}).

\subsection{The Klebanov-Strassler Cascade} \label{JEKS}

In the previous subsection we saw that the K-theory classification led to the solution of an old problem.  In this subsection we will describe a more mild success of the K-theory classification, we will argue that K-theory, twisted this time by the RR 3-form field strength $dC_2$, classifies universality classes of a class of cascading $SU(N+M)\times SU(N)$ gauge theories with a fixed step size $M$.  In this section we will be using the S-dual of the usual K-theory classification in which one classifies F-strings, D3-branes and NS5-branes.  The action of S-duality on higher branes and D-instantons is still a matter of debate in the literature and we will fortunately not need these branes in this example.

We have seen that homology groups are much larger than K-groups.  In particular multiple homology classes correspond to the same K-class.  We will now see that each of the homology classes corresponding to a single K-class describes one particular step in the cascade, and gives the ranks of the worldvolume gauge bundles at that step.  In the S-duality covariant K-theory one needs to classify not only D-branes but also their duals.  Now we are already using the S-dual of the ordinary K-theory, so in our case the S-duality covariant twisted K-theory is augmented by adding D-strings and D5-branes.  The number of D5-branes will be equal to the step size $M$, and so the S-duality covariant twisted K-theory will also classify $M$ and so will classify all of universality classes of all of the theories of this type. 

In the mathematics literature the word \textit{conifold} is used to describe any space which is a continuous manifold everywhere except for isolated ordinary regular double point singularities, which are locally the $2n-2$ real-dimensional solutions of the complex equation
\beq
\sum_{i=1}^n z_i^2=0 \label{JECONI}
\eeq
in $\C^n$.  We will instead refer to \textit{the conifold}, which in the physics literature has come to mean the 6-manifold which is the solution of Eq.~(\ref{JECONI}) in $\C^4$.  Topologically the conifold is a cone over the product $S^2\times S^3$.  

String theory on the conifold crossed with $\R^{3,1}$ was first considered in Ref.~\cite{KW}.  They found that a stack of $N$ D3-branes at the tip of the cone are described by a particular $N=1$ supersymmetric gauge theory with gauge group $U(N)\times U(N)$ with charged chiral multiplets and a given superpotential.  They were interested in the IR fixed point of the theory, in which the $U(1)$ factors decouple leaving an $SU(N)\times SU(N)$ conformal theory.  

Later in Ref.~\cite{GK} it was discovered that one can engineer the gauge group $SU(N+M)\times SU(N)$ by adding $M$ D5-branes wrapped around the 2-cycle in the $S^2\times S^3$.  The normal bundle to this $S^2$ is nontrivial and as a result the statistics of some of the worldvolume fields change \cite{MN}.  As in \cite{MM}, this nontrivial normal bundle couples D5-branes to the lower dimensional RR field $C_4$ so that the D5-brane carries one half of a Dirac unit of D3-brane charge.  This gravitational charge also couples to the fundamental string worldsheet, and so it is often considered to be a part of the $B$ field.

The half-integrality of the D3 charge does not violate Dirac's quantization argument, which ordinarily implies that D3-brane charge quantization is required for the well-definedness of the partition function of another D3-brane whose trajectory is deformed over an $S^5$ linking the original D3.  Dirac's argument fails because the linking D3 would need to pass through the D5, at which point there would be a Hanany-Witten transition \cite{HanWit} creating a D1-brane.  The contribution of this D1-brane to the partition function renders it well-defined.  In general Dirac's quantization condition does not apply straightforwardly to branes that are dissolved in other branes.

In Ref.~\cite{KS} the authors discovered that in a particular geometry, corresponding to the baryonic branch of the gauge theory, the gauge group $SU(N+M)\times SU(N)$ becomes strongly coupled in the infrared and admits a dual weakly coupled description with gauge group $SU(N)\times SU(N-M)$.  This duality is similar to Seiberg duality, which would occur if the $SU(N)$ symmetry were not gauged.  In their geometry the $SU(N)\times SU(N-M)$ gauge theory is also at its baryonic root, and so at weak coupling there is an effective gauge group $SU(N-M)\times SU(N-2M)$.  This process continues until $N$ is less than $M$.  They named this series of Seiberg dualities the cascade.

The one-half unit of D3-brane charge will not be essential in what follows.  The important feature of the $M$ D5-branes will be that, by Gauss' Law, if $S^3$ is a 3-sphere linking all of the D5's then
\beq
\int_{S^3} dC_2=M. \label{JE3FLUSSO}
\eeq
This means that the S-dual twisted K-theory, which we recall classifies fundamental strings, D3-branes and NS5-branes, is twisted by $M$ units. 

This argument is a bit too fast.  These lectures have been about smooth manifolds, and now we are considering a compactification with a singularity.  We have not defined K-theory on singular spaces and in fact inequivalent definitions exist.  In this example we are saved by the fact that the singularity can be eliminated in two different ways, each of which has little effect on the geometry far from the singularity.  First, it may be blown up, which topologically means that the singularity is replaced by an $S^2$ which is homotopic to the $S^2$ in the base, intuitively only the $S^3$ collapses and the space is nonsingular.  The second possibility is that the singularity may be deformed.  Algebraically this may be done by replacing the defining equation (\ref{JECONI}) with
\beq
\sum_{i=1}^n z_i^2=\xi 
\eeq
for some constant deformation parameter $\xi$.  In this case the singularity is replaced by a 3-sphere which is homotopic to the 3-sphere on the base and again the space becomes nonsingular.  Locality imposes that the physics far away from the singularity does not know whether or in which way the singularity was eliminated, and so when considering distant physics, like the K-theory of the base, one can often consider whichever smoothing is more convenient.  We want a third cohomology class on which there is $dC_2$ flux and so we will consider the deformed conifold.

On the deformed conifold the $S^2$ is contractible.  This means that the D5-branes may shrink to points, but they can never completely decay as they carry D3-brane charge.  The fact that we have chosen the deformed conifold cannot affect the physics far away, and so in particular (\ref{JE3FLUSSO}) still implies that the 3-form RR field strength is topologically nontrivial, and is the element $M$ in the third cohomology group with compact support of the deformed conifold, which is $\Z$.

The cohomology with compact support of the deformed conifold consists of only two nontrivial classes
\beq
\H^3=\Z\hsp \H^6=\Z \label{JECO}
\eeq
which are Poincar\'e dual to a brane wrapping the $S^3$ and to pointlike branes respectively.  The $N$ D3-branes are charged under the later, as are the $M$ half D3-branes.  We are interested in the S-dual of the K-theory classification, which we have stressed classifies only F1's, D3's and NS5's.  This means that we will only try to classify the D3-branes and not the half D3-branes, which are really D5-branes.  Thus cohomology classifies the number $N$ of D3-branes, which is the same number $N$ that appears in the rank of the gauge group $SU(N+M)\times SU(N)$.

The cohomology group $\H^3=\Z$ classifies a different charge, that of branes wrapping the 3-sphere.  However this 3-sphere has nontrivial $dC_2$ flux (\ref{JE3FLUSSO}) which couples to the D3-brane worldvolume via, for example, the term
\beq
\mathcal{L}_{D3}\supset C_2\wedge B.
\eeq
This implies that $C_2$ flux carries fundamental string charge, and so $dC_2$ is the endpoint of a fundamental string.  In this case the presence of $M$ units of $dC_2$ implies that the D3-brane is a baryon on which $M$ fundamental strings end, in line with the S-dual Freed-Witten anomaly.  In the K-theory classification one does not assign a charge to baryonic configurations, and so these branes will not carry K-theory charges.

One can now calculate the twisted K-theory of the deformed conifold by applying the AHSS with twist $dC_2$ to the cohomology (\ref{JECO}).  The conifold itself is both $spin$ and $spin^c$ and the $S^3$ is also $spin^c$ and so the $Sq^3$ term is equal to zero.  Also there is no spacing between cohomology groups of more than 3, and so one only need consider the differential
\beq
d_3=dC_2\cup:x^3\mapsto Mx^6
\eeq
where $x^3$ and $x^6$ are the generators of $\H^3$ and $\H^6$ respectively.  The twisted K-theory is then the quotient of kernel of $d_3$, which is $\H^6=\Z$, by its image, which is $M\H^6=M\Z$.  This yields
\beq
\K^0_{dC_2}=\frac{\Ker(d_{3}:\HE\longrightarrow\HO)}{\Im(d_3:\HO\longrightarrow\HE)}=\frac{\Z}{M\Z}=\Z_M\hsp  \K^1_{dC_2}=0
\eeq
where $\K^1_{dC_2}$ vanishes because no elements of the odd cohomology are in the kernel of $d_3$.

The group $\K^0_{dC_2}$ has $M$ elements, of which the $J$th is the lift of all of the cohomology classes corresponding to numbers $N$ of D3-branes which are equal to $J$ modulo $M$.  In other words, the element $J$ of K-theory corresponds to all of the gauge theories with gauge groups
\beq
SU(J+(K+1)M)\times SU(J+KM). \label{JEGRUPI} 
\eeq
The gauge groups (\ref{JEGRUPI}) are the set of gauge groups in a given cascade.  Thus $\K^0_{dC_2}$ parametrizes the possible cascades with step size $M$, or equivalent the endpoints of the cascade, which are the universality classes of the gauge theory when one restricts attention to the baryonic root vacua.

\section* {Acknowledgement}
Jarah would like to thank the Modave Summer school for inviting him to give these lectures in an environment with many flammable materials.
The work of J.E. is partially supported by IISN - Belgium (convention 4.4505.86), 
by the ``Interuniversity Attraction Poles Programme -- Belgian Science Policy''
and by the European Commission programme MRTN-CT-2004-005104, in which he is
associated to V. U. Brussel.

%%%%%%%%%%%%%%%%%%%%%%%%%%%%%%%

\end{document}

\bibitem{}
,
{\it },
[{\tt arXiv:hep-th/}].

[{\tt arXiv:hep-th/}].

%%%%%%%%%%%%%%%%%%%%%%%%%%%%%%%%%%%%%%%%%%%
\section{The Atiyah-Hirzebruch Spectral Sequence}
%%%%%%%%%%%%%%%%%%%%%%%%%%%%%%%%%%%%%%%%%%%
\label{background}

A D-brane that wraps a nontrivial cycle carries a charge that corresponds to the homology
class of the cycle.  Diaconescu, Moore and Witten (DMW) ~\cite{DMW} have shown that not 
all of these charges are conserved, instead there are dynamical processes in which branes 
wrapping nontrivial cycles can decay.  In addition, in Ref.~\cite{FW} the authors have 
found that certain cycles cannot be wrapped by single branes. They argued that any brane 
wrapping such a cycle would be anomalous, and in fact evidence was presented in \cite{DMW} 
that their contributions to the partition function cancel.  Thus to compute the partition
function it suffices to restrict one's attention to equivalence classes of anomaly-free 
branes.  In other words, D-branes are classified by a quotient of a subset of homology.  

MMS have argued that this quotient of a subset is precisely twisted K-theory.  They used 
a mathematical algorithm known as the Atiyah-Hirzebruch spectral sequence (AHSS) to 
determine which homology classes lift to K-theory classes, that is, to determine which 
D-branes are unstable and which are not allowed.   While in their examples the anomalous 
branes suffered from the Freed-Witten (FW) anomaly, in general the AHSS construction 
eliminates some branes that are FW anomaly-free.  This leads to the question of whether 
the branes that are eliminated by the AHSS construction, but not by the FW anomaly, are 
allowed in the physical theory.  If such branes are allowed, they would provide counterexamples to the 
K-theory classification program and to the IIA version of the Sen conjecture.  On the other hand, if such branes are not allowed, 
they would be examples of a new anomaly.  In the 
present note we will adopt the more modest goal of providing a characterization of these branes.

The AHSS consists of a series of differential operators $d_{2p-1}$, $p\geq 2$ which map 
elements of the $q$th integral cohomology to elements of the 
$(q+2p-1)$th cohomology
\beq
d_{2p-1}:\H^q\longrightarrow\H^{q+2p-1}\hsp p=2,3,\cdots .
\eeq
In general the differential $d_{2p-1}$ consists of cohomology operations on the free
parts of the integral cohomology and also on cyclic subgroups of prime order less than 
or equal to $p$.  In particular, in the case of untwisted K-theory and when $p$ is prime, 
$d_{2p-1}$ contains a primary cohomology operation known as the first Milnor primitive
\footnote{Higher Milnor primitives appear in the differentials of `higher' 
generalized cohomology theories, for example $Q_2$ appears in Morava 
K-theory and elliptic cohomology \cite{KS1}.}  
$Q_1$ whose image is a $p$ torsion class $\Z_p$, and also it may contain secondary 
operations whose images are torsion at lower primes.    A secondary cohomology operation 
is an operation that is not defined on the entire cohomology, but is defined on the 
kernels of the preceding differentials.

We will use Poincar\'e duality to identify a D$(9-q)$-brane wrapping a $(10-q)$-cycle 
in the integral homology group $\H_{10-q}$ with its dual cocycle in $\H^q$, which in 
terms of supergravity fields corresponds to the Ramond-Ramond source $dG_{q-1}$.  The 
homology class of a D-brane wrapping the cycle $N_{10-q}$ lifts to a twisted K-theory 
class if and only if its dual cohomology class PD($N_{10-q})$ is in the kernel of all of the differentials
\beq
d_{2p-1} (\textup{PD}(N_{10-q}))=0 \hspace{.3cm} \textup{\ for\ all\ }p.
\eeq

For example, the first nontrivial differential contains a primary operation at prime 2 
and can be explicitly written
\beq
d_3 x=Q_1x+H\cup x=Sq^3x+H\cup x
\eeq
where $H$ is the NSNS 3-class and $\cup$ is the cup product, the integral version of the wedge product.  
The Milnor primitive $Q_1$ at prime 2 is often denoted $Sq^3$ and is called a Steenrod square or more precisely square 3.  $Sq^3$, like the cup product with $H$, increases the degree of a cohomology class by three.  DMW have explained that 
if $d_3$ does not annihilate the class of a brane then the brane suffers from an FW 
anomaly.  The converse is not true since some FW anomalous branes are annihilated 
by $d_3$.  MMS have found an example of this phenomenon in the SU(3) 
Wess-Zumino-Witten WZW model, and in that case the offending class was not in the 
kernel of $d_5$ and so, as expected, did not lift to twisted K-theory.  

More concretely, consider a brane with worldvolume $N$ in the spacetime $M$.  Let 
$i:N\hookrightarrow M$ be the inclusion map of the brane into the spacetime.  Then the 
FW anomaly is \cite{FW}
\beq
W_3+H=0
\eeq
where $W_3$ is the third integral Stiefel-Whitney class of the normal bundle of $N$ in $M$ and 
$H$ is the pullback of the NSNS 3-form to the brane worldvolume $N$.  The pushforward of 
$W_3+H$ to the spacetime $M$ is 
\beq
i_*(W_3+H)=Sq^3({\textup{PD}}(N))+H\cup\textup{PD}(N)=d_3(\textup{PD}(N)).
\eeq
In the aforementioned SU(3) example $W_3+H$ is nontrivial but it is in the kernel of the 
pushforward.  This example suggests that the role of the secondary operations is to pick 
up the anomalies that were in the kernel of the pushforward.  In particular one may 
conjecture that the secondary operations do not imply the existence of any new anomalies, for example all of the two-torsion operations encode the FW anomaly.

\section* {Acknowledgement}
We would like to thank Peter Bouwknegt, Alan Carey, Varghese Mathai, Tony Pantev, 
Eric Sharpe and Bai-Ling Wang for useful discussions. We also thank the organizers of the 
workshop ``Gerbes, Groupoids, and Quantum Field Theory" at the Erwin Schr\"odinger 
Institute for the invitation and the stimulating environment. 
The work of J.E. is partially supported by IISN - Belgium (convention 4.4505.86), 
by the ``Interuniversity Attraction Poles Programme -- Belgian Science Policy''
and by the European Commission programme MRTN-CT-2004-005104, in which he is
associated to V. U. Brussel.

%%%%%%%%%%%%%%%%%%%%%%%%%%%%%%%

\end{document}

\begin{table}[h]
\begin{center}
\begin{tabular}{|l|l|l|l|}
\hline \textbf{BRST} & \textbf{K-theory} & \textbf{IIA Supergravity} & \textbf{IIA String Theory}\\
\hline fields & $\oplus_k \H^{2k}(M)$ & RR diff. forms $G_{2k}$ & RR int. classes $G_{2k}$\\
\hline ghosts&$\oplus_k \H^{2k+1}(M)$&gauge xforms/branes&gauge xforms/branes\\
\hline BRST operator&$d_{2p+1}$&$d_3=H\wedge$&$d_3=Sq^3+H\cup,\ d_5$\\
\hline Gauge xforms&$d:\HO\rightarrow\HE$&Wilson loop shift&FW monodromy\\
\hline Constraints&$d:\HE\rightarrow\HO$&Bianchi identities&source FW anomaly\\
\hline physical fields&$\K^0_H(M)$&orbits of Bianchi solns&orbits of FW solns\\
\hline anomalies&$\K^1_H(M)$&stable p-branes&stable D-branes\\
\hline
\end{tabular}
\end{center}
\caption{K-theory vs BRST in type IIA supergravity and in type IIA string theory}
\end{table}

\subsection{What kind of objects are twisted K-groups?}

The integral cohomology $\textup{H}^*(M)$ of a space $M$ is a collection of abelian groups $\textup{H}^k(M)$ indexed by a nonnegative integer $k$ and endowed with a product
\beq
\cup:\textup{H}^j(M)\otimes_\Z \textup{H}^k(M)\rightarrow \textup{H}^{j+k}(M):x\otimes y\mapsto x\cup y \label{mult}
\eeq
called the cup product.  The free part $\Z^k$ of the integral cohomology group $\Z^k\oplus_i\Z_{q_i}$ is isomorphic to an integral lattice of de Rham cohomology $\R^k$.  On this integral lattice the cup product $x\cup y$ reduces to the wedge product of differential forms $x\wedge y$.  Thus we may guess that the wedge products of the previous section will, when we include torsion, be written as cup products plus torsion corrections.

The twisted K-theory $\K_H^*(M)$, with twist $H$, of a space $M$ is another collection of abelian groups $\K_H^k(M)$, this time indexed by a general integer $k$ and depending on an integral three-class $H$.  However, unlike the cohomology groups, the $\K$ groups are not independent.  Instead they are related by the Bott periodicity relation
\beq
\K_H^j(M)\cong \K_H^{j+2}(M)
\eeq
and thus it will suffice to compute $\K_H^0(M)$ and $\K_H^1(M)$.  This structure better mimics the charge structure of D-branes.  For example there are dielectric and fractional D$p$-branes that carry a half unit of D$(p-2)$-brane charge, meaning that twice the generator of D$p$ charge should be equal to the generator of D$(p-2)$ charge.  Such a relation, between generators of different degrees, would be impossible in cohomology.  However in K-theory it is automatic, a single group classifies all even degrees, while another classifies all odd degrees.  Thus, in each string theory, one K-group classifies all of the D-branes simultaneously.  As D-branes source fluxes, one can arrive at a similar story for fluxes.

While ordinary K-theory admits a multiplication similar to (\ref{mult}), the K group twisted by a fixed $H$ admits no such multiplication.  Instead when one multiplies two twisted K-groups the twists add.  However the twist corresponds to a fixed NS 3-form flux, and so this multiplication changes the $H$ flux and so does not correspond to any obvious physical process in string theory.  Setting the twist of one of the factors to zero, one finds that the twisted K-theory $\K^*_H(M)$ is a module of the untwisted K group $\K^0_H$, but no physical interpretation of this fact in string theory has appeared to date.

\subsection{Constructing twisted K-theory}

We will now describe the AHSS, which is an algorithm for computing the twisted K groups $\K^*_H(M)$ from the $H$ flux and the integral cohomology $\H^*(M)$.  In type IIB string theory RR field strengths correspond to odd cohomology classes, and so we shall see they will be classified by $\K^1_H(M)$.  Similarly in IIA string theory the fluxes are even classes and so will be described by $\K^0_H(M)$.

We first assemble the even and odd cohomologies into two big groups $\HE$ and $\HO$
\beq
\HE(M)=\oplus_k \H^{2k}(M)\hsp \HO(M)=\oplus_k \H^{2k+1}(M).
\eeq
We want to construct the twisted K-group $\K^i_H(M)$ by finding a finite set of improving approximations $E^i_j$ starting with the odd cohomology and finishing with a set $E^i_n$ of the same cardinality as $\K^i_H(M)$
\beq
E^0_1=\HE(M)\hsp E^1_1=\HO(M)\hsp |E^i_n|=|\K^i_H(M)|.
\eeq
In general $E^i_n$ will not be isomorphic to $\K^i_H(M)$ as a group, that is, the addition rule will be different.  The addition rules are related by an extension problem.  However, as there is no addition rule for fluxes in the S-duality covariant case anyway, since the sum of two sets of fields that satisfy the nonlinear supergravity equations of motion generically is not another solution, we will not concern ourselves with reproducing the addition rule in this note.

To get from $E^i_1$ to $E^i_n$ we will need to introduce a series of differential operators $d_{2j+1}$ which are degree $(2j+1)$ cohomology operations, in other words
\beq
d_{2j+1}:\H^k(M)\rightarrow\H^{k+2j+1}(M)\hsp d_{2j+1}d_{2j+1}=0.
\eeq
As we will soon see, only the first differential operator, $d_3$, needs to be well-defined on the full integral cohomology.  To pass from $E^i_j$ to the next approximation $E^i_{j+1}$ we need to take the cohomology with respect to the differential operator $d_{2j+1}$
\beq
E^i_{j+1}=\frac{\Ker(d_{2j+1}:E^i_j\rightarrow E^{i+1}_j)}{\Im(d_{2j+1}:E^{i+1}_j\rightarrow E^{i}_j)}.    
\eeq
In particular, we see that while the differentials do not need to be defined on all of the cohomology classes, as some classes will be eliminated earlier in the procedure by the kernel operations, they do need to be well-defined on the equivalence classes obtained by quotienting by the images of their predecessors.  This identifies the differentials beyond $d_3$ not as ordinary cohomology operations, but as secondary cohomology operations obtained perhaps from Toda brackets of primary cohomology operations.  However the details of the constructions of these $d$'s will not concern us, it will suffice to use the claims of Diaconescu, Moore and Witten \cite{DMW} and then Maldacena, Moore and Seiberg \cite{MMS} that they compute Freed-Witten anomalies.

In known examples, the first differential $d_3$ is the only differential that does not vanish in the absence of torsion cohomology.  In fact, the torsion free parts of all of the differentials have been computed in \cite{AS}, where they have been seen to be Massey products of $H$.  This implies in particular that they will always be trivial on compact Kahler manifolds.  However in principle they could appear in supergravity, and it would be interesting to find the corresponding gauge transformations.  $d_3$ may be written
\beq
d_3x=Sq^3x+H\cup x
\eeq
where $H$ is our familiar NS 3-class and $Sq^3$ is a cohomology operation known as a Steenrod square which takes an integral class in the $k$th cohomology to a class in the $(k+3)$rd cohomology, as does the cup product with $H$.  Unlike the cup product with $H$, however, $Sq^3$ is only nontrivial when acting on $\Z_2$ torsion components of $\H^k(M)$, and the image is likewise always a $\Z_2$ torsion component of $\H^{k+3}(M)$.  This means in particular that in the supergravity limit such torsion components disappear because there is no Dirac quantization and so $\Z_2$ torsion fields may always be written as two times another field and so torsion fields are even and therefore equal to zero modulo 2.  Therefore in the classical supergravity limit the $Sq^3$ term vanishes and $d_3$ reduces to the operator in Eq.~(\ref{solutions}).  

\subsection{An example: The twisted K-theory of $S^3$} \label{wzwsec}

The supersymmetric SU(2) WZW model at affine level $k-2$ describes string theory on the three-sphere $S^3$ with $k\neq 0$ units of NS 3-form flux
\beq
\int_{S^3} H=k.
\eeq
The integer cohomology of the three-sphere is
\beq
\H^0(S^3)=\H^3(S^3)=\Z\hsp \H^1(S^3)=\H^2(S^3)=0
\eeq
where $0$ is the trivial group, which contains only the identity element.  In particular, the cohomology contains no torsion and so $Sq^3$ acts trivially.  As the maximum difference in the dimensions of two elements of the cohomology is equal to three, the operators $d_{2j+1}$ are trivial for $j>1$, leaving only $d_3$, which contains only the cup product with $H$.  

If $e_0$ is the generator of $\H^0(S^3)=\Z$ and $e_3$ is the generator of $\H^3(S^3)=\Z$ then
\beq
d_3 e_0=H\cup e_0 = k e_3\hsp d_3 e_3=H\cup e_3=0
\eeq
where $H$ kills $e_3$ because the cup product of two 3-classes is a 6-class, but the 6-cohomology is trivial.  Thus $d_3$ acts nontrivially on all of $\H^0(S^3)$, but annihilates all of $\H^3(S^3)$.  In other words, the kernel of $d_3$ is the third cohomology group, $\Z$.  Similarly the image of $d_3$ consists of those elements $k\Z\subset\Z$ of $\H^3$ which are multiples of $k$.  Note that both the kernel and the image lie in $\H^3$, which is in $\HO$ because $3$ is odd.  $\HE$, on the other hand, contains no elements of the kernel and so $\K^0_H(S^3)$ is trivial.  Summarizing, we have found that
\beq
\K^1_H(S^3)=\frac{\Ker(d_3:\HO(M)=\Z\rightarrow \HE(M)=\Z)}{\Im(d_{3}:\HE=\Z\rightarrow \HO=\Z)}=\frac{\H^3(M)=\Z}{k\H^3(M)=k\Z}=\Z_k. \label{wzw}
\eeq
As was shown in Ref.~\cite{FS}, the twisted K-groups (\ref{wzw}) reproduce the known symmetric D-branes in the supersymmetric SU(2) WZW model.
    
Physically the nontrivial element $j\in\Z_k=K^1_H(S^3)$ corresponds to a RR 3-form field strength $G_3$ such that $\int_{S^3}G_3=j$ in an embedding of the SU(2) WZW model into IIB string theory, for example on a 3-sphere linking an NS5-brane.  We recall that $G_3$ is related to the gauge-invariant field strength $F_3$ via
\beq
G_3=F_3-C_0H
\eeq
and so the large gauge transformation corresponds to $C_0\mapsto C_0+1$, which is the generator $T$ of the S-duality group $SL(2,\Z)$.

The Bianchi identity $G_0H=0$ implies that $G_0=0$ and so there is no braneless embedding of the SU(2) WZW model in massive IIA.  Instead such an embedding will require $k$ D6-branes intersecting the 3-sphere.  In the above NS5-brane realization, this reflects the familiar fact that an NS5-brane is confined by $k$ D6's when the Romans' mass is equal to $k$.

\section{Torsion corrections from the Freed-Witten anomaly} \label{fwsec}

\subsection{The Freed-Witten anomaly}

In Ref.~\cite{FW} Freed and Witten have demonstrated that global worldsheet anomaly cancellation dictates the condition
\beq
W_3+H=dF \label{fw}
\eeq
on the worldvolume of a D$p$-brane wrapping a compact cycle $N\subset M$ in the type II string theory spacetime $M$.  Here $W_3$ is the third Stiefel-Whitney class of the normal bundle of $N$ in $M$, which vanishes if $N$ is $spin^c$ as $M$ and $N$ are both orientable in type II if the D-brane is to carry a RR charge and $M$ is $spin^c$.  More generally $W_3$ is a $\Z_2$-valued class in the third integral cohomology of $N$.  $H$ is, as usual, the NSNS 3-form field strength pulled back to $N$, or rather the associated class in integral cohomology.  $dF$ is, as a differential form, the exterior derivative of the 2-form field strength of the D-brane worldvolume's $U(1)$ gauge field, which is the magnetic monopole charge and so is Poincar\'e dual, in $N$, to boundary of a D$(p-2)$ brane that ends on our D$p$-brane.  

All three terms in Eq.~(\ref{fw}) are integral cohomology classes, thus as usual only the free part of $dF$ may be locally interpreted as the derivative of a field strength 2-form.  In this context the magnetic monopole charge $dF$ is still Poincar\'e dual to a codimension 3 submanifold of $N$ on which a lower-dimensional D-brane ends, however it may be that this submanifold is homologically trivial when included into the spacetime $M$.  An example of such a phenomenon occurs in the SU(3) WZW model and is described in Ref.~\cite{MMS}.  In this instance one finds that, instead of a D$(p-2)$-brane ending on the D$p$, a D$(p-4)$-brane ends on the D$p$.  In general the differential $d_{2j+1}$ will give us information about D$(p-2j)$-brane insertions on the D$p$-brane.  The inserted branes extend outward from the D$p$-brane until either they hit another brane on which Eq.~(\ref{fw}) permits them to end, or else until they reach the end of the spacetime.  Such configurations of branes ending on branes are referred to as baryons in Ref.~\cite{bbads}, as in some examples they represent baryonic vertices in a dual conformal field theory.

\subsection{Quantum corrections to the Bianchi identities}

We are now ready to use the Freed-Witten anomaly to compute the torsion corrections to the BRST operator.  In particular, we need to know the quantum corrections to the constraints, which determine the action of the BRST operator on the physical fields, and we need to know the quantum corrections to the gauge transformations, which determine the action of of the BRST operator on the ghosts.

While we are searching for a classification of fluxes, the Freed-Witten anomaly is a condition on branes.  To convert a condition on branes into a condition on fluxes we use an argument based on Gauss' law that has appeared, for example, in Ref.~\cite{m2quant}.  As a differential form in de Rham cohomology, a field strength is determined by its integral over the cycles which represent various homology classes.  This notion is easily extended to the full integral cohomology by replacing integration with the homology/cohomology pairing.  Thus when we speak of a $p$-flux on a $p$-cycle, we will really be referring to the pairing of the corresponding cohomology and homology classes.  As has been argued in Ref.~\cite{MooreWitten}, the RR $p$-flux on a topologically trivial $p$-cycle measures the D$(8-p)$-brane charge that is linked by the trivial cycle.  In particular, D-brane charge in the full quantum theory is classified by integral homology, and so this pairing also includes torsion terms.  

Witten has argued in Ref.~\cite{m2quant} that even when a cycle is topologically nontrivial, the consistency conditions for the fluxes supported on the cycle, being local, cannot depend on whether at some far away place the cycle degenerates and the flux is sourced by a D-brane.  Thus the only consistent RR $p$-fluxes, even on noncontractible cycles, are supposedly those fluxes that could be sourced by a D($8-p$)-brane had the cycle been contractible.

The twisted K-theory classification only applies to fluxes in the absence of D-brane charges.  This is because D-brane charges shift the Bianchi identities and so shift the set of consistent fluxes away from twisted K-theory.  Thus the $p$-cycle on which we are measuring the $p$-flux must not intersect any D-branes.  However, if $W_3+H$ is nonzero on the worldvolume of the D$(8-p)$-brane that may source this flux, then there will be lower D-brane insertions ending on the D$(8-p)$-brane.  The other insertions may either end on other D-branes, whose total $H+W_3$ cancels that of our original brane, or they            may continue to the end of spacetime.  In the first case, the total $H+W_3$ vanishes, where this addition is defined after the various forms are pushed forward from the brane worldvolumes $N$ on to the spacetime $M$, alternately one may think of both D-branes as a single disconnected D-brane which is therefore subject to Freed-Witten.  In the second case, the semi-infinite inserted branes will intersect our $p$-cycle, no matter how distant the D($8-p$) is, and so invalidate our initial hypothesis that our spacetime in fact contains no branes.  Thus the Freed-Witten anomaly leads to quantum corrections, that is torsion corrections, to the Bianchi identity which impose that the $p$-flux on any cycle can be generated by a brane for which $W_3+H$ vanishes.

A necessary, but not sufficient, condition for the vanishing of $W_3+H$ on this D-brane worldvolume is that the $p$-flux be in the kernel of the AHSS differential
\beq
d_3=Sq^3+H\cup.
\eeq
To see this, we simplify matters by considering a spacetime which is homogenous in the radial direction from the D($8-p$), which is the direction followed by any brane insertions.  In particular if one is classifying fluxes on time slices then one may identify this radial direction with time, as in Ref.~\cite{MMS}.  On each 9-dimensional radial slice Gauss' law then ensures that the $p$-flux is Poincar\'e dual to the $(9-p)$-dimensional D($8-p$)-brane worldvolume.  The $H$ flux in the Freed-Witten anomaly (\ref{fw}) is the pullback of the $H$ flux in the bulk spacetime, thus its pushforward back to the bulk reproduces the $H$ term in $d_3$.  $Sq^3$ on the other hand is defined to be the pushforward of the third Stiefel-Whitney class, $W_3$, of the normal bundle of the cycle Poincar\'e dual to our $p$-flux $G_p$, which is precisely our D-brane worldvolume.  Thus the image of $d_3G_p$ is precisely the pushforward of $W_3+H$ under the inclusion $i:N\hookrightarrow M$, which must vanish if $W_3+H$ is indeed zero as the pushforward map is linear
\beq
d_3G_ p=Sq^3G_p+H\cup G_p=i_*(W_3+H)=0.
\eeq

\section* {Acknowledgement}

I would like to thank Allan Adams, Glenn Barnich, Nazim Bouatta, Christopher Douglas, Michael Douglas, Marc Henneaux, Mike Hill, Stanislav Kuperstein and even Daniel Persson for speaking with me.  I would like to thank the organizers of Problemi Attuali di Fisica Teorica at Vietri sul Mare for inviting me to give this talk, as it motivated me to finally write this up after four years. 

My work is partially supported by IISN - Belgium (convention 4.4505.86), by the ``Interuniversity Attraction Poles Programme -- Belgian Science Policy'' and by the European Commission programme MRTN-CT-2004-005104, in which I am associated to V. U. Brussel.

\end{document}

\bibitem{}
,
{\it},
[{\tt arXiv:hep-th/}].

\bibitem{Manjarin}
J.~J.~Manjarin,
{\it Topics on D-brane Charges with B-fields},
[{\tt arXiv:hep-th/0405074}].

\bibitem{Baryons}
E.~Witten,
{\it Baryons and Branes in Anti de Sitter Space},
[{\tt arXiv:hep-th/9805112}].

\bibitem{MMS}
J.~Maldacena, G.~Moore and N.~Seiberg,
{\it D-Brane Instantons and K-Theory Charges},
%J. High Energy Phys. {\bf 11} (2001) 062,
[{\tt arXiv:hep-th/0108100}].

\bibitem{Feb}
J.~Evslin,
{\it Twisted K-Theory from Monodromies},
%J. High Energy Phys. {\bf 05} (2003) 030, 
[{\tt arXiv:hep-th/0302081}].

\bibitem{KS}
I.~R.~Klebanov and M.~J.~Strassler,
{\it Supergravity and a Confining Gauge Theory: Duality Cascades and $\chi$SB-Resolution of Naked Singularities},
[{\tt arXiv:hep-th/0007191}].

\bibitem{Nov}
J.~Evslin,
{\it IIB Soliton Spectra with All Fluxes Actived},
[{\tt arXiv:hep-th/0211172}].

\bibitem{Ftheory}
C.~Vafa,
{\it Evidence for F-Theory},
%Nucl. Phys. {\bf B469} (1996) 403-418, \newline
[{\tt arXiv:hep-th/9602022}].

\bibitem{Oscar}
O.~Loaiza-Brito,
{\it Instantonic Branes, Atiyah-Hirzebruch Spectral Sequence, and SL(2,Z) Duality of N=4 SYM},
[{\tt arXiv:hep-th/0311028}].

\bibitem{KPV}
S.~Kachru, J.~Pearson and H.~Verlinde,
{\it Brane/Flux Annihilation and the String Dual of a Non-Supersymmetric Field Theory},
[{\tt arXiv:hep-th/0112197}].

\bibitem{BCMMS}
P. Bouwknegt, A. Carey, V. Mathai, M. Murray and D. Stevenson,
{\it Twisted K-theory and K-theory of bundle gerbes},
Comm. Math. Phys. {\bf 228} (2002) 17-45,
[{\tt arXiv:hep-th/0106194}].

\bibitem{MathStev}
V.~Mathai and D.~Stevenson,
{\it On a Generalized Connes-Hochschild-Kostant-Rosenberg Theorem},
[{\tt arXiv:hep-th/0404329}].

\bibitem{Marolf}
D.~Marolf,
{\it Chern-Simons Terms and the Three Notions of Charge},
[{\tt arXiv:hep-th/0006117}].

%\bibitem{TwistChern}
%V.~Mathai and D.~Stevenson,
%,
%{\it Chern Character in Twisted K-Theory: Equivariant and Holomorphic Cases},
%[{\tt arXiv:hep-th/0201010}].

\bibitem{PetrIIA}
P.~ Ho$\check{\textup{r}}$ava,
{\it Type IIA D-Branes, K-Theory and Matrix Theory},
[{\tt arXiv:hep-th/9812135}].

\bibitem{MN}
J.~M.~Maldacena and C.~Nunez,
{\it Towards the large N limit of pure N=1 super Yang Mills},
[{\tt arXiv:hep-th/0008001}].

\bibitem{HanWit}
A.~ Hanany and E.~Witten,
{\it Type \twob\ Superstrings, BPS Monopoles and Three-Dimensional Gauge Dynamics},
[{\tt arXiv:hep-th/9611230}].

\bibitem{WittenMQCD}
E.~Witten,
{\it Solutions of Four-Dimensional Field Theories Via M-Theory},
[{\tt arXiv:hep-th/9703166}].

\bibitem{Hitoshi}
J.~Evslin, H.~Murayama, U.~Varadarajan and J.~.E.~Wang,
{\it Dial M for Flavor Symmetry Breaking},
[{\tt arXiv:hep-th/0107072}].

\bibitem{SeibergDuality}
N.~Seiberg,
{\it Electric-Magnetic Duality in Supersymmetric Non-Abelian Gauge Theories},
[{\tt arXiv:hep-th/9411149}].

\bibitem{Ken}
G.~Carlino, K.~Konishi and H.~Murayama,
{\it Dynamics of Supersymmetric $SU(n_c)$ and $USp(2n_c)$ Gauge Theories},
[{\tt arXiv:hep-th/0001036}].

\bibitem{OT}
K.~Oh and R.~Tatar,
{\it Duality and Confinement in N=1 Supersymmetric Theories from Geometric Transitions},
[{\tt arXiv:hep-th/0112040}].
%OhTatar 0112040

\bibitem{AtiyahWitten}
M.~Atiyah and E.~Witten,
{\it M-Theory Dynamics on a Manifold of $G_2$ Holonomy},
[{\tt arXiv:hep-th/0107177}].

\bibitem{AD}
P.~C.~Argyres and M.~R.~Douglas,
{\it New Phenomena in SU(3) Supersymmetric Gauge Theory},
[{\tt arXiv:hep-th/9505062}].

\bibitem{KN}
I.~R.~Klebanov and N.~Nekrasov,
{\it Gravity Duals of Fractional Branes and Logarithmic RG Flow},
[{\tt arXiv:hep-th/9911096}].

\bibitem{BT}
J.~D.~Brown and C.~Teitelboim,
{\it Neutralization of the Cosmological Constant by Membrane Creation},
Nucl. Phys. {\bf B297}, 787, (1988).

\bibitem{bion}
N.~R.~Constable, R.~C.~Myers and O.~Tafjord,
{\it The Noncommutative Bion Core},
[{\tt arXiv:hep-th/9911136}].

\bibitem{Gibbons}
G.~W.~Gibbons,
{\it Branes as BIons},
[{\tt arXiv:hep-th/9803203}].

\bibitem{MW}
G.~Moore and E.~Witten,
{\it Self duality, Ramond-Ramond fields, and K-theory},
J. High Energy Phys. {\bf 05} (2000) 032,
[{\tt arXiv:hep-th/9912279}].

\bibitem{Townsend}
M. B. Green, C. M. Hull and P. K. Townsend,
{\it D-brane Wess-Zumino Actions, T-duality and the Cosmological Constant},
[{\tt arXiv:hep-th/9604119}].

\bibitem{VanProeyen}
E. Bergshoeff, R. Kallosh, T. Ortin, D. Roest and A. Van Proeyen,
{\it New Formulations of D=10 Supersymmetry and D8-O8 Domain Walls},
[{\tt arXiv:hep-th/0103233}].

\end{thebibliography}
% =========================================================================
\end{document}

\bibitem{}
,
{\it},
[{\tt arXiv:hep-th/}].

\bibitem{}
,
{\it},
[{\tt arXiv:hep-th/}].

\bibitem{NdW}
B. de Wit and H. Nicolai, Phys.\ Lett.\ {\bf 155B}, 47 (1985); Nucl. Phys. \textbf{B274}, 363 (1986).

\bibitem{Nic}
H. Nicolai, Phys.\ Lett.\ {\bf 187B}, 363 (1987). 

\bibitem{HW}
P.~ Ho$\check{\textup{r}}$ava and E.~ Witten,
{\sl Heterotic and Type I String Dynamics from Eleven Dimensions},
Nucl. Phys. {\bf B460}, 506, (1996), 
[{\tt arXiv:hep-th/9510209}].

\bibitem{FH}
M. Fabinger and P.~ Ho$\check{\textup{r}}$ava
{\it Casimir Effect between World-Branes in Heterotic M-Theory},
Nucl Phys. {\bf B580}, 243, (2000), 
[{\tt arXiv:hep-th/0002073}].

\bibitem{BEM}
P. Bouwknegt, J.~Evslin, and V. Mathai, 
{\it T-duality: Topology Change from H flux}, 
[{\tt arXiv:hep-th/0306062}].

\bibitem{FluxQuant}
E.~ Witten, {\sl On Flux Quantization in M-Theory and the Effective 
Action},
J. Geom. Phys. {\bf 22},1 , (1997), 
[{\tt arXiv:hep-th/9609122}].

\bibitem{GM}
C.~ Gomez, J.~ J.~ Manjarin,
{\it Dyons, K-theory and M-theory}, 
[{\tt arXiv:hep-th/0111169}].

\bibitem{allan}
A.~Adams and J.~Evslin, {\it The Loop Group of $E_8$ and K-Theory from $11d$},
JHEP {\bf 02}, 29 (2003), 
[{\tt arXiv:hep-th/0203218}]. 

\bibitem{Morrison}
D.~R.~Morrison, 
{\it "Half $K3$ surfaces"}
talk at Strings 2002, Cambridge,
\\
http://www.damtp.cam.ac.uk/strings02/avt/morrison/

\bibitem{MS}
G.~Moore and N.~Saulina,
{\it T-duality, and the K-theoretic partition function of Type \twoa
superstring theory},
[{\tt arXiv:hep-th/0206092}].

\bibitem{Kentaro}
K.~Hori
{\sl Consistency Conditions for Five-Brane Theory in M-Theory on $R^5/\Z_2$ Orbifold},
Nucl. Phys. {\bf B539}, 35, (1999), hep-th/9805141.

\bibitem{slow}
A.~ Adams, J.~ Evslin, and U.~ Varadarajan,
(To Appear).

\bibitem{Stong}
R.~Stong,
{\it Calculation of $\Omega_{11}^{\textup{spin}}({\mathbf{K}(\Z,4))})$},
in Unified String Theories, eds. M.~B.~Green and D.~J.~Gross, World 
Scientific, 1986.

\bibitem{Horava}
P.~Ho\v rava, 
(Unpublished).

\bibitem{Hull}
C.~ M.~ Hull,
{\it Massive String Theories from M-Theory and F-Theory},
JHEP {\bf 9811} (1998) 027, 
[{\tt arXiv:hep-th/9811021}].

\bibitem{Sethi}
S.~ Sethi,
(Unpublished).

\bibitem{Manjarin}
C.~ Gomez and J. J. Manjarin,
(To Appear).

\bibitem{KentaroIndexTDuality}
K.~Hori, {\it D-branes, T-duality, and index theory},
Adv. Theor. Math. Phys. {\bf 3} (1999) 281-342,
[{\tt arXiv:hep-th/9902102}].

\bibitem{AABL}
E.~\'Alvarez, L.~\'Alvarez-Gaum\'e, J.L.F.~Barb\'on and Y.~Lozano,
{\it Some global aspects of duality in string theory},
Nucl. Phys. {\bf B415} (1994) 71-100,
[{\tt arXiv:hep-th/9309039}].

\bibitem{RV}
M.~Ro\v cek and E.~Verlinde,
{\it Duality, quotients, and currents},
Nucl. Phys. {\bf 373} (1992) 630-646,
[{\tt arXiv:hep-th/9110053}].

\end{thebibliography}
% =========================================================================

\begin{thebibliography}{23}
%%%%%%%%%%%%%%%%%%%%%%%%%%%%%%%%

\bibitem{MM}
R.~Minasian and G.~Moore,
{\it K-theory and Ramond-Ramond charge},
%J. High Energy Phys. {\bf 11} (1997) 002,
[{\tt arXiv:hep-th/9710230}].

\bibitem{WittenK}
E.~Witten,
{\it D-Branes and K-Theory},
%JHEP {\bf 12} (1998) 019, \newline
[{\tt arXiv:hep-th/9810188}].

\bibitem{Sen}
A. Sen,
{\it Tachyon Condensation on the Brane Anti-Brane System},
%JHEP {\bf 9808} (1998) 012,
[{\tt arXiv:hep-th/9805170}].


\bibitem{Kapustin}
A.~Kapustin,
{\it D-Branes in a Topologically Nontrivial B-Field},
[{\tt arXiv:hep-th/9909089}].

\bibitem{MW}
G. Moore and E. Witten,
{\it Self-Duality, Ramond-Ramond Fields, and K-Theory},
[{\tt arXiv:hep-th/9912279}].

\bibitem{DMW}
E.~ Diaconescu, G.~ Moore and E.~ Witten,
{\it $E_8$ Gauge Theory, and a Derivation of K-Theory from M-Theory},
%Adv. Theor. Math. Phys. {\bf 6} (2003) 1031,
[{\tt arXiv:hep-th/0005090}].

\bibitem{MMS}
J.~Maldacena, G.~Moore and N.~Seiberg,
{\it D-Brane Instantons and K-Theory Charges},
%JHEP {\bf 11} (2001) 062,
[{\tt arXiv:hep-th/0108100}].

\bibitem{BM}
P. Bouwknegt and V. Mathai,
{\it D-branes, B fields and twisted K theory},
%JHEP {\bf 0003} (2000) 007,
[{\tt arXiv:hep-th/0002023}].

\bibitem{Townsend}
P. K. Townsend,
{\it p-Brane Democracy},
[{\tt arXiv:hep-th/9507048}].

\bibitem{GHT}
M. B. Green, C. M. Hull and P. K. Townsend,
{\it D-Brane Wess-Zumino Actions, T-Duality and the Cosmological Constant},
[{\tt arXiv:hep-th/9604119}].


\bibitem{VanProeyen}
E. Bergshoeff, R. Kallosh, T. Ortin, D. Roest and A. Van Proeyen,
{\it New Formulations of D=10 Supersymmetry and D8-O8 Domain Walls},
[{\tt arXiv:hep-th/0103233}].


\bibitem{MS}
G. W. Moore and G. Segal,
{\it D-Branes and K-Theory in 2D Topological Field Theory},
[{\tt arXiv:hep-th/0609042}].

\bibitem{BBA}
E.~Witten,
{\it Baryons and Branes in Anti de Sitter Space},
[{\tt arXiv:hep-th/9805112}].


\bibitem{FW}
D.~S.~Freed and E.~Witten,
{\it Anomalies in String Theory with D-Branes},
%Asian J. Math. {\bf 3} (1999) 819,
[{\tt arXiv:hep-th/9907189}].

\bibitem{DS}
M. J. Douglas and S. H. Shenker,
{\it Dynamics of SU(N) Supersymmetric Gauge Theory},
[{\tt arXiv:hep-th/9503163}].

\bibitem{BHK}
C. Bohr, B. Hanke, and D. Kotschick,
{\it Cycles, submanifolds, and structures on normal bundles},
%Manuscripta Math. {\bf 108} (2002) 483, 
[{\tt arXiv:math.GT/0011178}].


\bibitem{Szabo}
R. M. G. Reis and R. J. Szabo,
{\it Geometric K-Homology of Flat D-Branes},
[{\tt arXiv:hep-th/0507043}].


\bibitem{ES}
J. Evslin and H. Sati,
{\it Can D-Branes Wrap Nonrepresentable Cycles?}
[{\tt arXiv:hep-th/0607045}].

\bibitem{BDS}
C.~Bachas, M.~Douglas and C.~Schweigert,
{\it Flux Stabilization of D-Branes},
[{\tt arXiv:hep-th/0003037}].

\bibitem{Wati}
W.~ Taylor,
{\it D-Branes in B Fields},
[{\tt arXiv:hep-th/0004141}].

\bibitem{GHM}
M. Green, J. A. Harvey and G. Moore,
{\it I-Brane Inflow and Anomalous Couplings on D-Branes},
[{\tt arXiv:hep-th/9605033}].


\bibitem{APS}
A.~Adams, J. Polchinski and E. Silverstein,
{\it Don't Panic! Closed String Tachyons in ALE Spacetimes},
[{\tt arXiv:hep-th/0108075}].

\bibitem{FH}
M. Fabinger and P.~ Ho$\check{\textup{r}}$ava
{\it Casimir Effect between World-Branes in Heterotic M-Theory},
%Nucl Phys. {\bf B580}, 243, (2000), 
[{\tt arXiv:hep-th/0002073}].

\bibitem{Petr}
P. Ho{\v{r}}ava,
{\it Type IIA D-Branes, K-Theory, and Matrix Theory},
%Adv. Theor. Math. Phys. {\bf 2} (1999) 1373,
[{\tt arXiv:hep-th/9812135}].

\bibitem{BRST}
J. Evslin, {\it  Twisted K-Theory as a BRST Cohomology},
[{\tt arXiv:hep-th/0605049}].

\bibitem{aussy2005}
P. Bouwknegt, J. Evslin, B. Jurco, V. Mathai and H. Sati,
{\it Flux Compactifications on Projective Spaces and the S-Duality Puzzle},
%Adv. Theor. Math. Phys. {\bf 10} (2006) 345,
[{\tt arXiv:hep-th/0501110}].

\bibitem{Thom}
R. Thom,
{\it Quelques propri\'et\'es globales des vari\'et\'es diff\'erentiables},
Comment. Math. Helv. {\bf 28} (1954) 17.

\bibitem{Sull}
D. Sullivan,
{\it Ren\'e Thom's work on geometric homology and bordism},
Bull. Amer. Math. Soc. (N.S.) {\bf 41} (2004) 341.

\bibitem{Myers}
R.~ Myers, {\it Dielectric-Branes}, 
%JHEP {\bf 9912}, 22 (1999), 
[{\tt arXiv:hep-th/9910053}].


\bibitem{AS}
M. Atiyah and G. Segal,
{\it Twisted K-theory and Cohomology},
[{\tt arXiv:math.KT/0510674}].

\bibitem{Uday}
J.~Evslin and U.~Varadarajan, 
{\it K-Theory and S-Duality: Starting over from Square 3},
J. High Energy Phys. {\bf 03} (2003) 026, 
[{\tt arXiv:hep-th/0112084}].

\bibitem{CavalcantiFormale}
G. R. Cavalcanti,
{\it Formality in Generalized Kahler Geometry},
[{\tt arXiv:math.DG/0603596}].

\bibitem{Cavalcanti}
G. R. Cavalcanti,
{\it New Aspects of the $dd^c$ Lemma},
[{\tt arXiv:math.DG/0501406}].

\bibitem{Luca1}
L. Martucci and P. Smyth,
{\it Supersymmetric D-branes and Calibrations on General $N=1$ Backgrounds},
[{\tt arXiv:hep-th/0507099}].

\bibitem{Luca2}
L. Martucci,
{\it D-Branes on Generalized $N=1$ Backgrounds: Superpotentials and D-terms},
[{\tt arXiv:hep-th/0602129}].

\bibitem{Gualtieri}
M. Gualtieri,
{\it Generalized Complex Geometry},
[{\tt arXiv:math.DG/0401221}].

\bibitem{WitFluxQuant}
E. Witten,
{\it On Flux Quantization in M-Theory and the Effective Action},
[{\tt arXiv:hep-th/9609122}].

\bibitem{MEK}
J. Evslin,
{\it From $E_8$ to F via T},
[{\tt arXiv:hep-th/0311235}].

\bibitem{MEMONO}
J.~ Evslin,
{\it Twisted K-Theory from Monodromies},
[{\tt arXiv:hep-th/0302081}].

\bibitem{Allan}
A.~Adams and J.~Evslin, 
{\it The Loop Group of $E_8$ and K-Theory from $11d$},
%JHEP {\bf 02}, 29 (2003), 
[{\tt arXiv:hep-th/0203218}]. 

\bibitem{Harvey}
J. A. Harvey,
{\it Topology of the Gauge Group in Noncommutative Gauge Theory},
[{\tt arXiv:hep-th/0105242}].


\bibitem{Carlo}
C. Maccaferri,
{\it Chan-Paton Factors and Higgsing from Vacuum String Field Theory},
[{\tt arXiv:hep-th/0506213}]. 


\bibitem{Freed}
D. S. Freed,
{\it Dirac Charge Quantization and Generalized Differential Cohomology},
[{\tt arXiv:hep-th/0011220}].

\bibitem{AD}
P. S. Aspinwall and R. Y. Donagi,
{\it The Heterotic String, the Tangent Bundle, and Derived Categories},
[{\tt arXiv:hep-th/9806094}].

\bibitem{Aspinwall}
P. S. Aspinwall,
{\it D-Branes on Calabi-Yau Manifolds},
[{\tt arXiv:hep-th/0403166}].

\bibitem{FS}
S. Fredenhagen and V. Schomerus,
{\it Branes on Group Manifolds, Gluon Condensates and Twisted K-Theory},
[{\tt arXiv:hep-th/0012164}].

\bibitem{AS2}
A. Yu. Alekseev and V. Schomerus,
{\it D-branes in the WZW Model},
[{\tt arXiv:hep-th/9812193}].

\bibitem{ABE}
R. Auzzi, S. Bolognesi and J. Evslin,
{\it Monopoles Can be Confined by 0, 1 or 2 Vortices},
[{\tt arXiv:hep-th/0411074}].

\bibitem{WittenMQCD2}
E. Witten,
{\it Branes and the Dynamics of QCD},
[{\tt arXiv:hep-th/9706109}].

\bibitem{Stefano}
S. Bolognesi and J. Evslin,
{\it Stable vs Unstable Vortices in SQCD},
[{\tt arXiv:hep-th/0506174}].

\bibitem{WitTop}
E. Witten,
{\it Duality Relations Among Topological Effects in String Theory},
[{\tt arXiv:hep-th/9912086}].

\bibitem{HW}
P.~ Ho$\check{\textup{r}}$ava and E.~ Witten,
{\it Heterotic and Type I String Dynamics from Eleven Dimensions},
Nucl. Phys. {\bf B460}, 506, (1996), 
[{\tt arXiv:hep-th/9510209}].

\bibitem{SW}
N. Seiberg and E. Witten,
{\it String Theory and Noncommutative Geometry},
[{\tt arXiv:hep-th/9908142}].

\bibitem{BEM}
P. Bouwknegt, J. Evslin and V. Mathai,
{\it T-Duality: Topology Change from $H$-Flux},
[{\tt arXiv:hep-th/0306062}].

\bibitem{Hisham}
I. Kriz and H. Sati,
{\it M Theory, Type IIA Superstrings, and Elliptic Cohomology},
[{\tt arXiv:hep-th/0404013}].

\bibitem{Hisham2}
I. Kriz and H. Sati,
{\it Type IIB String Theory, S-Duality, and Generalized Cohomology},
[{\tt arXiv:hep-th/0410293}].

\bibitem{Hisham3}
I. Kriz and H. Sati,
{\it Type II String Theory and Modularity},
[{\tt arXiv:hep-th/0501060}].

\bibitem{Hisham4}
H. Sati,
{\it Flux Quantization and the M-Theoretic Characters},
[{\tt arXiv:hep-th/0507106}].

\bibitem{Hisham5}
I. Kriz,
{\it Toward a derivation of E-theory from F-theory},
[{\tt arXiv:hep-th/0511011}].

\bibitem{Hisham6}
H. Sati,
{\it The Elliptic Curves in Gauge Theory, String Theory, and Cohomology},
[{\tt arXiv:hep-th/0511087}].

\bibitem{DFM}
E.~ Diaconescu, D.~ Freed, and G.~ Moore,
{\it The M-Theory 3-form and $E_8$ Gauge Theory},
[{\tt arXiv:hep-th/0312069}].

\bibitem{Uranga}
A. M. Uranga,
{\it D-Brane Probes, RR Tadpole Cancellation and K-Theory Charge},
[{\tt arXiv:hep-th/0011048}].

\bibitem{Buscher}
T. H. Buscher,
{\it Path Integral Derivation of Quantum Duality in Nonlinear Sigma Models},
Phys. Lett. {\bf B201} (1988) 466.

\bibitem{AABL}
E. Alvarez, L. Alvarez-Gaume, J. L. F. Barbon and Y. Lozano,
{\it Some Global Aspects of Duality in String Theory},
[{\tt arXiv:hep-th/9309039}].  %Palindrome

\bibitem{KSTT}
S. Kachru, M. B. Schulz, P. K. Tripathy and S. P. Trivedi,
{\it New Supersymmetric String Compactifications},
[{\tt arXiv:hep-th/0211182}].

\bibitem{RR}
I.~Raeburn and J. Rosenberg,
{\it Crossed products of continuous-trace $C^*$-algebras by
smooth actions}, 
Trans. Amer. Math. Soc. {\bf 305} (1988) 1-45.

\bibitem{KW}
I. R. Klebanov and E. Witten,
{\it Superconformal Field Theory on Threebranes at a Calabi-Yau Singularity},
[{\tt arXiv:hep-th/9807080}].

\bibitem{GK}
S. S. Gubser and I. R. Klebanov,
{\it Baryons and Domain Walls in an $N$=1 Superconformal Gauge Theory},
[{\tt arXiv:hep-th/9808075}].

\bibitem{MN}
J. M. Maldacena and C. Nunez,
{\it Towards the Large $N$ Limit of Pure $N=1$ Super Yang-Mills},
[{\tt arXiv:hep-th/0008001}].

\bibitem{HanWit}
A.~ Hanany and E.~Witten,
{\it Type IIB Superstrings, BPS Monopoles and Three-Dimensional Gauge Dynamics},
[{\tt arXiv:hep-th/9611230}].

\bibitem{KS}
I. R. Klebanov and M. J. Strassler,
{\it Supergravity and a Confining Gauge Theory: Duality Cascades and $\chi$SB-Resolution of Naked Singularities},
[{\tt arXiv:hep-th/0007191}].

\end{thebibliography}

\begin{thebibliography}{23}
%%%%%%%%%%%%%%%%%%%%%%%%%%%%%%%%
\bibitem{MM}
R.~Minasian and G.~Moore,
{\it K-theory and Ramond-Ramond charge},
JHEP {\bf 11} (1997) 002,
[{\tt arXiv:hep-th/9710230}].

\bibitem{WittenK}
E.~Witten,
{\it D-Branes and K-Theory},
JHEP {\bf 12} (1998) 019, \newline
[{\tt arXiv:hep-th/9810188}].

\bibitem{BHK}
C. Bohr, B. Hanke, and D. Kotschick,
{\it Cycles, submanifolds, and structures on normal bundles},
Manuscripta Math. {\bf 108} (2002) 483, [{\tt arXiv:math.GT/0011178}].

\bibitem{Bryan}
A. L. Carey and B. Wang,
{\it Fusion of Symmetric D-branes and Verlinde Rings}, 
[{\tt arXiv:math-ph/0505040}].



\bibitem{Hisham}
H. Sati, {\it unpublished}.

\bibitem{KS1}
I.~Kriz and H.~Sati, {\it M Theory, type IIA superstrings, and
elliptic cohomology}, Adv. Theor. Math. Phys. {\bf 8} (2004) 345,
[{\tt arXiv:hep-th/0404013}].

\bibitem{Marchesano}
F. Marchesano,
{\it D6-Branes and Torsion},
JHEP {\bf 0605} (2006) 019,
[{\tt arXiv:hep-th/0603210}].

\end{thebibliography}

\begin{thebibliography}{23}

\bibitem{FS}
S. Fredenhagen and V.~Schomerus,
{\it Branes on Group Manifolds, Gluon Condensates and Twisted K-Theory},
[{\tt arXiv:hep-th/0012164}].

\bibitem{WittenK}
E.~Witten,
{\it D-Branes and K-Theory},
%J. High Energy Phys. {\bf 12} (1998) 019, \newline
[{\tt arXiv:hep-th/9810188}].

\bibitem{FW}
D.~S.~Freed and E.~Witten,
{\it Anomalies in String Theory with D-Branes},
[{\tt arXiv:hep-th/9907189}].

\bibitem{DMW}
E.~ Diaconescu, G.~ Moore and E.~ Witten,
{\it $E_8$ Gauge Theory, and a Derivation of K-Theory from M-Theory}, 
[{\tt arXiv:hep-th/0005090}].

\bibitem{Sen}
A.~Sen,
{\it Tachyon Condensation on the Brane Antibrane System},
[{\tt arXiv:hep-th/9805170}].

\bibitem{me}
J.~Evslin and U.~Varadarajan, {\it K-Theory and S-Duality: Starting over from Square 3},
%J. High Energy Phys. {\bf 03} (2003) 026, 
[{\tt arXiv:hep-th/0112084}].
J.~ Evslin,
{\it IIB Soliton Spectrum with All Fluxes Activated},
[{\tt arXiv:hep-th/0211172}].
J.~ Evslin,
{\it Twisted K-Theory from Monodromies},
[{\tt arXiv:hep-th/0302081}].

\bibitem{DFM}
E.~ Diaconescu, D.~ Freed, and G.~ Moore,
{\it The M-Theory 3-form and $E_8$ Gauge Theory},
[{\tt arXiv:hep-th/0312069}].

\bibitem{aussy2005}
P. Bouwknegt, J. Evslin, B. Jurco, V. Mathai and H. Sati,
{\it Flux Compactifications on Projective Spaces and the S-Duality Puzzle},
[{\tt arXiv:hep-th/0501110}].

\bibitem{Hisham}
I. Kriz and H. Sati,
{\it M Theory, Type IIA Superstrings, and Elliptic Cohomology},
[{\tt arXiv:hep-th/0404013}].
I. Kriz and H. Sati,
{\it Type IIB String Theory, S-Duality, and Generalized Cohomology},
[{\tt arXiv:hep-th/0410293}].
I. Kriz and H. Sati,
{\it Type II String Theory and Modularity},
[{\tt arXiv:hep-th/0501060}].
H. Sati,
{\it Flux Quantization and the M-Theoretic Characters},
[{\tt arXiv:hep-th/0507106}].
I. Kriz,
{\it Toward a derivation of E-theory from F-theory},
[{\tt arXiv:hep-th/0511011}].
H. Sati,
{\it The Elliptic Curves in Gauge Theory, String Theory, and Cohomology},
[{\tt arXiv:hep-th/0511087}].


\bibitem{mecascade}
J. Evslin,
{\it The Cascade is a MMS Instanton},
[{\tt arXiv:hep-th/0405210}].

\bibitem{strings2000notes}
E. Witten,
{\it Overview of K-theory Applied to Strings},
[{\tt arXiv:hep-th/0007175}].

\bibitem{Freed}
D. Freed,
{\it K-Theory in Quantum Field Theory},
[{\tt arXiv:hep-th/0206031}].

\bibitem{EF}
T. Eguchi and P. Freund,
{\it Quantum Gravity and World Topology},
Phys. Rev. Lett. {\bf 37} (1976) 1251.

\bibitem{MMS}
J.~Maldacena, G.~Moore and N.~Seiberg,
{\it D-Brane Instantons and K-Theory Charges},
%J. High Energy Phys. {\bf 11} (2001) 062,
[{\tt arXiv:hep-th/0108100}].


\bibitem{T}
P. K. Townsend,
{\it p-Brane Democracy},
[{\tt arXiv:hep-th/9507048}].

\bibitem{Townsend}
M. B. Green, C. M. Hull and P. K. Townsend,
{\it D-brane Wess-Zumino Actions, T-duality and the Cosmological Constant},
[{\tt arXiv:hep-th/9604119}].

\bibitem{VanProeyen}
E. Bergshoeff, R. Kallosh, T. Ortin, D. Roest and A. Van Proeyen,
{\it New Formulations of D=10 Supersymmetry and D8-O8 Domain Walls},
[{\tt arXiv:hep-th/0103233}].

\bibitem{Alvarez}
O.~Alvarez,
{\it Topological Quantization and Cohomology}
Commun. Math. Phys. {\bf 100} (1985) 279.

\bibitem{AS}
M. Atiyah and G. Segal,
{\it Twisted K-theory and Cohomology},
[{\tt arXiv:math.KT/0510674}].

\bibitem{bundlegerbes}
P. Bouwknegt, A. L. Carey, V. Mathai, M. K. Murray and D. Stevenson,
{\it Twisted K-Theory and K-Theory of Bundle Gerbes},
[{\tt arXiv:hep-th/0106194}].

\bibitem{m2quant}
E. Witten,
{\it On Flux Quantization in M-Theory and the Effective Action},
[{\tt arXiv:hep-th/9609122}].

\bibitem{Chris}
Mark Behrens, Chris Douglas, and Mike Hill, private communication.

\bibitem{tfs}
J. de Boer, R. Dijkgraaf, K. Hori, A.Keurentjes, J. Morgan, D. R. Morrison and S. Sethi,
{\it Triples, Fluxes, and Strings},
[{\tt arXiv:hep-th/0103170}].

\bibitem{FluxQuant}
E.~ Witten, {\sl On Flux Quantization in M-Theory and the Effective 
Action},
[{\tt arXiv:hep-th/9609122}].


\end{thebibliography}
\end{document}